\documentclass[structabstract]{aa}  
\usepackage{graphicx}
\usepackage{txfonts}
\usepackage{natbib}
\bibpunct[]{(}{)}{;}{a}{}{,}
\begin{document}
   \title{Will the starless cores in Chamaeleon~I and III turn prestellar?
   \thanks{Based on observations carried out with the Atacama Pathfinder 
   Experiment telescope (APEX). APEX is a collaboration between the 
   Max-Planck Institut f\"ur Radioastronomie, the European Southern 
   Observatory, and the Onsala Space Observatory.}$^,$
   \thanks{The FITS file of Fig.~\ref{f:labocamap} is available in electronic
   form at the CDS via anonymous ftp to cdsarc.u-strasbg.fr (130.79.128.5).}
   }
   %\subtitle{}

   \author{A. Belloche
          \inst{1}
          \and
          B. Parise
          \inst{1}
          \and
          F. Schuller
          \inst{1}
          \and
          Ph. Andr{\'e}
          \inst{2}
          \and
          S. Bontemps
          \inst{3}
          \and
          K.~M. Menten
          \inst{1}
          }

   \institute{Max-Planck Institut f\"ur Radioastronomie, Auf dem H\"ugel 69,
              53121 Bonn, Germany\\
              \email{belloche@mpifr-bonn.mpg.de}
         \and
             {Laboratoire AIM, CEA/DSM-CNRS-Universit{\'e} Paris Diderot, 
              IRFU/Service d'Astrophysique, CEA Saclay, 91191 Gif-sur-Yvette, 
              France}
         \and
             {Universit{\'e} de Bordeaux, Laboratoire d'Astrophysique 
              de Bordeaux, CNRS/INSU, UMR 5804, 
              BP 89, 33271 Floirac cedex, France}
             }

   \date{Received 17 May 2011; accepted 16 June 2011}

% \abstract{}{}{}{}{} 
% 5 {} token are mandatory
 
  \abstract
   {The nearby Chamaeleon molecular cloud complex is a good laboratory to 
   study the process of low-mass star formation since it consists of three 
   clouds with very different properties. Chamaeleon~III does not show any 
   sign of star formation, while star formation has been very active in 
   Chamaeleon~I and may already be finishing.}
   {Our goal is to determine whether star formation can proceed in Cha~III by 
   searching for prestellar cores, and to compare the results to our recent
   survey of Cha~I.}
   {We used the Large APEX Bolometer Array (LABOCA) to map Cha~III in dust 
   continuum emission at 870~$\mu$m. The map is compared with a 2MASS
   extinction map and decomposed with a multiresolution algorithm. 
   The extracted sources are analysed by carefully taking into account the 
   spatial filtering inherent in the data reduction process.}
   {29 sources are extracted from the 870~$\mu$m map, all of them being 
   starless. The estimated 90$\%$ completeness limit is 0.18~M$_\odot$. The 
   starless cores are found down to a visual extinction of 1.9~mag, in marked
   contrast with other molecular clouds, including Cha~I. Apart from this
   difference, the Cha~III starless cores share very similar properties with 
   those found in Cha~I. They are less dense than those detected in continuum 
   emission in other clouds by a factor of a few. At most two sources ($< 7\%$) 
   have a mass larger than the critical Bonnor-Ebert mass, which suggests that
   the fraction of prestellar cores is very low, even lower than in Cha~I 
   ($< 17\%$). Only the most massive sources are candidate prestellar cores,
   in agreement with the correlation found earlier in the Pipe nebula. The mass 
   distribution of the 85 starless cores of Cha~I and III that are not 
   candidate prestellar cores is consistent with a single power law down to 
   the 90$\%$ completeness limit, with an exponent close to the Salpeter value.
   A fraction of the starless cores detected with LABOCA in Cha~I and III may 
   still grow in mass and become gravitationally unstable. Based on 
   predictions of numerical simulations of turbulent molecular clouds, we 
   estimate that at most 50$\%$ and 20$\%$ of the starless cores of Cha~I and 
   III, respectively, may form stars.}
   {The LABOCA survey reveals that Cha~III, and Cha~I to some extent too, is a 
   prime target to study the \textit{formation} of prestellar cores, and thus 
   the onset of star formation. Getting observational constraints on the 
   duration of the core-building phase prior to gravitational collapse will be 
   necessary to make further progress.
   }


   \maketitle
%
%________________________________________________________________

\section{Introduction}
\label{s:intro}

With the advent of large (sub)mm bolometer arrays, the search for cold dense 
cores in molecular clouds is becoming more efficient. These dense cores 
correspond to the earliest stages of the birth of stars and their 
study is essential to understand the process of star formation. In particular, 
\textit{unbiased} searches for prestellar cores\footnote{A prestellar core is 
usually defined as a starless core that is gravitationally bound 
\citep[][]{Andre00,DiFrancesco07}. Starless means that does not 
contain any young stellar object.} and protostars are 
needed to better understand their formation and derive their lifetimes. 

In this respect, the Chamaeleon molecular cloud complex is of particular 
interest. It is one of the nearest low-mass star forming regions 
\citep[150--180~pc,][; see Appendix~B1 of Belloche et al. 2011 for more 
details]{Whittet97,Knude98} and mainly consists of three molecular clouds, 
Chamaeleon~I, II, and III (hereafter Cha~I, II, and III), that have very 
different degrees of star formation activity. Their populations of young 
stellar objects (YSOs) have been well studied from the infrared to X-rays:
nearly one order of magnitude more YSOs have been found in \object{Cha~I} 
compared to Cha~II -- while \object{Cha~III} does not seem to contain any YSO 
observable at these wavelengths 
\citep[][]{Persi03,Luhman08}. The three clouds have similar masses as traced 
with $^{13}$CO~1--0 ($\sim 1000$~M$_\odot$) but the fraction of denser gas 
traced with C$^{18}$O~1--0 is the highest in Cha~I (24$\%$) while it is the 
lowest in Cha~III (4$\%$) \citep[][]{Mizuno99,Mizuno01}. Finally, several 
indications of jets/outflows were found in Cha~I 
\citep[][]{Mattila89,Gomez04,Wang06,Belloche06}. Only three Herbig-Haro 
objects are known in Cha~II and none  has been found in Cha~III 
\citep[][]{Schwartz77}. Therefore, Cha~III is the least active region in the 
Chamaeleon cloud complex. 

Among the three clouds of the Chamaeleon complex, Cha~III stands out with a 
prominent filamentary structure as revealed by the InfraRed Astronomical 
Satellite (IRAS) at 100~$\mu$m \citep[see Figs.~5 and 7 of][]{Boulanger98}. It 
consists of a system of interwoven filaments that can be disentangled with 
velocity information derived from molecular line emission \citep[][]{Gahm02}.
With an angular resolution of 45$\arcsec$, better than that of the 
$^{13}$CO/C$^{18}$O surveys mentioned above, 38 small clumps embedded in these 
filaments were detected in C$^{18}$O~1--0 with the Swedish ESO Submillimeter 
Telescope (SEST) in the course of a survey targetting the column density peaks 
of cold dust emission based on IRAS data \citep[][]{Gahm02}. The clumps have 
mean densities in the range 1--8~$\times 10^4$~cm$^{-3}$ and their internal 
kinetic energy is dominated by turbulence. Most of them are far from virial 
equilibrium, suggesting that they are not sites of current star formation.

Since Cha~III was mapped in the 1--0 transition of CO and its isotopologues, 
a transition tracing low densities and a molecule suffering from depletion at
high density, only the distribution of its low density gas is relatively well 
known. However, in contrast to Cha~I \citep[][, hereafter Paper~I]{Belloche11} 
and in a limited way 
Cha~II \citep[low-sensitivity 1.3~mm survey of][]{Young05}, no systematic 
survey for (sub)mm dust continuum emission has been performed in Cha~III. 
Therefore, the population of dense, \textit{prestellar} cores in Cha~III is 
completely unknown. On the one hand, the absence of signposts of active star 
formation in the protostellar and pre-main-sequence phases in Cha~III could be 
the result of environmental conditions different from those prevailing in, 
e.g., Cha~I, and unfavorable to star formation. On the other hand, star 
formation could just be starting in Cha~III and finding \textit{prestellar} 
cores would favor this interpretation.

To unveil the present status of the earliest stages of star formation in 
Cha~III and compare this cloud to Cha~I, we carried out a deep, unbiased dust 
continuum survey of Cha~III at 870~$\mu$m with the Large APEX Bolometer 
Camera at the Atacama Pathfinder Experiment (APEX). The observations and 
data reduction are described in Sect.~\ref{s:obs}. Section~\ref{s:results} 
presents the maps and the method used to extract the detected sources. The 
properties of these sources are analysed in Sect.~\ref{s:analysis}. The 
implications are discussed in Sect.~\ref{s:discussion}. 
Section~\ref{s:conclusions} gives a summary of our results and conclusions.

%__________________________________________________________________

\section{Observations and data reduction}
\label{s:obs}

\begin{figure}
%\centerline{\resizebox{1.0\hsize}{!}{\includegraphics[angle=270]{/homes/belloche/Chamaeleon/Continuum/Cha3/Analysis/Extinction/fields_av_artcha3cont.eps}}}
\centerline{\resizebox{1.0\hsize}{!}{\includegraphics[angle=270]{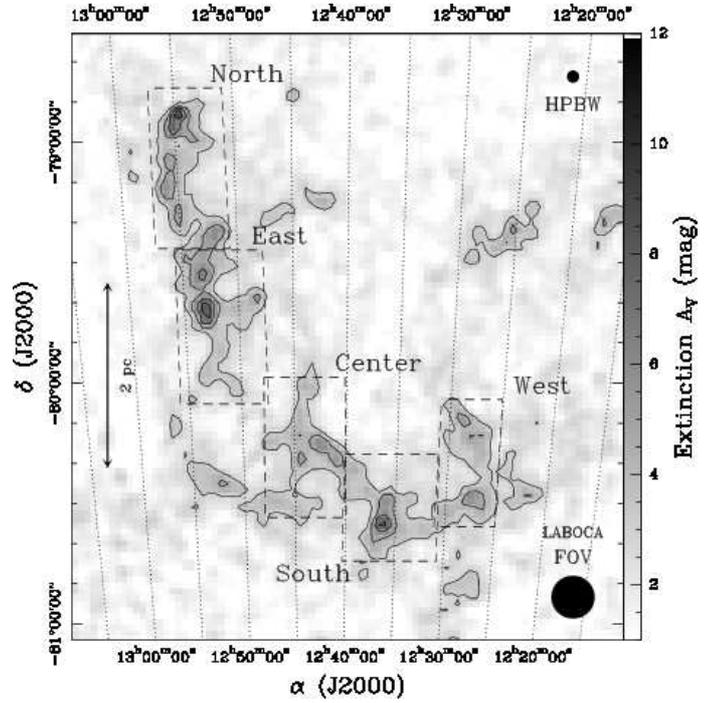}}}
\caption{Extinction map of Cha~III derived from 2MASS in radio projection.
The projection center is at ($\alpha, \delta$)$_{\mathrm{J2000}}$ =
($12^{\mathrm{h}}42^{\mathrm{m}}24^{\mathrm{s}}$, $-79^\circ43\arcmin48\arcsec$). 
The contours start at $A_V = 3$ mag and increase by steps of 1.5~mag. The 
dotted lines are lines of constant right ascension. The angular resolution of 
the map ($HPBW = 3\arcmin$) is shown in the upper right corner. 
The five fields selected for mapping with LABOCA are delimited with
dashed lines. The field of 
view of LABOCA is displayed in the lower right corner.}
\label{f:av}
\end{figure}

\subsection{Extinction map from 2MASS}
\label{ss:2mass}

We derived an extinction map of Cha~III from the publicly available 
2MASS\footnote{The Two Micron All Sky Survey (2MASS) is a joint project of the 
University of Massachusetts and the Infrared Processing and Analysis 
Center/California Institute of Technology, funded by the National Aeronautics 
and Space Administration and the National Science Foundation.}
point source catalog in the same way as we did for Cha~I 
(see Paper~I), except that no source filtering was applied since 
there are no known embedded YSOs in Cha~III. We used the same resolution of 
3$\arcmin$ ($FWHM$) with a pixel size of 1.5$\arcmin$. With these parameters, 
most pixels have at least 10 stars within a radius equal to $FWHM/2$. Only a 
few pixels contain fewer stars, the minimum being 5 stars for 6 pixels. The 
resulting map is shown in Fig.~\ref{f:av}. The typical rms noise level in the 
outer parts of the map is 0.4~mag, corresponding to a $3\sigma$ detection 
level of 1.2~mag for an $FWHM$ of 3$\arcmin$. This
rms noise level is however expected to increase towards the higher-extinction 
regions because of the decreasing number of stars per element of resolution.

\subsection{870 $\mu$m continuum observations with APEX}
\label{ss:laboca}

The region of Cha~III with a visual extinction higher than 3~mag was selected 
on the basis of the extinction map derived from 2MASS as described above
(Sect.~\ref{ss:2mass}). It was divided into five contiguous fields labeled 
Cha3-North, East, Center, South, and West with a total angular area of 
0.93~deg$^2$ (see Fig.~\ref{f:av}). The five fields were mapped in continuum 
emission with the Large APEX BOlometer CAmera \citep[LABOCA,][]{Siringo09} 
operating with about 250 working pixels in the 870~$\mu$m atmospheric window 
at the APEX 12~m submillimeter telescope \citep[][]{Guesten06}. The central 
frequency of LABOCA is 345~GHz and its angular resolution is 19.2$\arcsec$ 
($HPBW$). The observations were carried out for a total of 88 hours in 
September and November 2010, under excellent 
($\tau_{\mathrm{zenith}}^{\mathrm{870}\,\mu\mathrm{m}}$ = 0.12) to moderate 
($\tau_{\mathrm{zenith}}^{\mathrm{870}\,\mu\mathrm{m}}$ = 0.43) 
atmospheric conditions. The sky opacity was measured every 1 to 2 hours with
skydips. The focus was optimised on $\eta$ Carina, Mars, or 
G34.26+0.15 at
least once per day/night. The pointing of the telescope was checked every 1
to 1.5 hour on the nearby quasar \hbox{PKS1057-79} and was found to be 
accurate within 2.5$\arcsec$ (rms). The 
calibration was performed with the secondary calibrators IRAS~13134-6264, 
G5.89-0.39, G34.26+0.15, or NGC 2071 that were observed every 1 to 2 hours 
\citep[see Table A.1 of][]{Siringo09}. Measurements on the primary calibrator 
Mars were also used.

The observations were performed on the fly with a rectangular pattern
(``OTF''). The OTF maps were performed with a scanning speed of 
\hbox{2~arcmin~s$^{-1}$} and were alternately scanned in right ascension and 
declination, with a random position angle between $-12^\circ$ and $+12^\circ$ 
to improve the sampling and reduce striping effects. 

\subsection{LABOCA data reduction}
\label{ss:reduction}

The LABOCA data were reduced with the BoA software\footnote{See 
http://www.mpifr-bonn.mpg.de/div/submmtech/software/boa/ boa\_main.html.} 
following the iterative procedure described in Paper~I. The only
difference is that the baselines of the individual OTF maps scanned in 
declination had to be removed subscan-wise rather than scan-wise to improve 
the flatness of the background level.
The gridding was done with a cell size of 6.1$\arcsec$ 
and the map was smoothed with a Gaussian kernel of size 9$\arcsec$ ($FWHM$). 
The angular resolution of the final map is 21.2$\arcsec$ ($HPBW$) and the rms 
noise level is 11.5~mJy/21.2$\arcsec$-beam (see Sect.~\ref{ss:labocamap}).

The spatial filtering properties of the data reduction and the convergence of
the iterative process are analysed in Appendix~\ref{s:reduction_appendix} in
the same way as was done for Cha~I (Paper~I). In short, the Cha~III dataset 
that was obtained with rectangular scanning patterns seems to suffer slightly 
more from spatial filtering than the Cha~I dataset that combined rectangular 
and more compact spiral scanning patterns. The reasons for this 
counterintuitive result are unclear. It may be related to the slightly smaller 
number of well-working pixels for the Cha~III dataset compared to the Cha~I one.

\begin{figure*}
%\centerline{\resizebox{1.0\hsize}{!}{\includegraphics[angle=270]{/homes/belloche/Chamaeleon/Continuum/Cha3/Analysis/contmap_artcha3cont.eps}}}
\centerline{\resizebox{1.0\hsize}{!}{\includegraphics[angle=270]{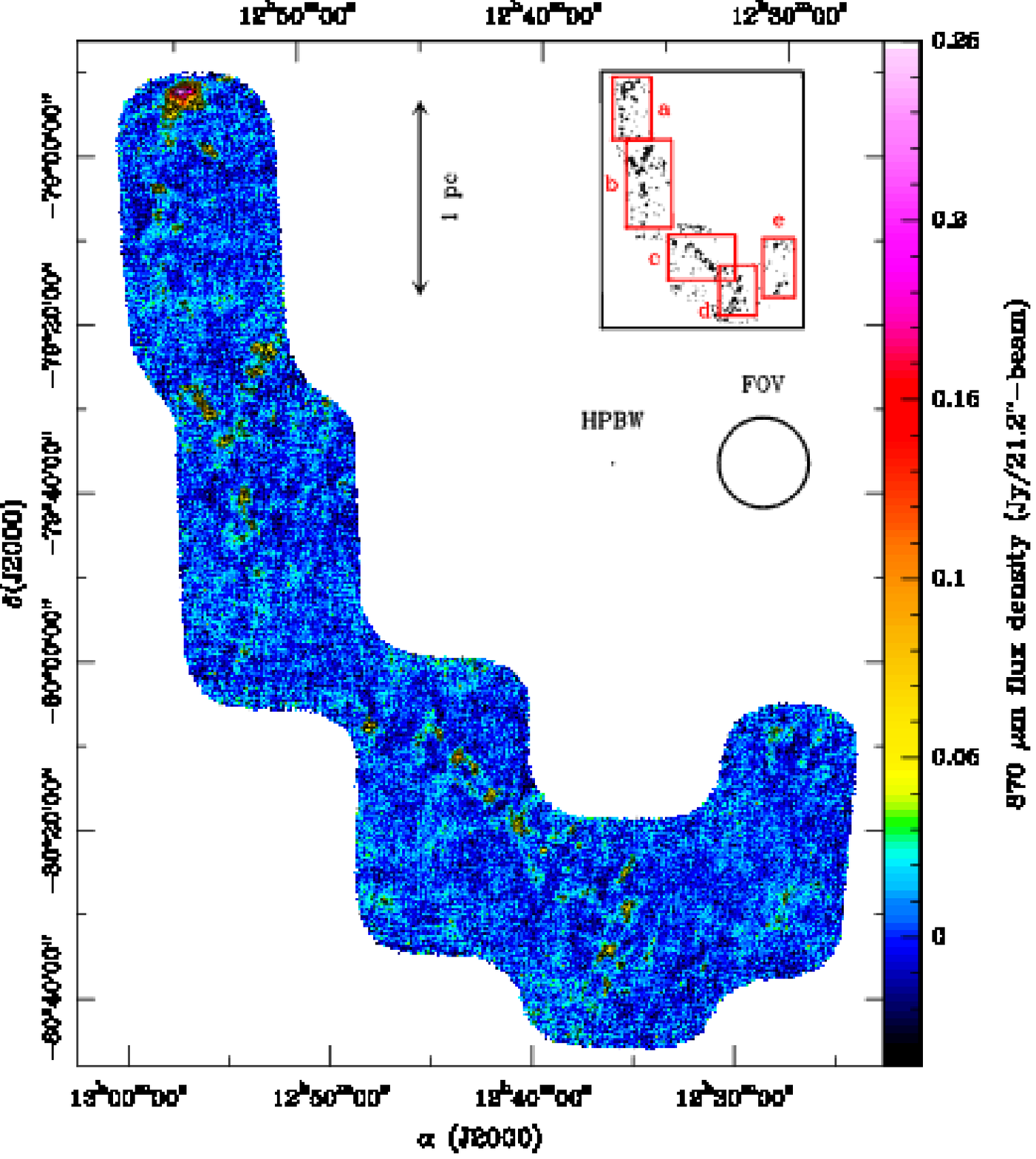}}}
\caption{870 $\mu$m continuum emission map of Cha~III obtained with LABOCA at 
APEX. The projection type and center are the same as in Fig~\ref{f:av}. The 
contour levels are $a$, $2a$, $4a$, and $6a$ with 
$a = 34.5$~mJy/21.2$\arcsec$-beam, i.e. 3 times the rms noise level. The flux 
density color scale is shown on the right. The field of view of LABOCA 
(10.7$\arcmin$) and the angular resolution of the map ($HPBW = 21.2\arcsec$) 
are shown on the right. The red boxes in the insert are labeled like 
Figs.~\ref{f:labocamapdet}a--e and show their limits overlaid on the first 
870 $\mu$m contour.
}
\label{f:labocamap}
\end{figure*}

%
%________________________________________________________________

\section{Basic results and source extraction}
\label{s:results}

In the following, we make the same assumptions as in Paper~I (see its 
Appendix~B) to derive the physical properties of the detected 
sources, in particular we assume a uniform dust temperature of 12~K 
\citep[see][]{Toth00}, a distance of 150~pc \citep[][]{Whittet97}, a dust 
mass opacity, $\kappa_{870}$, of 0.01~cm$^{2}$ per gram of gas+dust, a 
gas-to-dust mass ratio of 100, and a  mean molecular weight per free 
particle, $\mu$, of 2.37. These assumptions are not repeated in the 
following, except in the few cases where there could be an ambiguity.

\subsection{Maps of dust continuum emission in Cha~III}
\label{ss:labocamap}

The final 870~$\mu$m continuum emission map of Cha~III obtained with LABOCA is 
shown in Fig.~\ref{f:labocamap}. Pixels with a weight ($1/\sigma^2$) smaller 
than 3500~beam$^2/$Jy$^2$ are masked. The resulting map contains 
0.32~megapixels (out of 0.76 that contain some signal), corresponding to a 
total area of 0.92~deg$^2$ (6.3~pc$^2$). The mean and median weights are 5717 
and 5798~beam$^2/$Jy$^2$, respectively. The noise distribution is fairly 
uniform and Gaussian. The average noise level is 
11.5~mJy/21.2$\arcsec$-beam. This translates into an H$_2$ column density of 
$1.0 \times 10^{21}$~cm$^{-2}$, and corresponds to a visual extinction 
$A_V \sim$~1.1~mag with $R_V = 3.1$ 
(see the other assumptions in Appendix~ B of Paper~I).

The dust continuum emission map of Cha~III reveals many weak, spatially 
resolved sources. In contrast to Cha~I, not a single unresolved, compact source 
is detected, which is consistent with the absence of signposts of star 
formation at other wavelengths in Cha~III. 
Figure~\ref{f:labocamapdet} presents all the detected structures in more 
detail. The most prominent one is the dense core in the northern part of 
field Cha3-North (Fig.~\ref{f:labocamapdet}a). All other detected structures 
are much fainter. Although there is no direct detection of any large-scale, 
filamentary structure (see Fig.~\ref{f:mrmed}b), the distribution of detected 
sources is much reminiscent of the filamentary structures seen in the 
far-infrared (see Sect.~\ref{s:intro}). In field Cha3-East, most 
detected sources (Cha3-C21, 7, 17, 12, 
and fainter $3\sigma$ compact structures) are distributed along a 1.7~pc long 
filament (Fig.~\ref{f:labocamapdet}b). A second, shorter (0.3~pc), filamentary 
structure that may connect to the latter is suggested by the spatial 
distribution of Cha3-C19, 8, and a fainter $3\sigma$ compact structure to the 
south-west. In field Cha3-Center, Cha3-C3, 4, 13, 5, 16, and 22 are remarkably 
aligned and nearly equally distributed along a straight line of length 0.9~pc
(Fig.~\ref{f:labocamapdet}c). Finally, the sources detected in field Cha3-South 
also suggest the existence of a filament of length 0.8~pc 
(Fig.~\ref{f:labocamapdet}d). 

Even if the filamentary structure of Cha~III is not directly seen 
with LABOCA, we expect that it will be detected with the 
\textit{Herschel} Space Observatory in the frame of the Gould Belt Survey 
\citep[][]{Andre10}. Filaments have been detected in all clouds analysed 
from this survey so far, and a close connection between these filaments and the 
formation of dense cores has been established 
\citep[][]{Andre10,Menshchikov10,Arzoumanian11}.

\begin{figure}
%\centerline{\resizebox{0.87\hsize}{!}{\includegraphics[angle=270]{/homes/belloche/Chamaeleon/Continuum/Cha3/Analysis/contmapdet1_artcha3cont.eps}}}
\centerline{\resizebox{0.87\hsize}{!}{\includegraphics[angle=270]{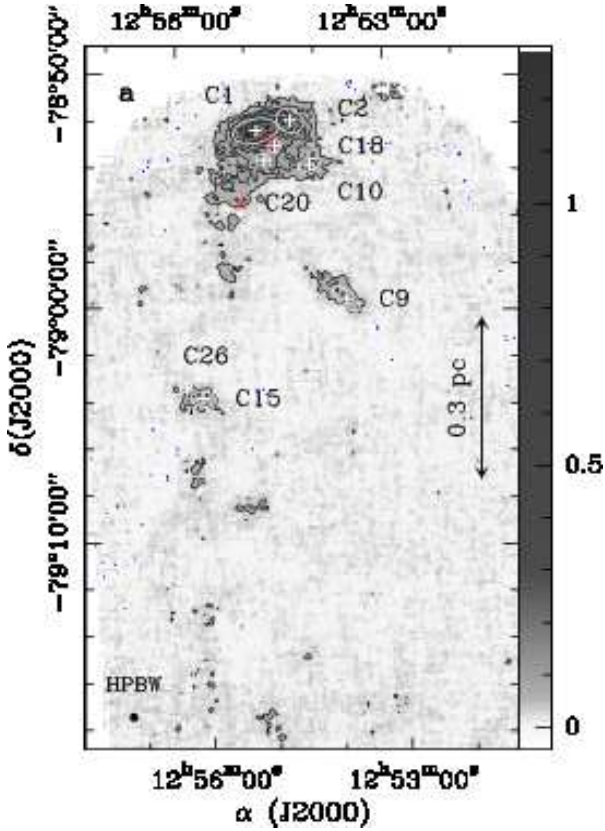}}}
\caption{Detailed 870 $\mu$m continuum emission maps of Cha~III extracted from 
the map shown in Fig.~\ref{f:labocamap}. The flux density greyscale is shown 
on the right of each panel and labeled in Jy/21.2$\arcsec$-beam. It has been 
optimized to reveal the faint emission with a better contrast. The contours 
start at $a$ and increase with a step of $a$, with 
$a = 34.5$~mJy/21.2$\arcsec$-beam, i.e. 3 times the rms noise level. The 
dotted blue contour corresponds to $-a$. The angular resolution of the map is 
shown in the lower left corner of each panel ($HPBW = 21.2\arcsec$). The white 
plus symbols and ellipses show the positions, sizes ($FWHM$), and orientations 
of the Gaussian sources extracted with \textit{Gaussclumps} in the filtered 
map shown in Fig.~\ref{f:mrmed}a. The sources are labeled like in the first 
column of Table~\ref{t:starless}. The (red) crosses show the peak position of 
the 38 clumps detected in C$^{18}$O~1--0 with SEST 
\citep[][]{Gahm02}.
\textbf{a} Field Cha3-North.}
\label{f:labocamapdet}
\end{figure}

\addtocounter{figure}{-1}
\begin{figure}
%\centerline{\resizebox{1.0\hsize}{!}{\includegraphics[angle=270]{/homes/belloche/Chamaeleon/Continuum/Cha3/Analysis/contmapdet2_artcha3cont.eps}}}
\centerline{\resizebox{1.0\hsize}{!}{\includegraphics[angle=270]{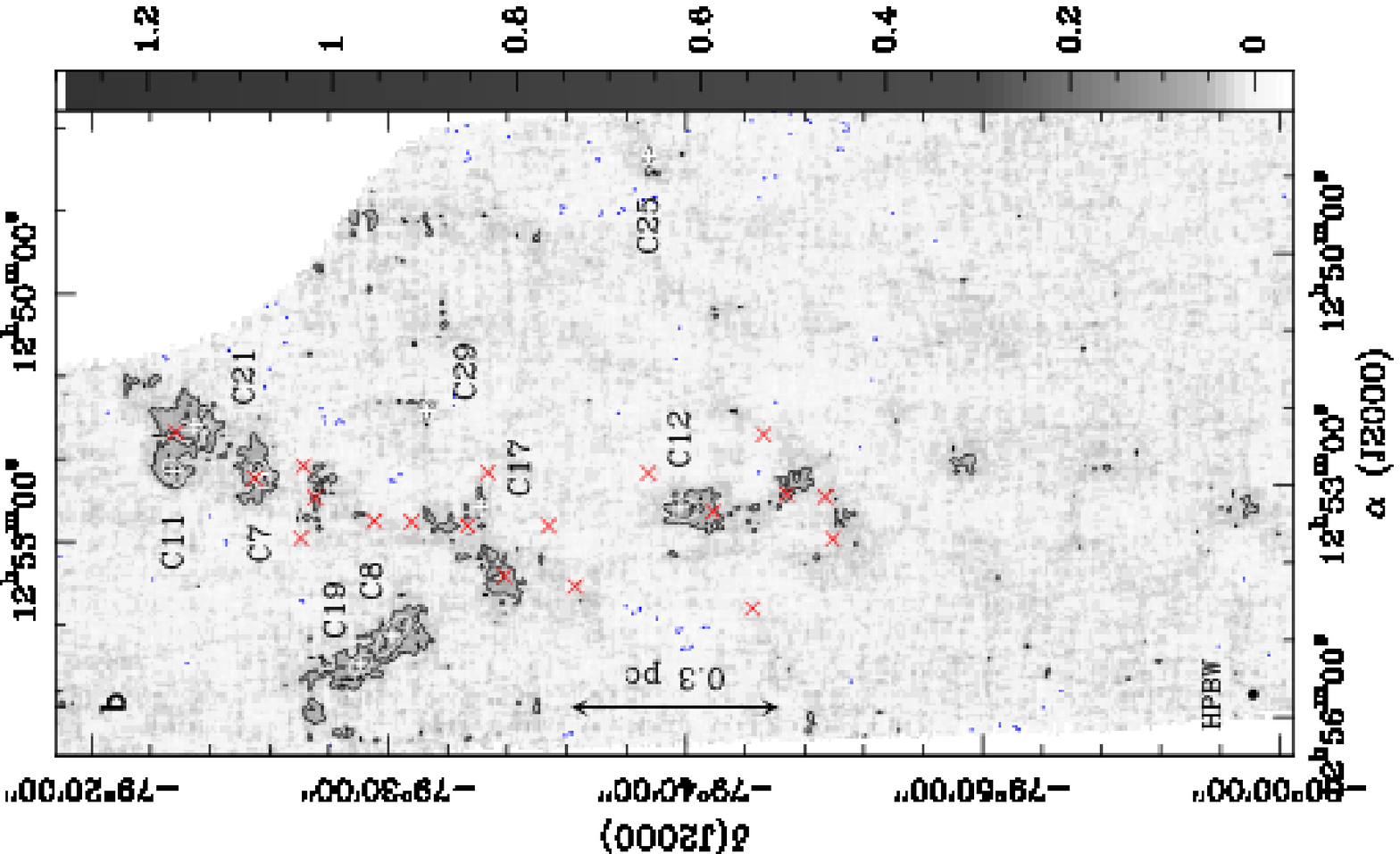}}}
\caption{(continued) \textbf{b} Field Cha3 -East.} 
\end{figure}

\addtocounter{figure}{-1}
\begin{figure*}
%\centerline{\resizebox{0.66\hsize}{!}{\includegraphics[angle=270]{/homes/belloche/Chamaeleon/Continuum/Cha3/Analysis/contmapdet3_artcha3cont.eps}}}
\centerline{\resizebox{0.66\hsize}{!}{\includegraphics[angle=270]{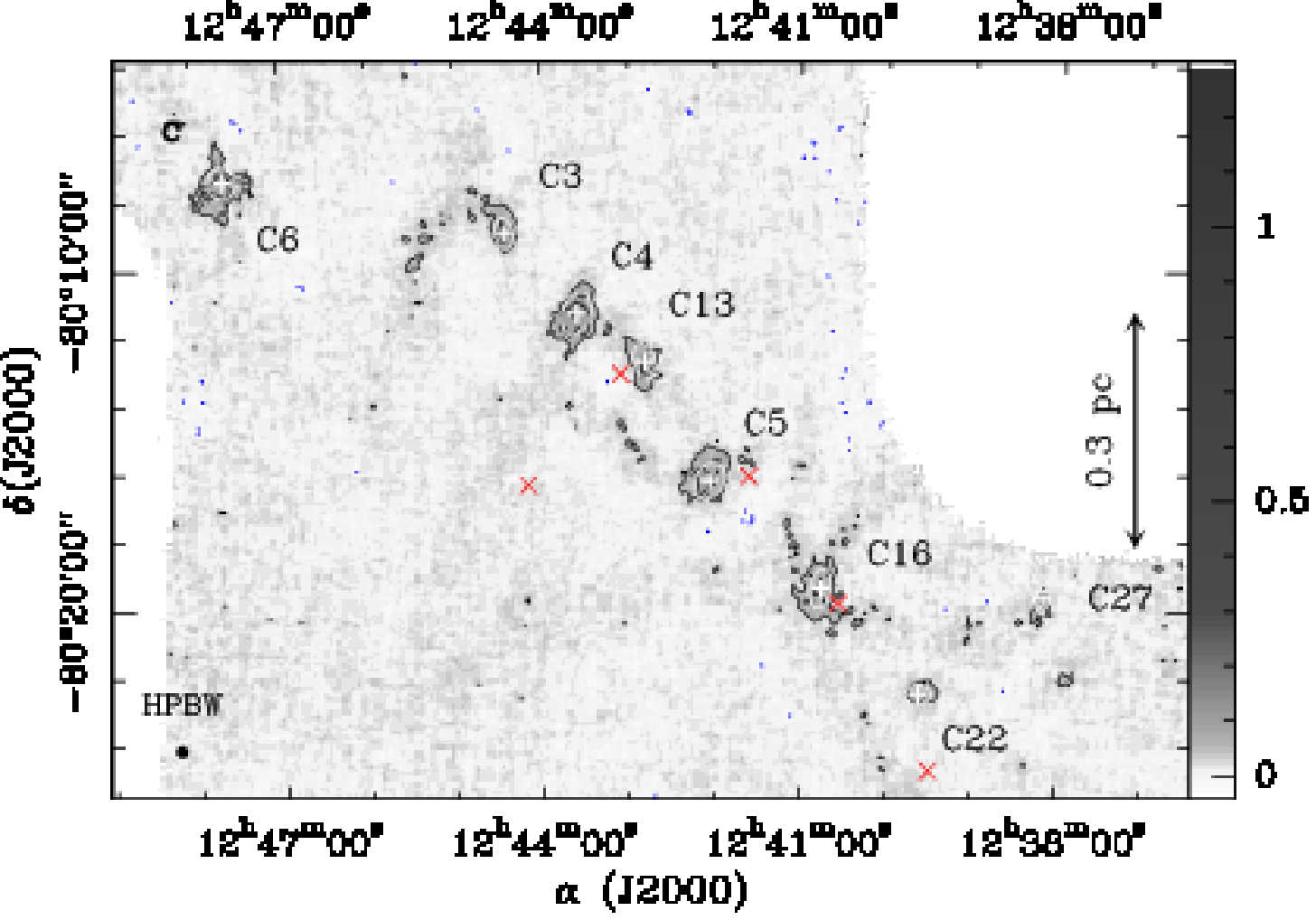}}}
\caption{(continued) \textbf{c} Field Cha3-Center.}
\end{figure*}

\addtocounter{figure}{-1}
\begin{figure}
%\centerline{\resizebox{0.84\hsize}{!}{\includegraphics[angle=270]{/homes/belloche/Chamaeleon/Continuum/Cha3/Analysis/contmapdet4_artcha3cont.eps}}}
\centerline{\resizebox{0.84\hsize}{!}{\includegraphics[angle=270]{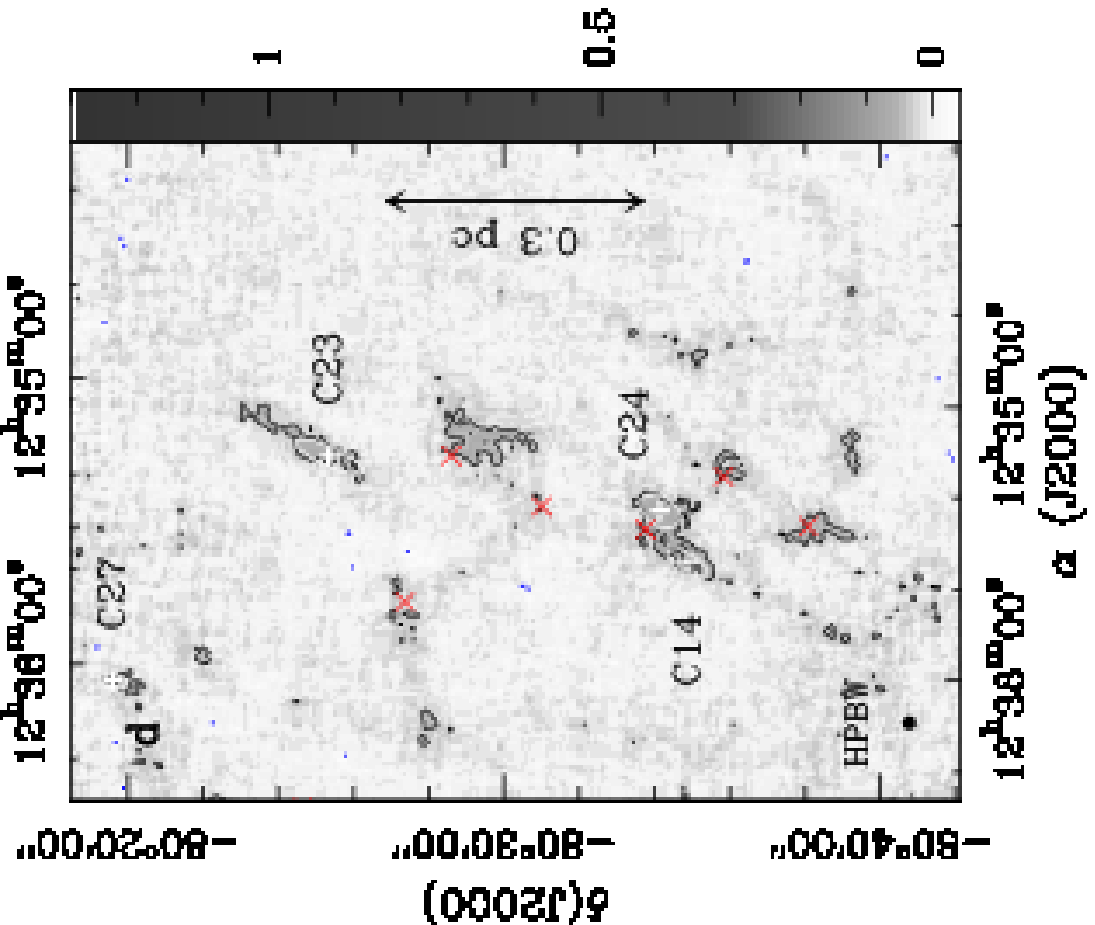}}}
\caption{(continued) \textbf{d} Field Cha3-South.}
\end{figure}

\addtocounter{figure}{-1}
\begin{figure}
%\centerline{\resizebox{0.75\hsize}{!}{\includegraphics[angle=270]{/homes/belloche/Chamaeleon/Continuum/Cha3/Analysis/contmapdet5_artcha3cont.eps}}}
\centerline{\resizebox{0.75\hsize}{!}{\includegraphics[angle=270]{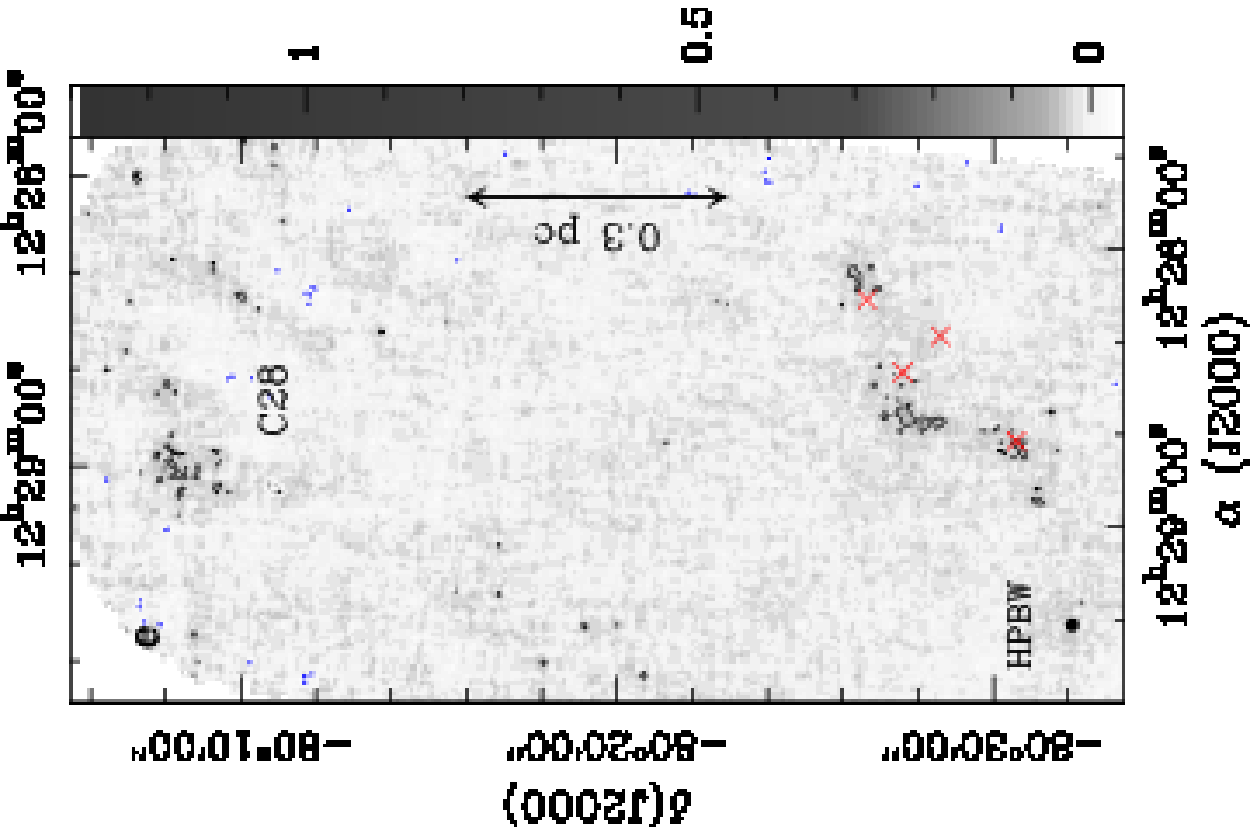}}}
\caption{(continued) \textbf{e} Field Cha3-West.}
\end{figure}

\subsection{Masses traced with LABOCA and 2MASS}
\label{ss:cha3masses}

The total 870~$\mu$m flux in the whole map of Cha~III is about 42.7~Jy. 
This translates into a cloud mass of 22.6~M$_\odot$. It corresponds to 1.2$\%$ 
of the total mass traced by CO in Cha~III 
\citep[1890~M$_\odot$,][]{Mizuno01}, 2.0$\%$ of the mass traced by $^{13}$CO 
\citep[1100~M$_\odot$,][]{Mizuno99}, and 54$\%$ of the mass traced by 
C$^{18}$O \citep[42~M$_\odot$,][]{Mizuno99}. In Cha~I, these fractions were 
5.9$\%$, 7.7$\%$, and 27--32$\%$, respectively (Paper~I).

The extinction map shown in Fig.~\ref{f:av} traces larger scales than
the 870~$\mu$m dust emission map. The median and mean extinctions over the 
0.92~deg$^2$ covered with LABOCA are 2.8 
and 3.0~mag, respectively. Although our survey was designed to cover the 
regions above 3~mag, a significant fraction of the resulting map that is based 
on rectangular fields comprises regions below 3~mag, which explains these low 
median and mean extinctions. Assuming an extinction to H$_2$ column density 
conversion factor of $9.4 \times 10^{20}$~cm$^{-2}$~mag$^{-1}$ 
(for $R_V = 3.1$, see Appendix~B.3 of Paper~I), we derive a total 
gas+dust mass of 401~M$_\odot$.
However, 96$\%$ of this mass, i.e. 386~M$_\odot$, is at $A_V < 6$~mag. 
With the appropriate conversion factor for $A_V > 6$~mag 
(see Appendix~B of Paper~I), the remaining mass is reduced to 
9.3~M$_\odot$, yielding a more accurate estimate of 395~M$_\odot$ for the 
total mass of Cha~III traced with the extinction. It is much lower than the 
masses traced by CO and $^{13}$CO mentioned above. Since the latter mass values 
(at least those derived from CO emission) were based on integrations over much 
larger areas where CO and $^{13}$CO still emit significantly\footnote{This 
caveat does not concern the mass derived from C$^{18}$O~1--0 since the map
of \citet{Mizuno99} has more or less the same size as the LABOCA map.} 
\citep[see region VI in Fig.~2 of][]{Mizuno01}, we consider the 
mass derived from the extinction map as the best estimate to compare with. 
Thus the mass traced with LABOCA represents about 5.7$\%$ of the cloud mass, 
which is slightly less than in Cha~I (7.5$\%$, see Paper~I). Given 
that the median extinction is 2.8~mag in the extinction map, i.e. about 2.5 
times the rms sensitivity achieved with LABOCA, the missing 94$\%$ were lost 
not only because of a lack of sensitivity but also because of the spatial 
filtering due to the correlated noise removal (see Sect.~\ref{ss:filtering}). 
Finally, we estimate the average density of free particles in the field covered
with LABOCA. We assume that the depth of the cloud along the line of sight is 
equal to the square root of its projected surface, i.e. 2.5~pc. This yields an 
average density of $\sim$~420~cm$^{-3}$. This value is very similar to the 
average density estimated for Cha~I (380 cm$^{-3}$, see Paper~I). 
Alternatively, if we assume that Cha~III is filamentary and that its 
depth is rather similar to its typical minor size in the plane of the sky 
(roughly 1 pc), then its average density becomes $1.1 \times 10^3$ cm$^{-3}$.

The average density derived above corresponds to an average thermal 
pressure $P_{\mathrm{th}}/k_{\mathrm{B}}$ of 
0.5--$1.3 \times 10^4$~cm$^{-3}$~K, assuming 
a kinetic temperature of 12~K. The C$^{18}$O and CO~1--0 linewidths measured 
in Cha~III are on the order of 0.9 and 2.6~km~s~$^{-1}$, respectively 
\citep[][]{Boulanger98,Mizuno99,Mizuno01}. The $^{13}$CO~1--0 emission traces
gas of densities on the order of $1 \times 10^3$ cm$^{-3}$ in Cha~III 
\citep[][]{Mizuno98} and its typical linewidth must be between those of
C$^{18}$O and CO~1--0. We take 1~km~s$^{-1}$ as a representative value, which 
is the typical value measured by \citet{Mizuno98} in the small clouds of the 
Chamaeleon complex. The turbulent pressure is defined as 
$P_{\mathrm{turb}} = \mu m_{\mathrm{H}} n \sigma_{\mathrm{NT}}^2$, with 
$\mu$ the mean molecular weight per free particle, $m_{\mathrm{H}}$ the 
atomic mass of hydrogen, $n$ the average free-particle density,
and $\sigma_{\mathrm{NT}}$ the non-thermal rms velocity dispersion derived 
from the linewidth. We obtain a turbulent pressure 
$P_{\mathrm{turb}}/k_{\mathrm{B}}$ of \hbox{2.1--$5.6 \times 10^{4}$~cm$^{-3}$~K}.
The turbulent pressure dominates by a factor $\sim 4$ over the thermal pressure
in Cha~III. It is on the same order as the total pressure of the ISM in the
mid-plane of the Galaxy \citep[$2 \times 10^4$~cm$^{-3}$~K, see][]{Cox05}.
The situation is similar in Cha~I.

\subsection{Source extraction and classification}
\label{ss:sourceid}

\begin{figure*}
%\centerline{\resizebox{1.00\hsize}{!}{\includegraphics[angle=270]{/homes/belloche/Chamaeleon/Continuum/Cha3/Analysis/contmapfilt_artcha3cont_1.eps}\hspace*{3ex}\includegraphics[angle=270]{/homes/belloche/Chamaeleon/Continuum/Cha3/Analysis/contmapfilt_artcha3cont_2.eps}}}
\centerline{\resizebox{1.00\hsize}{!}{\includegraphics[angle=270]{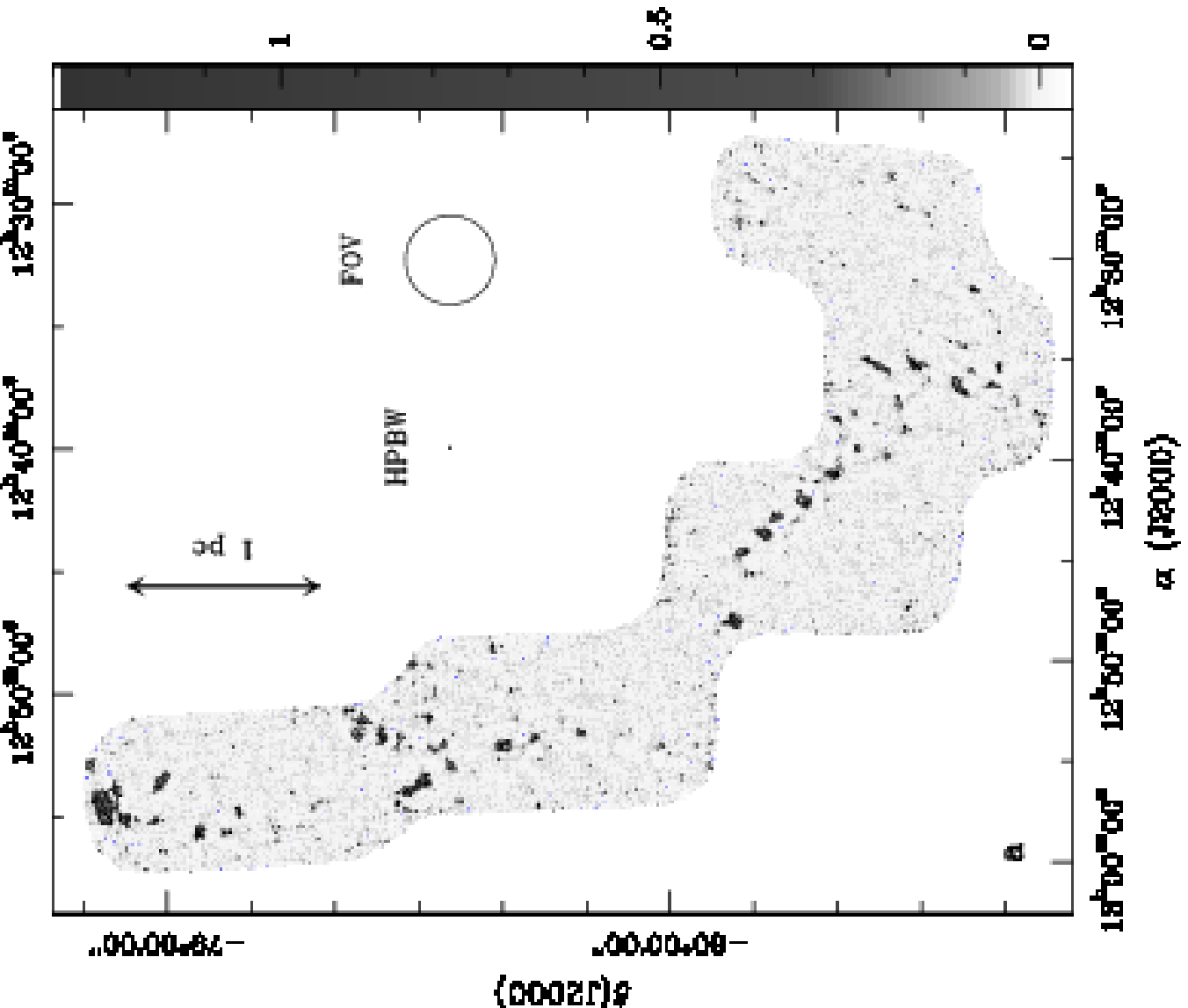}\hspace*{3ex}\includegraphics[angle=270]{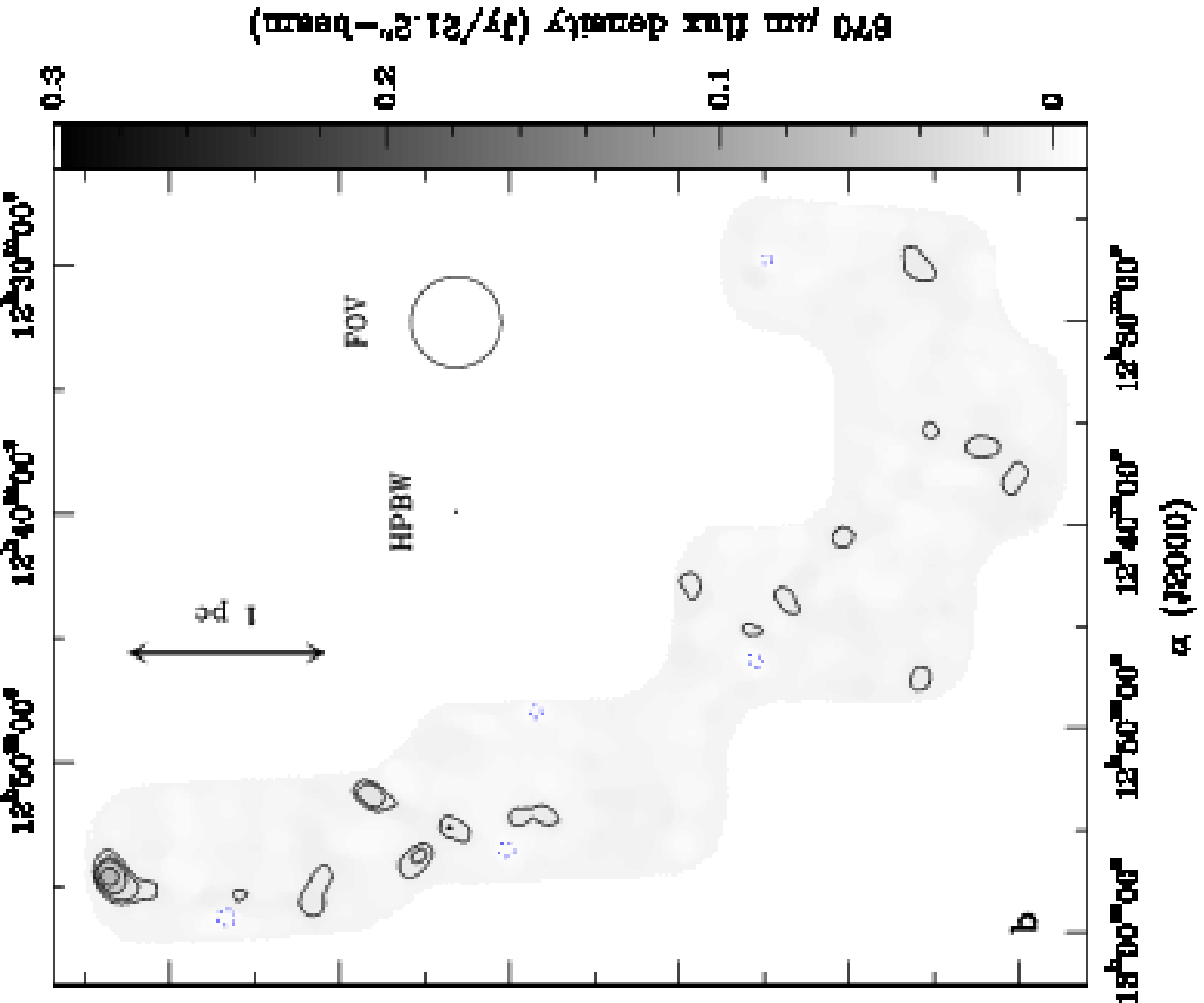}}}
\caption{\textbf{a} 870 $\mu$m continuum emission \textit{sum} map of Cha~III 
at scale 5. The contour levels are $-a$ (in dotted blue), $a$, $2a$, and $4a$, 
with $a = 34.5$~mJy/21.2$\arcsec$-beam, i.e. about 3 times the rms noise 
level. 
\textbf{b} Smoothed map, i.e. residuals, at scale 5. The contour levels are
$-c$ (in dotted blue), $c$, $2c$, $4c$, and $8c$, with 
$c = 8.1$~mJy/21.2$\arcsec$-beam, i.e. about 3 times the rms noise level in 
this map.
The greyscales of both maps are different. The sum of these two 
maps is strictly equal to the original map (Fig.~\ref{f:labocamap}).}
\label{f:mrmed}
\end{figure*}

The source extraction from the 870~$\mu$m map is performed in the same way as 
for Cha~I (Paper~I). The map is first decomposed into different scales with
our multiresolution program based on a median filter. 
The total fluxes measured in the \textit{sum} maps at scales 3 to 7 
(see definition in Appendix~C of Paper~I) are listed in 
Table~\ref{t:flux_mrmed}, as well as the corresponding masses. About half of 
the total flux is emitted by structures smaller than $\sim 200\arcsec$ 
($FWHM$), and only 8$\%$ by structures smaller than $\sim 60\arcsec$. 
Table~\ref{t:flux_mrmed} shows that the fraction of continuum flux at small 
scales ($< 60\arcsec$) is slightly smaller than in Cha~I. This may suggest 
that the structures in Cha~III are less centrally peaked or it could simply be 
a bias due to the slightly higher sensitivity of the Cha~III survey. The 
\textit{sum} map at scale 5 and its associated smoothed map are shown in 
Fig.~\ref{f:mrmed}. The sum of these two maps is strictly equal to the 
original map of Cha~III shown in Fig.~\ref{f:labocamap}.

The sources are extracted with the Gaussian fitting program 
\textit{Gaussclumps} with the same parameters as for Cha~I (see Paper~I). The 
\textit{sum} maps at scales 1 to 7 were decomposed into 0, 0, 2, 16, 29, 38, 
and 39 Gaussian  sources, respectively, and the full map was decomposed into 39 
Gaussian sources. These counts do not include the sources found too close to 
the noisier map edges (weight $<$ 4400 beam$^2$/Jy$^2$), which we consider as 
artefacts. 

We now consider the results obtained with \textit{Gaussclumps} for the 
\textit{sum} map at scale 5 (i.e. the map shown in Fig.~\ref{f:mrmed}a), which 
is a good scale to characterize sources with $FWHM < 120\arcsec$ as shown in 
Appendix~C of Paper~I. The positions, sizes, orientations, and indexes of the 
29 extracted Gaussian sources are listed in Table~\ref{t:id_gcl_simbad} in the 
order in which \textit{Gaussclumps} found them. We looked for associations 
with sources in the SIMBAD astronomical database. We used SIMBAD4 (release 
1.171) as of February 10$^{\mathrm{th}}$, 2011. None of the 29 
\textit{Gaussclumps} sources is associated with a SIMBAD object within its 
$FWHM$ ellipse. As a result, we consider that all these 29 sources are 
starless cores.

\begin{table}
 \caption{Continuum flux distribution in Cha~III and comparison to Cha~I.}
 \label{t:flux_mrmed}
 \centering
 \begin{tabular}{cccccc}
  \hline\hline
  \multicolumn{1}{c}{Scale} & \multicolumn{1}{c}{Typical size} & \multicolumn{1}{c}{Flux} & \multicolumn{1}{c}{Mass} & \multicolumn{1}{c}{$F/F_{\mathrm{tot}}$} &  \multicolumn{1}{c}{$F/F_{\mathrm{tot}}$(Cha~I)} \\
   \multicolumn{1}{c}{} &   &  \multicolumn{1}{c}{\scriptsize (Jy)} & \multicolumn{1}{c}{\scriptsize (M$_\odot$)} & \multicolumn{1}{c}{\scriptsize ($\%$)} & \multicolumn{1}{c}{\scriptsize ($\%$)}   \\
  \multicolumn{1}{c}{(1)} & \multicolumn{1}{c}{(2)} & \multicolumn{1}{c}{(3)} & \multicolumn{1}{c}{(4)} & \multicolumn{1}{c}{(5)} & \multicolumn{1}{c}{(6)} \\
  \hline
  3 & $<60\arcsec$  &  3.5 &  1.9 &   8 &  11 \\
  4 & $<120\arcsec$ & 10.3 &  5.4 &  24 &  22 \\
  5 & $<200\arcsec$ & 23.0 & 12.2 &  54 &  49 \\
  6 & $<300\arcsec$ & 36.9 & 19.5 &  86 &  82 \\
  7 & $\sim$ all    & 43.2 & 22.8 & 101 &  99 \\ 
  -- & all          & 42.7 & 22.6 & 100 & 100 \\
  \hline
 \end{tabular}
 \tablefoot{The last row corresponds to 
the full map, while rows 1 to 5 correspond to the sum of the filtered maps up 
to scale $i$ listed in the first column (i.e. the \textit{sum} map at scale 
$i$). Column 2 gives the range of sizes of the sources that significantly 
contribute to the emission with more than $40\%$ of their peak flux density 
(see Col.~2 of Table~C.1 of Paper~I). Columns~5 and 6 give the fraction of
flux detected in each map, for Cha~III and I, respectively.}
\end{table}

\begin{table*}
 \caption{
 Sources extracted with \textit{Gaussclumps} in the 870~$\mu$m continuum \textit{sum} map of Cha~III at scale 5.
 }
 \label{t:id_gcl_simbad}
 \centering
 \begin{tabular}{ccccccccc}
 \hline\hline
 \multicolumn{1}{c}{$N_{\mathrm{gcl}}$} & \multicolumn{1}{c}{R.A.} & \multicolumn{1}{c}{Decl.} & \multicolumn{1}{c}{${f_{\mathrm{peak}}}$\tablefootmark{a}} & \multicolumn{1}{c}{${f_{\mathrm{tot}}}$\tablefootmark{a}} & \multicolumn{1}{c}{maj.\tablefootmark{a}} & \multicolumn{1}{c}{min.\tablefootmark{a}} & \multicolumn{1}{c}{P.A.\tablefootmark{a}} & \multicolumn{1}{c}{$S$\tablefootmark{b}} \\ 
  & \multicolumn{1}{c}{\scriptsize (J2000)} & \multicolumn{1}{c}{\scriptsize (J2000)} & \multicolumn{1}{c}{\scriptsize (Jy/beam)} & \multicolumn{1}{c}{\scriptsize (Jy)} & \multicolumn{1}{c}{\scriptsize ($\arcsec$)} & \multicolumn{1}{c}{\scriptsize ($\arcsec$)} & \multicolumn{1}{c}{\scriptsize ($^\circ$)} & \multicolumn{1}{c}{\scriptsize ($\arcsec$)} \\ 
 \multicolumn{1}{c}{(1)} & \multicolumn{1}{c}{(2)} & \multicolumn{1}{c}{(3)} & \multicolumn{1}{c}{(4)} & \multicolumn{1}{c}{(5)} & \multicolumn{1}{c}{(6)} & \multicolumn{1}{c}{(7)} & \multicolumn{1}{c}{(8)} & \multicolumn{1}{c}{(9)} \\ 
 \hline
  1 & 12:54:53.58 & -78:52:24.2 &  0.183 &    2.928 & 116.0 &  62.1 & -77.9 &  84.8  \\ 
  2 & 12:54:23.02 & -78:51:58.0 &  0.116 &    1.070 &  71.9 &  57.7 &  55.2 &  64.4  \\ 
  3 & 12:44:24.79 & -80:08:50.2 &  0.100 &    0.380 &  56.2 &  30.4 &  11.2 &  41.3  \\ 
  4 & 12:43:35.51 & -80:11:11.7 &  0.099 &    0.831 &  77.5 &  48.9 & -20.1 &  61.6  \\ 
  5 & 12:42:03.29 & -80:16:06.6 &  0.093 &    0.699 &  67.4 &  50.4 & -36.7 &  58.3  \\ 
  6 & 12:47:38.53 & -80:07:23.2 &  0.089 &    0.562 &  55.7 &  50.9 &  21.7 &  53.2  \\ 
  7 & 12:52:17.60 & -79:25:42.1 &  0.087 &    0.434 &  59.2 &  38.0 & -69.4 &  47.4  \\ 
  8 & 12:54:20.65 & -79:30:08.0 &  0.085 &    0.716 &  96.2 &  39.4 &  52.0 &  61.5  \\ 
  9 & 12:53:40.41 & -78:59:34.7 &  0.085 &    0.607 &  86.1 &  37.4 &  48.5 &  56.8  \\ 
 10 & 12:54:08.10 & -78:53:51.0 &  0.083 &    0.301 &  59.8 &  27.4 & -76.5 &  40.5  \\ 
 11 & 12:52:12.93 & -79:22:45.4 &  0.079 &    0.376 &  56.5 &  37.7 & -56.9 &  46.1  \\ 
 12 & 12:52:56.46 & -79:39:49.4 &  0.076 &    0.586 &  77.4 &  44.7 &  -5.1 &  58.8  \\ 
 13 & 12:42:48.27 & -80:12:36.8 &  0.077 &    0.421 &  65.3 &  37.5 &  17.1 &  49.5  \\ 
 14 & 12:36:46.21 & -80:34:53.5 &  0.075 &    0.326 &  72.7 &  26.8 & -32.8 &  44.1  \\ 
 15 & 12:55:53.07 & -79:03:51.3 &  0.073 &    0.406 &  63.0 &  39.5 &  83.1 &  49.9  \\ 
 16 & 12:40:45.58 & -80:19:17.0 &  0.069 &    0.384 &  67.6 &  36.9 &  -9.7 &  50.0  \\ 
 17 & 12:52:49.09 & -79:33:06.9 &  0.070 &    0.101 &  30.4 &  21.4 &  82.9 &  25.5  \\ 
 18 & 12:54:37.43 & -78:53:03.5 &  0.071 &    0.173 &  35.4 &  30.7 & -20.6 &  33.0  \\ 
 19 & 12:54:40.89 & -79:29:03.3 &  0.069 &    0.242 &  50.0 &  31.7 &  30.5 &  39.8  \\ 
 20 & 12:54:47.60 & -78:53:39.6 &  0.070 &    0.197 &  38.2 &  33.2 &  54.7 &  35.6  \\ 
 21 & 12:51:39.85 & -79:23:29.5 &  0.063 &    0.211 &  66.7 &  22.7 &  -1.9 &  38.9  \\ 
 22 & 12:39:36.53 & -80:22:24.5 &  0.065 &    0.211 &  44.4 &  32.9 & -85.1 &  38.2  \\ 
 23 & 12:35:44.30 & -80:25:21.6 &  0.064 &    0.498 & 120.0 &  29.4 & -23.8 &  59.4  \\ 
 24 & 12:36:13.68 & -80:34:09.3 &  0.063 &    0.432 &  67.6 &  45.5 & -34.9 &  55.5  \\ 
 25 & 12:48:31.42 & -79:38:46.7 &  0.061 &    0.153 &  52.4 &  21.4 & -89.9 &  33.5  \\ 
 26 & 12:56:07.33 & -79:03:22.2 &  0.059 &    0.073 &  26.2 &  21.2 & -79.9 &  23.6  \\ 
 27 & 12:38:10.69 & -80:19:35.8 &  0.058 &    0.135 &  49.3 &  21.2 & -10.0 &  32.3  \\ 
 28 & 12:29:06.05 & -80:10:53.3 &  0.058 &    0.106 &  38.7 &  21.2 & -30.6 &  28.7  \\ 
 29 & 12:51:36.11 & -79:31:17.1 &  0.058 &    0.155 &  41.2 &  29.3 &  86.1 &  34.7  \\ 
 \hline
 \end{tabular}
 \tablefoot{Scale 5 is defined in Sect.~\ref{ss:sourceid} and Table~\ref{t:flux_mrmed}.
 \tablefoottext{a}{Peak flux density (in Jy/21.2$\arcsec$-beam), total flux, $FWHM$ along the major and minor axes, and position angle (east from north) of the fitted Gaussian.}
 \tablefoottext{b}{Mean source size, equal to the geometrical mean of the major and minor $FWHM$.}
 }
\end{table*}

%
%________________________________________________________________

\section{Analysis}
\label{s:analysis}

\subsection{Comparison with the extinction map}
\label{ss:av_cont}

\begin{figure}
%\centerline{\resizebox{1.00\hsize}{!}{\includegraphics[angle=270]{/homes/belloche/Chamaeleon/Continuum/Cha3/Analysis/Extinction/overlay_avcont_artcha3cont.eps}}}
\centerline{\resizebox{1.00\hsize}{!}{\includegraphics[angle=270]{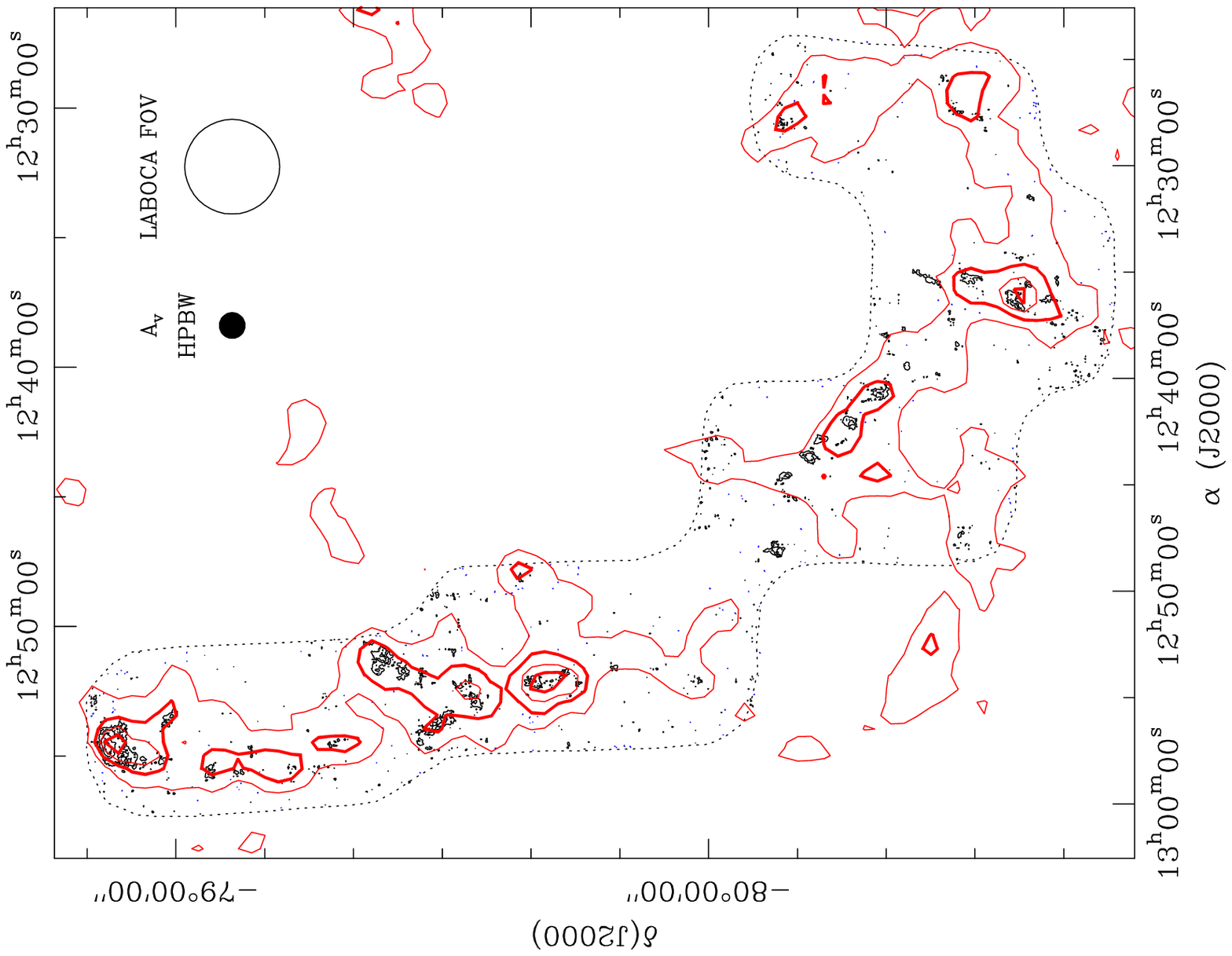}}}
\caption{Extinction map of Fig.~\ref{f:av} (red contours) overlaid on the 
870~$\mu$m continuum emission map of Cha~III (black contours). The contour 
levels of the extinction map start at 3~mag and increase by steps of 1.5~mag. 
The thicker red contours correspond to $A_V = 4.5$ and 7.5~mag. The contour 
levels of the 870~$\mu$m map are the same as in Fig.~\ref{f:labocamap}, plus a 
dotted blue contour at $-a$. The dotted 
line delimits the field mapped at 870~$\mu$m. The field of view of LABOCA and 
the angular resolution of the extinction map are shown in the upper right 
corner.}
\label{f:avcont}
\end{figure}

The extinction map derived from 2MASS is overlaid on the 870~$\mu$m dust 
continuum emission map of Cha~III in Fig.~\ref{f:avcont}. The overall 
correspondence is relatively good, most of the continuum emission coinciding
with the peaks of the extinction map.
The $3\sigma$ H$_2$ column density sensitivity limit of the 
870~$\mu$m map is $3.0 \times 10^{21}$~cm$^{-2}$, which corresponds to 
$A_V \sim 3.2$~mag. Most of the continuum emission detected at 
870~$\mu$m is above the contour level $A_V = 4.5$~mag, and most extended 
regions with $3 < A_V < 4.5$~mag traced by the extinction map are not detected
at 870~$\mu$m. This is partly due to a lack of sensitivity and to the spatial 
filtering related to the sky noise removal since these low-extinction 
regions have sizes on the order of 5$\arcmin$--10$\arcmin$, comparable to the 
field of view of LABOCA. On the other hand, there are also a few 870~$\mu$m 
sources detected below 4.5~mag that are not seen in the extinction map, most 
likely because of its poor angular resolution (e.g. Cha3-C6 and 23 in fields 
Cha3-Center and South, respectively). 

\subsection{Comparison with C$^{18}$O 1--0}
\label{ss:c18o}

The peak positions of the 38 clumps detected by \citet{Gahm02} in 
C$^{18}$O~1--0 emission are shown as (red) crosses in 
Figs.~\ref{f:labocamapdet}a--e. 
One caveat to keep in mind is that the C$^{18}$O~1--0 survey was biased and 
covered only parts of the field mapped with LABOCA. For instance, Cha3-C9, 15, 
and 26 in field Cha3-North, Cha3-C8 and 19 in field Cha3-East, and Cha3-C6 in 
field Cha3-Center were not covered. A second caveat is that the C$^{18}$O maps 
were undersampled by a factor of $\sim 2$, with a step of 1$\arcmin$ in fields 
Cha3-North and East (but 30$\arcsec$ in the brightest regions), and 40$\arcsec$ 
in fields Cha3-Center, South, and West. 

About twenty 870~$\mu$m sources only are detected in the fields covered by the 
C$^{18}$O observations, which implies a detection rate with LABOCA a factor of 
2 lower and shows that the C$^{18}$O observations were sensitive to lower 
density gas, as expected. Figures~\ref{f:labocamapdet}a--e show that there is 
no one-to-one 
correspondence between the peak positions of the 870~$\mu$m and C$^{18}$O 
sources. Only seven C$^{18}$O sources peak within 1$\arcmin$ from the 
peak position of a 870~$\mu$m source. This may partly be due to the 
undersampling of the C$^{18}$O maps, but most likely results from depletion 
that affects C$^{18}$O at high density. This confirms that the 870~$\mu$m 
emission traces the high density regions better than the C$^{18}$O~1--0 
emission. The 870~$\mu$m map should therefore give a better census of the 
potential future sites of star formation in Cha~III.

\subsection{Properties of the starless cores}
\label{ss:starless}

The properties of the 29 starless sources detected with LABOCA are listed in 
Table~\ref{t:starless} and their distribution is shown in Figs.~\ref{f:histo} 
and \ref{f:histom}. The column density (Col.~5) and masses (Cols.~7--9) are 
computed with the fluxes fitted with \textit{Gaussclumps} or 
directly measured in the \textit{sum} map at scale 5. As a caveat, we remind 
the reader that the assumption of a uniform temperature may be 
inaccurate and bias the 
measurements of the masses and column densities, as well as the mass 
concentration (or equivalently the density contrast). A dust temperature drop 
toward the center of starless dense cores is possible 
(see Appendix~B.2 of Paper~I and references therein).

\subsubsection{Extinction}
\label{sss:starless_extinction}

\begin{table*}
 \caption{
 Characteristics of starless sources extracted with \textit{Gaussclumps} in the 870~$\mu$m continuum \textit{sum} map of Cha~III at scale 5.
 }
 \label{t:starless}
 \centering
 \begin{tabular}{ccccccccccccccc}
 \hline\hline
  \multicolumn{1}{c}{Name} & \multicolumn{1}{c}{\hspace*{-2ex} ${N_{\mathrm{gcl}}}$\tablefootmark{a}} & \multicolumn{1}{c}{\hspace*{-2ex} $FWHM$\tablefootmark{b}} & \multicolumn{1}{c}{${R_{\mathrm{a}}}$\tablefootmark{b}} & \multicolumn{1}{c}{\hspace*{-3ex} ${N_{\mathrm{peak}}}$\tablefootmark{c}} & \multicolumn{1}{c}{\hspace*{-2.5ex} ${A_V}$\tablefootmark{d}} & \multicolumn{1}{c}{\hspace*{-1ex} ${M_{\mathrm{peak}}}$\tablefootmark{e}} & \multicolumn{1}{c}{${M_{\mathrm{tot}}}$\tablefootmark{e}} & \multicolumn{1}{c}{${M_{50\arcsec}}$\tablefootmark{e}} & \multicolumn{1}{c}{${C_M}$\tablefootmark{f}} & \multicolumn{1}{c}{$\alpha_{\mathrm{BE}}$\tablefootmark{g}} & \multicolumn{1}{c}{\hspace*{-3ex} ${n_{\mathrm{peak}}}$\tablefootmark{h}} & \multicolumn{1}{c}{\hspace*{-3ex} ${n_{\mathrm{mean}}}$\tablefootmark{h}} & \multicolumn{1}{c}{\hspace*{-3ex} ${n_{50\arcsec}}$\tablefootmark{h}} & \multicolumn{1}{c}{\hspace*{-3ex} ${c_n}$\tablefootmark{i}} \\ 
  & \hspace*{-2ex} & \multicolumn{1}{c}{\hspace*{-2ex} \scriptsize (1000 AU)$^2$} & & \multicolumn{1}{c}{\hspace*{-3ex} \scriptsize ($10^{21}$ cm$^{-2}$)} & \multicolumn{1}{c}{\hspace*{-2.5ex} \scriptsize (mag)} & \multicolumn{1}{c}{\hspace*{-1ex} \scriptsize (M$_\odot$)} & \multicolumn{1}{c}{\scriptsize (M$_\odot$)} & \multicolumn{1}{c}{\scriptsize (M$_\odot$)} &  \multicolumn{1}{c}{\scriptsize ($\%$)} & & \multicolumn{1}{c}{\hspace*{-3ex} \scriptsize ($10^{5}$ cm$^{-3}$)} & \multicolumn{1}{c}{\hspace*{-3ex} \scriptsize ($10^{4}$ cm$^{-3}$)} & \multicolumn{1}{c}{\hspace*{-3ex} \scriptsize ($10^{4}$ cm$^{-3}$)} & \hspace*{-3ex} \\ 
 \multicolumn{1}{c}{(1)} & \multicolumn{1}{c}{\hspace*{-2ex} (2)} & \multicolumn{1}{c}{\hspace*{-2ex} (3)} & \multicolumn{1}{c}{(4)} & \multicolumn{1}{c}{\hspace*{-3ex} (5)} & \multicolumn{1}{c}{\hspace*{-2.5ex} (6)} & \multicolumn{1}{c}{\hspace*{-1ex} (7)} & \multicolumn{1}{c}{(8)} & \multicolumn{1}{c}{(9)} & \multicolumn{1}{c}{(10)} & \multicolumn{1}{c}{(11)} & \multicolumn{1}{c}{\hspace*{-3ex} (12)} & \multicolumn{1}{c}{\hspace*{-3ex} (13)} & \multicolumn{1}{c}{\hspace*{-3ex} (14)} & \multicolumn{1}{c}{\hspace*{-3ex} (15)} \\ 
 \hline
 Cha3-C1 & \hspace*{-2ex} 1 & \hspace*{-2ex} 17.1 $\times$  8.8 &  2.0 & \hspace*{-3ex}  16 & \hspace*{-2.5ex}  5.0 & \hspace*{-1ex}  0.097 &  1.55 &  0.29 &  33( 2) & 1.12 & \hspace*{-3ex}  8.6 & \hspace*{-3ex}  3.0 & \hspace*{-3ex} 19.7 & \hspace*{-3ex} 4.4(0.3) \\  Cha3-C2 & \hspace*{-2ex} 2 & \hspace*{-2ex} 10.3 $\times$  8.0 &  1.3 & \hspace*{-3ex}  10 & \hspace*{-2.5ex}    8 & \hspace*{-1ex}  0.061 &  0.57 &  0.23 &  27( 3) & 0.55 & \hspace*{-3ex}  5.5 & \hspace*{-3ex}  2.7 & \hspace*{-3ex} 15.5 & \hspace*{-3ex} 3.5(0.4) \\  Cha3-C3 & \hspace*{-2ex} 3 & \hspace*{-2ex}  7.8 $\times$  3.3 &  2.4 & \hspace*{-3ex} 8.8 & \hspace*{-2.5ex}  4.2 & \hspace*{-1ex}  0.053 &  0.20 &  0.11 &  47( 7) & 0.35 & \hspace*{-3ex}  4.7 & \hspace*{-3ex}  5.6 & \hspace*{-3ex}  7.6 & \hspace*{-3ex} 6.2(0.9) \\  Cha3-C4 & \hspace*{-2ex} 4 & \hspace*{-2ex} 11.2 $\times$  6.6 &  1.7 & \hspace*{-3ex} 8.6 & \hspace*{-2.5ex}  3.7 & \hspace*{-1ex}  0.052 &  0.44 &  0.15 &  35( 5) & 0.45 & \hspace*{-3ex}  4.6 & \hspace*{-3ex}  2.5 & \hspace*{-3ex} 10.1 & \hspace*{-3ex} 4.6(0.6) \\  Cha3-C5 & \hspace*{-2ex} 5 & \hspace*{-2ex}  9.6 $\times$  6.9 &  1.4 & \hspace*{-3ex} 8.1 & \hspace*{-2.5ex}  5.7 & \hspace*{-1ex}  0.049 &  0.37 &  0.13 &  37( 5) & 0.40 & \hspace*{-3ex}  4.3 & \hspace*{-3ex}  2.5 & \hspace*{-3ex}  9.0 & \hspace*{-3ex} 4.9(0.7) \\  Cha3-C6 & \hspace*{-2ex} 6 & \hspace*{-2ex}  7.7 $\times$  6.9 &  1.1 & \hspace*{-3ex} 7.8 & \hspace*{-2.5ex}  1.9 & \hspace*{-1ex}  0.047 &  0.30 &  0.13 &  37( 6) & 0.36 & \hspace*{-3ex}  4.2 & \hspace*{-3ex}  2.7 & \hspace*{-3ex}  8.7 & \hspace*{-3ex} 4.8(0.7) \\  Cha3-C7 & \hspace*{-2ex} 7 & \hspace*{-2ex}  8.3 $\times$  4.7 &  1.8 & \hspace*{-3ex} 7.6 & \hspace*{-2.5ex}  5.1 & \hspace*{-1ex}  0.046 &  0.23 &  0.10 &  44( 7) & 0.32 & \hspace*{-3ex}  4.1 & \hspace*{-3ex}  3.4 & \hspace*{-3ex}  7.1 & \hspace*{-3ex} 5.8(0.9) \\  Cha3-C8 & \hspace*{-2ex} 8 & \hspace*{-2ex} 14.1 $\times$  5.0 &  2.8 & \hspace*{-3ex} 7.4 & \hspace*{-2.5ex}  4.5 & \hspace*{-1ex}  0.045 &  0.38 &  0.11 &  39( 6) & 0.40 & \hspace*{-3ex}  4.0 & \hspace*{-3ex}  2.3 & \hspace*{-3ex}  7.7 & \hspace*{-3ex} 5.2(0.8) \\  Cha3-C9 & \hspace*{-2ex} 9 & \hspace*{-2ex} 12.5 $\times$  4.6 &  2.7 & \hspace*{-3ex} 7.4 & \hspace*{-2.5ex}  4.9 & \hspace*{-1ex}  0.045 &  0.32 &  0.11 &  41( 7) & 0.37 & \hspace*{-3ex}  4.0 & \hspace*{-3ex}  2.6 & \hspace*{-3ex}  7.3 & \hspace*{-3ex} 5.4(0.9) \\  Cha3-C10 & \hspace*{-2ex} 10 & \hspace*{-2ex}  8.4 $\times$  2.6 &  3.2 & \hspace*{-3ex} 7.2 & \hspace*{-2.5ex}    6 & \hspace*{-1ex}  0.044 &  0.16 &  0.083 &  52(10) & 0.30 & \hspace*{-3ex}  3.9 & \hspace*{-3ex}  5.6 & \hspace*{-3ex}  5.7 & \hspace*{-3ex} 6.9(1.3) \\  Cha3-C11 & \hspace*{-2ex} 11 & \hspace*{-2ex}  7.9 $\times$  4.7 &  1.7 & \hspace*{-3ex} 7.0 & \hspace*{-2.5ex}  4.9 & \hspace*{-1ex}  0.042 &  0.20 &  0.10 &  42( 7) & 0.29 & \hspace*{-3ex}  3.7 & \hspace*{-3ex}  3.2 & \hspace*{-3ex}  6.8 & \hspace*{-3ex} 5.5(1.0) \\  Cha3-C12 & \hspace*{-2ex} 12 & \hspace*{-2ex} 11.2 $\times$  5.9 &  1.9 & \hspace*{-3ex} 6.7 & \hspace*{-2.5ex}    7 & \hspace*{-1ex}  0.040 &  0.31 &  0.10 &  40( 7) & 0.34 & \hspace*{-3ex}  3.6 & \hspace*{-3ex}  2.1 & \hspace*{-3ex}  6.8 & \hspace*{-3ex} 5.3(1.0) \\  Cha3-C13 & \hspace*{-2ex} 13 & \hspace*{-2ex}  9.3 $\times$  4.6 &  2.0 & \hspace*{-3ex} 6.8 & \hspace*{-2.5ex}  4.6 & \hspace*{-1ex}  0.041 &  0.22 &  0.10 &  41( 7) & 0.30 & \hspace*{-3ex}  3.6 & \hspace*{-3ex}  2.8 & \hspace*{-3ex}  6.8 & \hspace*{-3ex} 5.4(1.0) \\  Cha3-C14 & \hspace*{-2ex} 14 & \hspace*{-2ex} 10.4 $\times$  2.4 &  4.3 & \hspace*{-3ex} 6.6 & \hspace*{-2.5ex}    8 & \hspace*{-1ex}  0.040 &  0.17 &  0.083 &  48( 9) & 0.30 & \hspace*{-3ex}  3.5 & \hspace*{-3ex}  4.8 & \hspace*{-3ex}  5.6 & \hspace*{-3ex} 6.3(1.2) \\  Cha3-C15 & \hspace*{-2ex} 15 & \hspace*{-2ex}  8.9 $\times$  5.0 &  1.8 & \hspace*{-3ex} 6.4 & \hspace*{-2.5ex}  5.3 & \hspace*{-1ex}  0.039 &  0.21 &  0.088 &  44( 9) & 0.28 & \hspace*{-3ex}  3.5 & \hspace*{-3ex}  2.6 & \hspace*{-3ex}  6.0 & \hspace*{-3ex} 5.8(1.1) \\  Cha3-C16 & \hspace*{-2ex} 16 & \hspace*{-2ex}  9.6 $\times$  4.5 &  2.1 & \hspace*{-3ex} 6.0 & \hspace*{-2.5ex}  5.4 & \hspace*{-1ex}  0.037 &  0.20 &  0.082 &  44( 9) & 0.27 & \hspace*{-3ex}  3.2 & \hspace*{-3ex}  2.5 & \hspace*{-3ex}  5.6 & \hspace*{-3ex} 5.8(1.2) \\  Cha3-C17 & \hspace*{-2ex} 17 & \hspace*{-2ex}  3.3 $\times$  2.1 &  1.6 & \hspace*{-3ex} 6.1 & \hspace*{-2.5ex}  5.7 & \hspace*{-1ex}  0.037 &  0.053 &  0.048 &  77(21) & 0.18 & \hspace*{-3ex}  3.3 & \hspace*{-3ex} 10.7 & \hspace*{-3ex}  3.3 & \hspace*{-3ex} 10.0(2.7) \\  Cha3-C18 & \hspace*{-2ex} 18 & \hspace*{-2ex}  4.3 $\times$  3.3 &  1.3 & \hspace*{-3ex} 6.2 & \hspace*{-2.5ex}    9 & \hspace*{-1ex}  0.038 &  0.091 &  0.12 &  30( 5) & 0.21 & \hspace*{-3ex}  3.4 & \hspace*{-3ex}  6.1 & \hspace*{-3ex}  8.4 & \hspace*{-3ex} 4.0(0.7) \\  Cha3-C19 & \hspace*{-2ex} 19 & \hspace*{-2ex}  6.8 $\times$  3.5 &  1.9 & \hspace*{-3ex} 6.0 & \hspace*{-2.5ex}  4.8 & \hspace*{-1ex}  0.036 &  0.13 &  0.091 &  40( 8) & 0.23 & \hspace*{-3ex}  3.2 & \hspace*{-3ex}  3.9 & \hspace*{-3ex}  6.2 & \hspace*{-3ex} 5.2(1.1) \\  Cha3-C20 & \hspace*{-2ex} 20 & \hspace*{-2ex}  4.8 $\times$  3.8 &  1.2 & \hspace*{-3ex} 6.1 & \hspace*{-2.5ex}    9 & \hspace*{-1ex}  0.037 &  0.10 &  0.083 &  45( 9) & 0.22 & \hspace*{-3ex}  3.3 & \hspace*{-3ex}  4.8 & \hspace*{-3ex}  5.6 & \hspace*{-3ex} 5.9(1.2) \\  Cha3-C21 & \hspace*{-2ex} 21 & \hspace*{-2ex}  9.5 $\times$  2.1 &  4.5 & \hspace*{-3ex} 5.5 & \hspace*{-2.5ex}  5.9 & \hspace*{-1ex}  0.033 &  0.11 &  0.060 &  55(14) & 0.22 & \hspace*{-3ex}  3.0 & \hspace*{-3ex}  4.5 & \hspace*{-3ex}  4.0 & \hspace*{-3ex} 7.3(1.8) \\  Cha3-C22 & \hspace*{-2ex} 22 & \hspace*{-2ex}  5.8 $\times$  3.8 &  1.5 & \hspace*{-3ex} 5.7 & \hspace*{-2.5ex}  4.3 & \hspace*{-1ex}  0.034 &  0.11 &  0.073 &  47(10) & 0.21 & \hspace*{-3ex}  3.0 & \hspace*{-3ex}  3.8 & \hspace*{-3ex}  5.0 & \hspace*{-3ex} 6.1(1.4) \\  Cha3-C23 & \hspace*{-2ex} 23 & \hspace*{-2ex} 17.7 $\times$  3.1 &  5.8 & \hspace*{-3ex} 5.6 & \hspace*{-2.5ex}  2.8 & \hspace*{-1ex}  0.034 &  0.26 &  0.069 &  49(11) & 0.32 & \hspace*{-3ex}  3.0 & \hspace*{-3ex}  2.4 & \hspace*{-3ex}  4.7 & \hspace*{-3ex} 6.4(1.5) \\  Cha3-C24 & \hspace*{-2ex} 24 & \hspace*{-2ex}  9.6 $\times$  6.0 &  1.6 & \hspace*{-3ex} 5.5 & \hspace*{-2.5ex}    7 & \hspace*{-1ex}  0.033 &  0.23 &  0.091 &  37( 8) & 0.27 & \hspace*{-3ex}  3.0 & \hspace*{-3ex}  1.8 & \hspace*{-3ex}  6.2 & \hspace*{-3ex} 4.8(1.0) \\  Cha3-C25 & \hspace*{-2ex} 25 & \hspace*{-2ex}  7.2 $\times$  2.1 &  3.4 & \hspace*{-3ex} 5.4 & \hspace*{-2.5ex}  4.3 & \hspace*{-1ex}  0.032 &  0.081 &  0.052 &  62(17) & 0.18 & \hspace*{-3ex}  2.9 & \hspace*{-3ex}  4.9 & \hspace*{-3ex}  3.5 & \hspace*{-3ex} 8.1(2.2) \\  Cha3-C26 & \hspace*{-2ex} 26 & \hspace*{-2ex}  2.3 $\times$  2.1 &  1.1 & \hspace*{-3ex} 5.2 & \hspace*{-2.5ex}  3.7 & \hspace*{-1ex}  0.031 &  0.039 &  0.054 &  58(16) & 0.16 & \hspace*{-3ex}  2.8 & \hspace*{-3ex} 12.9 & \hspace*{-3ex}  3.6 & \hspace*{-3ex} 7.6(2.1) \\  Cha3-C27 & \hspace*{-2ex} 27 & \hspace*{-2ex}  6.7 $\times$  2.1 &  3.2 & \hspace*{-3ex} 5.1 & \hspace*{-2.5ex}  2.3 & \hspace*{-1ex}  0.031 &  0.071 &  0.049 &  63(18) & 0.17 & \hspace*{-3ex}  2.7 & \hspace*{-3ex}  4.9 & \hspace*{-3ex}  3.3 & \hspace*{-3ex} 8.2(2.4) \\  Cha3-C28 & \hspace*{-2ex} 28 & \hspace*{-2ex}  4.9 $\times$  2.1 &  2.3 & \hspace*{-3ex} 5.1 & \hspace*{-2.5ex}  3.2 & \hspace*{-1ex}  0.031 &  0.056 &  0.038 &  81(27) & 0.16 & \hspace*{-3ex}  2.7 & \hspace*{-3ex}  6.2 & \hspace*{-3ex}  2.6 & \hspace*{-3ex} 10.7(3.6) \\  Cha3-C29 & \hspace*{-2ex} 29 & \hspace*{-2ex}  5.3 $\times$  3.0 &  1.7 & \hspace*{-3ex} 5.1 & \hspace*{-2.5ex}  3.7 & \hspace*{-1ex}  0.031 &  0.082 &  0.061 &  50(13) & 0.18 & \hspace*{-3ex}  2.7 & \hspace*{-3ex}  4.6 & \hspace*{-3ex}  4.2 & \hspace*{-3ex} 6.5(1.7) \\  \hline
 \end{tabular}
 \tablefoot{
 \tablefoottext{a}{Numbering of \textit{Gaussclumps} sources like in Table~\ref{t:id_gcl_simbad}.}
 \tablefoottext{b}{Deconvolved physical source size ($FWHM$) and aspect ratio ($R_{\mathrm{a}}$) of the fitted Gaussian. The minimum size that can be measured is 2100~AU (see Sect.~\ref{sss:starless_sizes}). The aspect ratio is the ratio of the deconvolved sizes along the major and minor axes.}
 \tablefoottext{c}{Peak H$_2$ column density. The statistical rms uncertainty is $1.0 \times 10^{21}$~cm$^{-2}$.}
 \tablefoottext{d}{Visual extinction derived from 2MASS.}
 \tablefoottext{e}{Mass in the central beam ($HPBW = 21.2\arcsec$) (${M_{\mathrm{peak}}}$), total mass derived from the Gaussian fit (${M_{\mathrm{tot}}}$), and mass computed from the flux measured in an aperture of 50$\arcsec$ in diameter (${M_{50\arcsec}}$). The statistical rms uncertainties of $M_{\mathrm{peak}}$ and $M_{50\arcsec}$ are 0.006 and 0.010 M$_\odot$, respectively.}
 \tablefoottext{f}{Mass concentration $m_{\mathrm{peak}}/m_{50\arcsec}$. The statistical rms uncertainty is given in parentheses.}
 \tablefoottext{g}{Ratio $M_{\mathrm{tot}}/M_{\mathrm{BE}}$, with $M_{\mathrm{BE}}$ the critical Bonnor-Ebert mass (see Sect.~\ref{sss:starless_massvssize}).}
 \tablefoottext{h}{Beam-averaged free-particle density within the central beam (${n_{\mathrm{peak}}}$) and mean free-particle densities computed for the total mass (${n_{\mathrm{mean}}}$) and the mass $M_{50\arcsec}$ in the aperture of diameter 50\arcsec (${n_{50\arcsec}}$). The statistical rms uncertainties of $n_{\mathrm{peak}}$ and $n_{50\arcsec}$ are $5.4 \times 10^{4}$ and $6.9 \times 10^{3}$ cm$^{-3}$, respectively.}
 \tablefoottext{i}{Density contrast $n_{\mathrm{peak}}/n_{50\arcsec}$. The statistical rms uncertainty is given in parentheses.}
 }
\end{table*}

\begin{figure*}
%\centerline{\resizebox{0.85\hsize}{!}{\includegraphics[angle=270]{/homes/belloche/Chamaeleon/Continuum/Cha3/Analysis/histo_artcha3cont.eps}}}
\centerline{\resizebox{0.85\hsize}{!}{\includegraphics[angle=270]{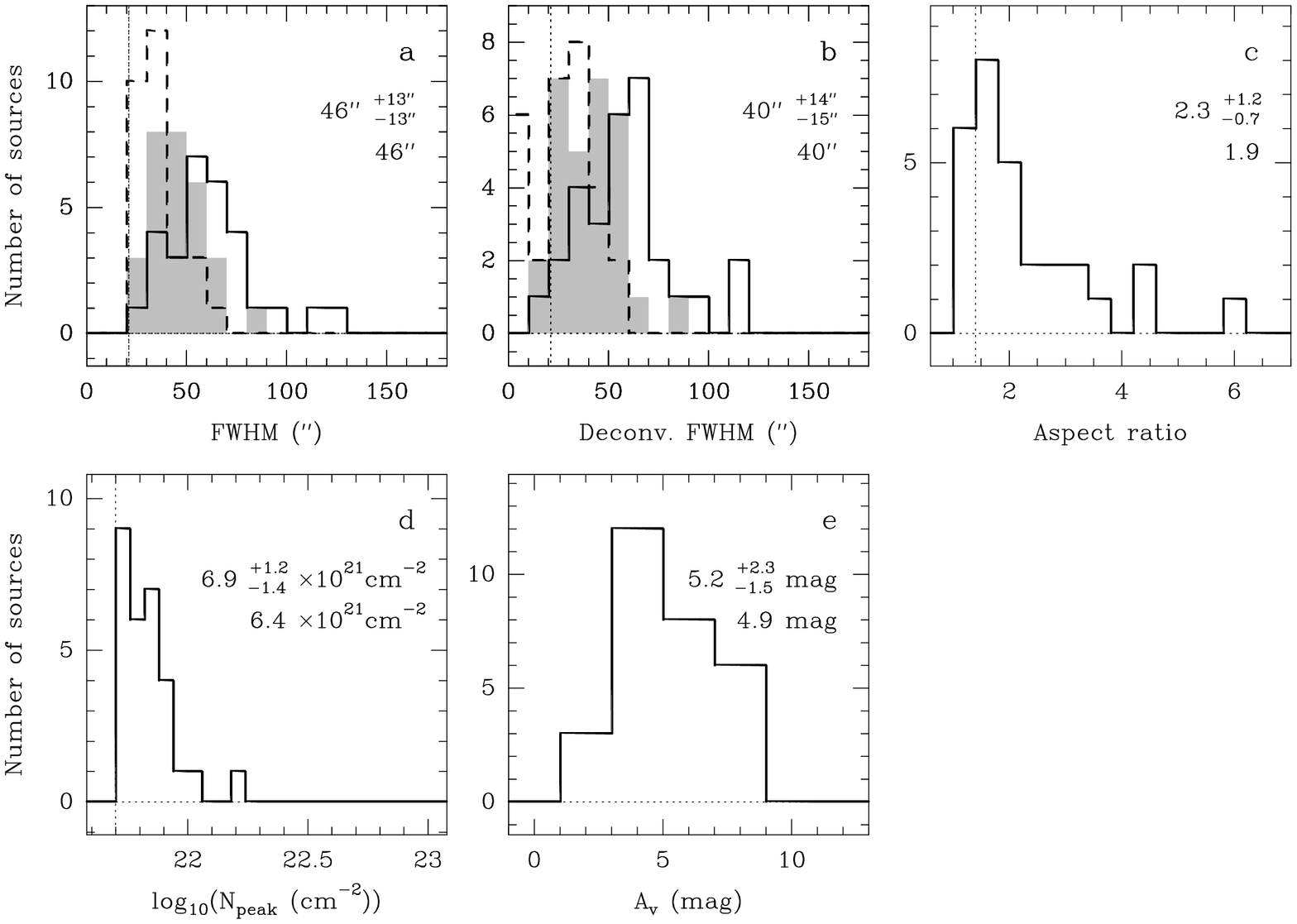}}}
\caption{Distributions of physical properties obtained for the 29 starless 
sources found with \textit{Gaussclumps} in the \textit{sum} map of 
Cha~III at scale 5. The mean, standard deviation, and median of the 
distribution are given in each panel. The asymmetric standard deviation defines
the range containing 68$\%$ of the sample.
\textbf{a} $FWHM$ sizes along the major (solid line) and minor (dashed 
line) axes. The filled histogram shows the distribution of geometrical mean 
of major and minor sizes. The mean and median values refer to the filled
histogram. The dotted line indicates the angular resolution 
(21.2$\arcsec$). 
\textbf{b} Same as \textbf{a} but for the deconvolved sizes.
\textbf{c} Aspect ratios computed with the deconvolved sizes. The dotted line 
at 1.4 shows the threshold above which the deviation from 1 (elongation) can 
be considered as significant.
\textbf{d} Peak H$_2$ column density. The dotted line at
$5.0 \times 10^{21}$~cm$^{-2}$ is the 5$\sigma$ sensitivity limit.
\textbf{e} Visual extinction derived from 2MASS.}
\label{f:histo}
\end{figure*}

The visual extinctions listed in Table~\ref{t:starless} and plotted in 
Fig.~\ref{f:histo}e are extracted from the extinction map derived from 2MASS 
(see Sect.~\ref{ss:2mass}). Given the lower resolution of this map 
($HPBW = 3\arcmin$) compared to the 870~$\mu$m map, it provides an estimate of 
the extinction of the environment in which the 870~$\mu$m sources are embedded.

The 870~$\mu$m sources are found down to a visual extinction, $A_V$, of 
$\sim 1.9$~mag
(as traced with 2MASS at low angular resolution). This is in marked contrast
with the threshold $A_V \sim$ 5--7~mag above which starless sources are found 
in other low-mass star forming regions like Cha~I (Paper~I), 
Ophiuchus \citep[][]{Johnstone04}, Perseus \citep[][]{Enoch06,Kirk06}, Taurus
\citep[][]{Goldsmith08}, and Aquila \citep[][]{Andre11}. We note
however that, with its high sensitivity, \textit{Herschel} revealed many
starless cores at low extinction in the nearby Polaris flare region, a  
non-star-forming molecular cloud \citep[][]{Andre10}.

About half of the
Cha~III sources are found at $A_V <$ 5~mag (median 4.9~mag), and the 
distribution peaks at $A_V \sim$ 4~mag. Both the median and mean as well as the
peak of the distribution are about a factor of 2 lower than in Cha~I 
(see Fig.~6e of Paper~I for comparison).

\subsubsection{Sizes}
\label{sss:starless_sizes}

The source sizes along the major and minor axes before and after deconvolution 
are listed in Cols.~6 and 7 of Table~\ref{t:id_gcl_simbad} and Col.~3 of 
Table~\ref{t:starless}, respectively. Their distributions are shown in 
Figs.~\ref{f:histo}a and b, respectively, along with the distribution of mean 
size (geometrical mean of major and minor sizes, i.e. 
$\sqrt{FWHM_{\mathrm{maj}} \times FWHM_{\mathrm{min}}}$). The average major, 
minor, and mean sizes are $62\,^{+15}_{-21}\,\arcsec$, 
$35\,^{+13}_{-13}\,\arcsec$, and $46\,^{+13}_{-13}\,\arcsec$, respectively.
These angular sizes correspond to physical sizes of 9300, 5300, and
6900~AU, respectively.
Only two sources have a major $FWHM$ size larger than 110$\arcsec$, and no 
source has a minor or mean $FWHM$ size larger than 90$\arcsec$. The results of 
the Monte-Carlo simulations of Paper~I in the elliptical case (see also 
Sect.~\ref{ss:reduction} and Appendix~\ref{ss:filtering}) imply that these 
sources, although nearly 
all of them are faint with a peak flux density lower than 150~mJy/beam, are 
not significantly affected by the spatial filtering due to the sky noise 
removal, with less than 15$\%$ loss of peak flux density and size.

Like for Cha~I (see Paper~I), the accuracy to which we can measure the 
size of a weak $\sim 5\sigma$ unresolved source is $4.2\arcsec$ 
($21.2/5$). Therefore, faint sources with a size smaller
than $\sim 25.4\arcsec$ cannot be reliably deconvolved and we artificially set
their size to $25.4\arcsec$ to perform the deconvolution. As a result the 
minimum deconvolved $FWHM$ size that we can measure is $\sim 14\arcsec$ 
(2100 AU). The average deconvolved mean $FWHM$ size is 
$40\,^{+14}_{-15}\,\arcsec$, i.e. $6000\,^{+2100}_{-2250}$~AU (see 
Fig.~\ref{f:histo}b). It is only $8\%$ smaller than the average deconvolved 
mean size of the Cha~I sources so the conclusions of Paper~I hold 
as well: the Cha~III sources have 
similar physical sizes as the Perseus cores, are probably larger than the 
Serpens cores (maybe by a factor of 1.5--2), are certainly larger than the
Ophiuchus cores (by a factor of 2--3), but are significantly smaller than the 
Taurus cores (by a factor of 3). 

The comparison to the population of dense cores in the Pipe nebula is less 
straightforward because these cores were extracted from an extinction map with 
\textit{Clumpfind} and only the radii of their lowest contours are available, 
not their \textit{FWHM} sizes \citep[][]{Alves07,Lada08,Rathborne09}. In 
addition, the resolution of the extinction map is $1\arcmin$, i.e. 7800~AU at 
a distance of 130~pc \citep[][]{Lombardi06}. Still, since their mean 
\textit{radius} is about 19000~AU \citep[][]{Rathborne09}, it seems very 
likely that the Pipe dense cores traced by the extinction are larger by a 
factor of a few compared to the Cha~III cores detected with LABOCA. 

\subsubsection{Aspect ratios and orientations}
\label{sss:aspect_ratios}

The distribution of aspect ratios computed with the deconvolved $FWHM$ sizes 
is shown in Fig.~\ref{f:histo}c. Based on the Monte Carlo simulations of 
Paper~I, we estimate that a faint source can reliably be
considered as intrinsically elongated when its aspect ratio is higher than 
1.4. 76$\%$ of the sources are above this threshold and can be 
considered as elongated, which is similar to Cha~I. The average aspect ratio 
is $2.3\,^{+1.2}_{-0.7}$. It is similar to the ones measured in Cha~I, 
Serpens, and Taurus, somewhat larger than in Perseus, and significantly larger 
than in Ophiuchus (see Table~7 of Paper~I).

The distribution of position angles (Col.~8 of Table~\ref{t:id_gcl_simbad}) 
does not show any preferred direction. An inspection by eye does not reveal
any particular alignment of the elongated sources with the putative 
filaments mentioned in Sect.~\ref{ss:labocamap} either, especially in field 
Cha3-Center (Fig.~\ref{f:labocamapdet}c).

\subsubsection{Column densities}
\label{sss:starless_coldens}

The median peak H$_2$ column density of the starless sources in Cha~III is 
$6.4 \times 10^{21}$~cm$^{-2}$ (Fig.~\ref{f:histo}d), nearly identical to the
median value found in Cha I 
($6.5 \times 10^{21}$~cm$^{-2}$, see Paper~I). However, the average 
peak H$_2$ column density of the starless sources in Cha~III 
($6.9\,^{+1.2}_{-1.4} \times 10^{21}$~cm$^{-2}$) is 1.4 times lower than the 
average peak column density of the starless cores in Cha~I. It is 5 times 
lower than in Perseus and Serpens, and 4 to 9 times lower than in Ophiuchus 
(see Table~7 of Paper~I). It appears to be significantly lower 
than in Taurus too (by a factor of 3), but since the Taurus sample is not 
complete and the source extraction methods differ, we cannot draw any firm 
conclusion.
 
\subsubsection{Masses and densities}
\label{sss:starless_masses}

\begin{figure*}
%\centerline{\resizebox{1.00\hsize}{!}{\includegraphics[angle=270]{/homes/belloche/Chamaeleon/Continuum/Cha3/Analysis/histom_artcha3cont.eps}}}
\centerline{\resizebox{1.00\hsize}{!}{\includegraphics[angle=270]{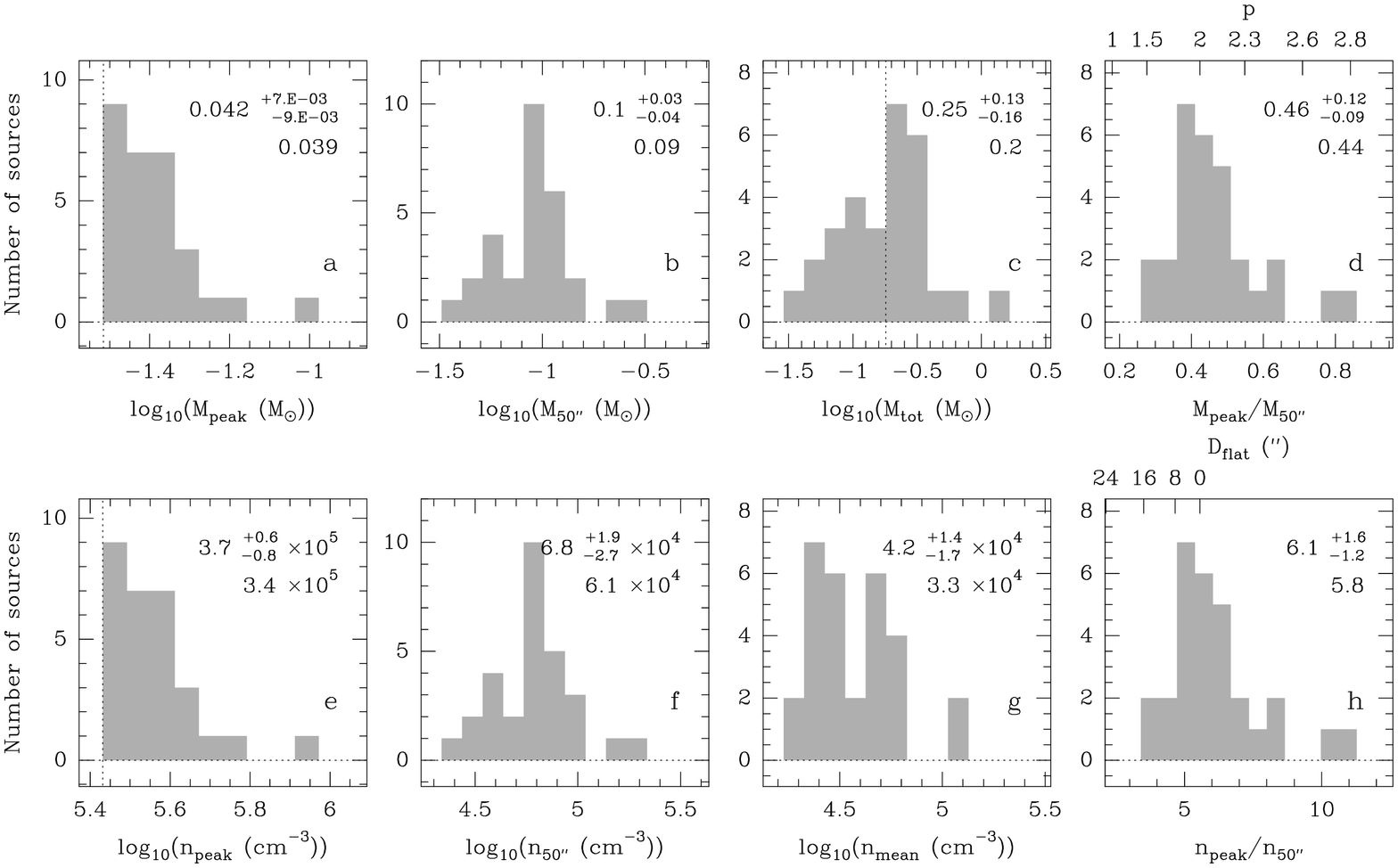}}}
\caption{Distributions of masses and free-particle densities obtained 
for the 29 starless
sources found with \textit{Gaussclumps} in the \textit{sum} map of 
Cha~III at scale 5. The mean, standard deviation, and median of the 
distribution are given in each panel. The asymmetric standard deviation defines
the range containing 68$\%$ of the sample.
\textbf{a} Peak mass of the fitted Gaussian.
\textbf{b} Mass within an aperture of diameter $50\arcsec$.
\textbf{c} Total mass of the fitted Gaussian. The dotted line indicates the 
estimated completeness limit at 90$\%$ for Gaussian sources corresponding to a 
6.3$\sigma$ peak detection limit for the average source size. 
\textbf{d} Mass concentration, ratio of the peak mass to the mass within an 
aperture of diameter $50\arcsec$. 
\textbf{e} Peak density.
\textbf{f} Mean density within an aperture of diameter $50\arcsec$.
\textbf{g} Mean density derived from the total mass.
\textbf{h} Density contrast, ratio of peak density to mean density.
In panels \textbf{a} and \textbf{e}, the dotted line indicates the 
$5\sigma$ sensitivity limit.
The upper axis of panel \textbf{d}, which can also be used for panel 
\textbf{h}, shows the power-law exponent $p$ derived assuming that the sources 
have a density profile proportional to $r^{-p}$ and a uniform dust temperature. 
Alternately, the upper axis of panel \textbf{h}, which can also be used for 
panel \textbf{d}, deals with the case where the density is uniform within a 
diameter $D_{\mathrm{flat}}$ and decreasing as $r^{-2}$ outside, still with 
the assumption of a uniform temperature. 
}
\label{f:histom}
\end{figure*}

The distribution of masses and free-particle densities listed in 
Table~\ref{t:starless} are displayed in Fig.~\ref{f:histom}. The $5\sigma$ 
sensitivity limit used to extract sources with \textit{Gaussclumps} 
corresponds to a peak mass of 0.030 M$_\odot$ and a peak density of 
$2.7 \times 10^5$~cm$^{-3}$, computed for a diameter of 21.2$\arcsec$ 
(3200~AU). The median of the peak mass distribution is 0.039~M$_\odot$, 
implying a median peak density of $3.4 \times 10^5$~cm$^{-3}$ 
(Figs.~\ref{f:histom}a and e). We give the mass integrated within an aperture 
of diameter 50$\arcsec$ (7500~AU) in Col.~9 of Table~\ref{t:starless}, which 
is the aperture used for Cha~I (Paper~I). This aperture is well adapted to the 
Cha~III sample too, for the same three reasons: it nearly corresponds to the 
average mean, undeconvolved $FWHM$ size (see Sect.~\ref{sss:starless_sizes} and 
Fig.~\ref{f:histo}a), it is not affected by the spatial filtering due to the 
sky noise removal (see Appendix~A.1 and Table~A.1 of Paper~I), and 
it is still preserved in the \textit{sum} map at scale 5 
(see Appendix~C and Table~C.1 of Paper~I). The median of the mass 
integrated within this aperture is 0.09~M$_\odot$, corresponding to a median 
mean density of $6 \times 10^4$~cm$^{-3}$ (Figs.~\ref{f:histom}b and f).
The median values of the peak and aperture masses are nearly the same as those
of the Cha~I sample, but the mean values are smaller by about 30$\%$. Within
the uncertainties, the mass distributions of Cha~I and III are very similar,
the only main difference being that the four most massive cores of Cha~I 
have no counterpart in Cha~III.

Figure~\ref{f:histom}c shows the distribution of total masses computed 
from the Gaussian fits (Col.~8 of Table~\ref{t:starless}). The 
completeness limit at 90$\%$ is estimated from a peak flux detection threshold 
at 6.3$\sigma$ for the average size of the source sample 
($FWHM = 46\arcsec$)\footnote{For a Gaussian distribution of mean value $m$ 
and standard deviation $\sigma$, the relative population below $m-1.28\sigma$ 
represents 10$\%$. Therefore our peak flux detection threshold at 5$\sigma$ 
implies a 90$\%$ completeness limit at $6.3\sigma$, with $\sigma$ the rms 
noise  level in the \textit{sum} map at scale 5.}. It corresponds to a total 
mass of 0.18~M$_\odot$, slightly lower than in Cha~I 
(0.22~M$_\odot$, see Paper~I). Like in Cha~I, the median total 
mass is very similar (0.20~M$_\odot$), which implies that only 50$\%$ of the 
detected sources are above the estimated $90\%$ completeness limit. The 
mass completeness limit is slightly better than that obtained by 
\citet{Konyves10} for their 11 deg$^2$ sensitive continuum survey of the 
Aquila Rift cloud complex (distance 260~pc) with \textit{Herschel}, and about a 
factor of 6 better than that obtained by 
\citet{Enoch08} in Perseus (see Paper~I for details). It is however a factor 
$\sim 20$ worse than that obtained with \textit{Herschel} for the Polaris 
flare region which is at the same distance as Cha~III \citep[][]{Andre10}.

The mean density of each source is estimated from its total mass as derived 
from the Gaussian fits and a \textit{radius} set equal to 
$\sqrt{FWHM_{\mathrm{maj}} \times FWHM_{\mathrm{min}}}$. It is given in Col.~13 of 
Table~\ref{t:starless} and the distribution for the full Cha~III sample is 
shown in Fig.~\ref{f:histom}g. The average and median mean densities are 
$4.2^{+1.4}_{-1.2} \times 10^4$ and $3.3 \times 10^4$~cm$^{-3}$, respectively. 
For Chamaeleon~I, the corresponding numbers are $4.8^{+2.4}_{-2.6} \times 10^4$ 
and $3.6 \times 10^4$~cm$^{-3}$, respectively 
(not given in Paper~I). The Cha~I and III sources have thus very 
similar mean densities, a factor of $\sim 5$ higher than the mean density of 
the cores of the Pipe nebula extracted from extinction maps \citep[][]{Lada08}. 

We estimate the mass concentration of the Cha~III sources from the ratio of 
the peak mass to the mass within an aperture of 50$\arcsec$ (Col.~10 of 
Table~\ref{t:starless}) which is relatively insensitive to the spatial 
filtering due to
the data reduction. A similar property is the density contrast measured as the 
ratio of the peak density to the mean density within this aperture (Col.~15 of 
Table~\ref{t:starless}). The statistical rms uncertainties on the peak mass 
and the mass within 50$\arcsec$ are 0.006 and 0.010~M$_\odot$, respectively, 
which means a relative uncertainty of up to 25$\%$ for the weakest source. 
The distributions of both ratios are shown in Figs.~\ref{f:histom}d and h and 
their rms uncertainties\footnote{The relative uncertainty of the ratio is 
equal to the square root of the quadratic sum of the relative uncertainties of 
its two terms, i.e. we assume both terms are uncorrelated.} are given in 
parentheses in Cols.~10 and 15 of Table~\ref{t:starless}. The two outliers 
with the largest ratios are also those with the highest relative uncertainty 
(about $30\%$).
The upper axis of Fig.~\ref{f:histom}d, which can also be used for 
Fig.~\ref{f:histom}h, displays the exponent of the density profile under the 
assumptions that the sources are spherically symmetric with a power-law 
density profile, i.e. $\rho \propto r^{-p}$, and that the dust temperature is 
uniform. The median mass concentration and density contrast are 0.44 and 5.8, 
respectively, similar to the Cha~I sample. This corresponds to 
$p \sim 2.0$, suggesting that most sources are significantly centrally-peaked 
(see Paper~I for the caveats of this estimate). It is similar to 
the exponent of the singular isothermal sphere.

The upper axis of Fig.~\ref{f:histom}h, which can also be used for 
Fig.~\ref{f:histom}d, deals with an alternate case where the sources have a 
constant density within a diameter $D_{\mathrm{flat}}$ and a density 
decreasing as $r^{-2}$ outside, still with the assumption of a uniform 
temperature. Under these assumptions, the measurements are consistent with a 
flat inner region of diameter 16$\arcsec$ \textit{at most} (2400~AU) for a few 
sources, but most sources have $D_{\mathrm{flat}} < 10\arcsec$ (1500~AU), or 
cannot be described with such a density profile.

\subsubsection{Mass versus size}
\label{sss:starless_massvssize}

\begin{figure*}
%\centerline{\resizebox{1.0\hsize}{!}{\includegraphics[angle=270]{/homes/belloche/Chamaeleon/Continuum/Cha3/Analysis/massvssize_artcha3cont.eps}}}
\centerline{\resizebox{1.0\hsize}{!}{\includegraphics[angle=270]{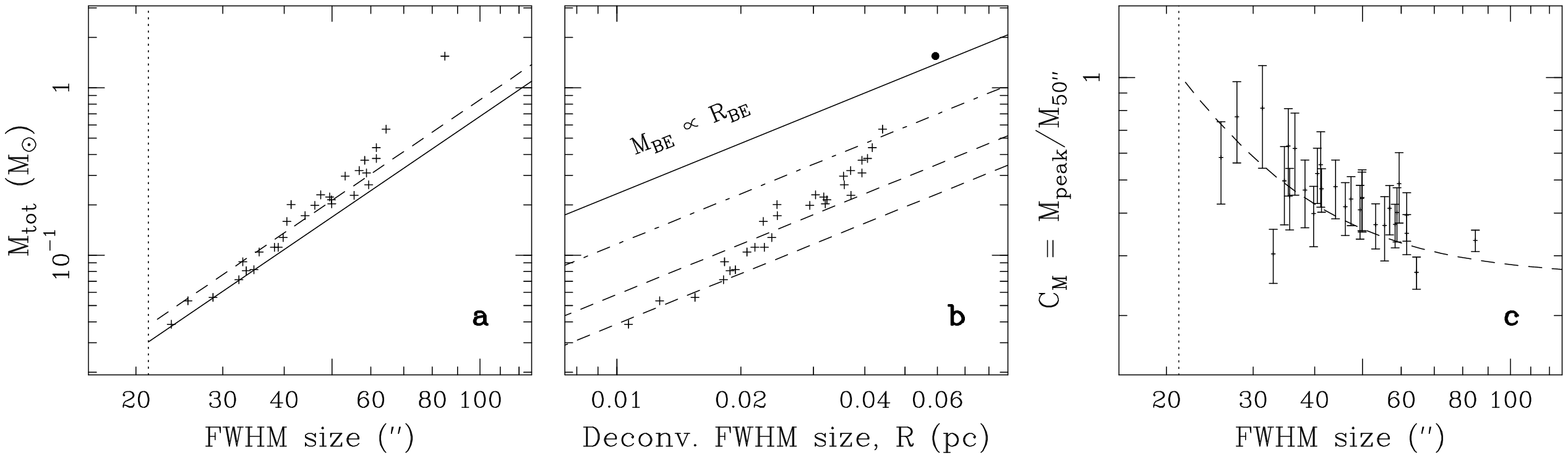}}}
\caption{\textbf{a} Total mass versus mean $FWHM$ size for the 29 starless
sources found with \textit{Gaussclumps} in the \textit{sum} map of 
Cha~III at scale 5. The angular resolution (21.2$\arcsec$) is marked by the 
dotted line. The solid line ($M \propto FWHM^2$) is the 5$\sigma$ peak 
sensitivity limit for Gaussian sources. The dashed line shows the 6.3$\sigma$ 
peak sensitivity limit which corresponds to a completeness limit of 90$\%$ for 
Gaussian sources. 
\textbf{b} Total mass versus mean deconvolved $FWHM$ size. Sizes smaller than
25.4$\arcsec$ were set to 25.4$\arcsec$ before deconvolution (see note b of 
Table~\ref{t:starless}). The solid line shows the relation 
$M = 2.4 \, Ra_{\mathrm{s}}^2/G$ that characterizes critical Bonnor-Ebert 
spheres (see Sect.~\ref{sss:starless_massvssize}). The dash-dotted line
shows the location of this relation when divided by 2, and the dashed lines 
when divided by 4 and 6. The source with a mass larger than the critical 
Bonnor-Ebert mass $M_{\mathrm{BE,}P_{\mathrm{ext}}}$ estimated from the 
ambient cloud pressure is shown with a filled circle. 
\textbf{c} Mass concentration versus mean $FWHM$ size. The dashed line is the 
expectation for a circular Gaussian flux density distribution.}
\label{f:massvssize}
\end{figure*}

The distribution of total masses versus source sizes derived from the Gaussian 
fits is shown in Fig.~\ref{f:massvssize}a. About 50$\%$ of the sources are 
located between the 5$\sigma$ detection limit (solid line) and the 
estimated 
$90\%$ completeness limit (dashed line), suggesting that we most likely miss a 
significant number of sources with a low peak column density. 
Figure~\ref{f:massvssize}b shows a similar diagram for the deconvolved source
size. If we assume that the deconvolved $FWHM$ size is a good estimate of the
external \textit{radius} of each source, then we can compare this distribution 
to the critical Bonnor-Ebert mass that characterizes the limit above which 
the hydrostatic equilibrium of an isothermal sphere with thermal support only 
is gravitationally unstable. This relation 
$M_{\mathrm{BE}}(R) = 2.4 \, Ra_{\mathrm{s}}^2/G$ \citep[][]{Bonnor56}, with 
$M_{\mathrm{BE}}(R)$ the total mass, $R$ the external radius, 
$a_{\mathrm{s}}$ the sound speed, and $G$ the gravitational constant, is drawn 
for a temperature of 12~K as a solid line in Fig.~\ref{f:massvssize}b. We 
define $\alpha_{\mathrm{BE}} = M_{\mathrm{tot}}/M_{\mathrm{BE}}$. Only one source 
(Cha3-C1) has $\alpha_{\mathrm{BE}} >1$, i.e. is located above the critical 
mass limit (see Fig.~\ref{f:massvssize}b and Col.~11 of 
Table~\ref{t:starless}). The threshold $\alpha_{\mathrm{BE}} = 0.5$ 
approximately defines the limit above which an isothermal sphere is 
gravitationally bound if it is only supported by thermal pressure and the 
confinement by the external pressure is negligible. Only one source in 
addition to Cha3-C1 falls above this limit and may be gravitationally bound 
(Cha3-C2, see dash-dotted line in Fig.~\ref{f:massvssize}b). Most 
sources, however, have a mass lower than the critical Bonnor-Ebert mass by a 
factor of 2 to 6. The uncertainty on the temperature 
(see Appendix~B.2 of Paper~I) does not influence these results 
much since, even in the unlikely case of the \textit{bulk} of the mass being 
at a temperature of 7~K, the measured masses would move upwards relative to 
the critical Bonnor-Ebert mass limit by a factor of 1.9 only, because the 
latter is also temperature dependent.

The critical Bonnor-Ebert mass  can also be estimated from the external 
pressure with $M_{\mathrm{BE}}(P_{\mathrm{ext}}) = 1.18 \, a_{\mathrm{s}}^4 \,
G^{-\frac{3}{2}} \, P^{-\frac{1}{2}}_{\mathrm{ext}}$
\citep{Bonnor56}, the external pressure being estimated from the extinction of 
the environment in which the sources are embedded 
(see Paper~I for the equations and references). The 
single source with a mass larger than $M_{\mathrm{BE}}(P_{\mathrm{ext}})$ is the 
same as for $M_{\mathrm{BE}}(R)$. The agreement between both estimates of 
$M_{\mathrm{BE}}$ suggests that our estimates of the external radius and 
external pressure are consistent. No additional source falls above 
the threshold $\alpha_{\mathrm{BE}} = 0.5$ based on 
$M_{\mathrm{BE}}(P_{\mathrm{ext}})$. In summary, only one source is 
likely above the critical Bonnor-Ebert mass limit (Cha3-C1), and one 
additional source may be gravitationally bound if it is supported by
thermal pressure only (Cha3-C2). The implications of this analysis will be 
discussed in Sect.~\ref{s:discussion}.

The mass concentration $C_M$ is plotted versus source size in 
Fig.~\ref{f:massvssize}c. $C_M$ is actually equal to the ratio of the 
peak flux to the flux integrated within the aperture of diameter 50$\arcsec$.
When the sources do not overlap, this ratio is nearly independent of the 
Gaussian fitting since the second and third 
stiffness parameters of \textit{Gaussclumps} were set to 1, i.e. 
\textit{Gaussclumps} was biased to keep the fitted peak amplitude close to the 
observed 
one and the fitted center position close to the position of the observed peak.
The dashed line shows the expected ratio if the (not deconvolved) sources were 
exactly Gaussian and circular and allows us to estimate the departure 
of the sources from being Gaussian within 50$\arcsec$. 
Most sources have a mass concentration consistent with the Gaussian 
expectation, but many of them have a significant uncertainty on $C_M$ that
prevents a more accurate analysis. The obvious outlier toward the lower 
left is source Cha3-C18, which has a strong neighbor significantly 
contaminating its flux within 50$\arcsec$ (source Cha3-C1).

\begin{figure}
%\centerline{\resizebox{1.00\hsize}{!}{\includegraphics[angle=270]{/homes/belloche/Chamaeleon/Continuum/Cha3/Analysis/massvsav_artcha3cont.eps}}}
\centerline{\resizebox{1.00\hsize}{!}{\includegraphics[angle=270]{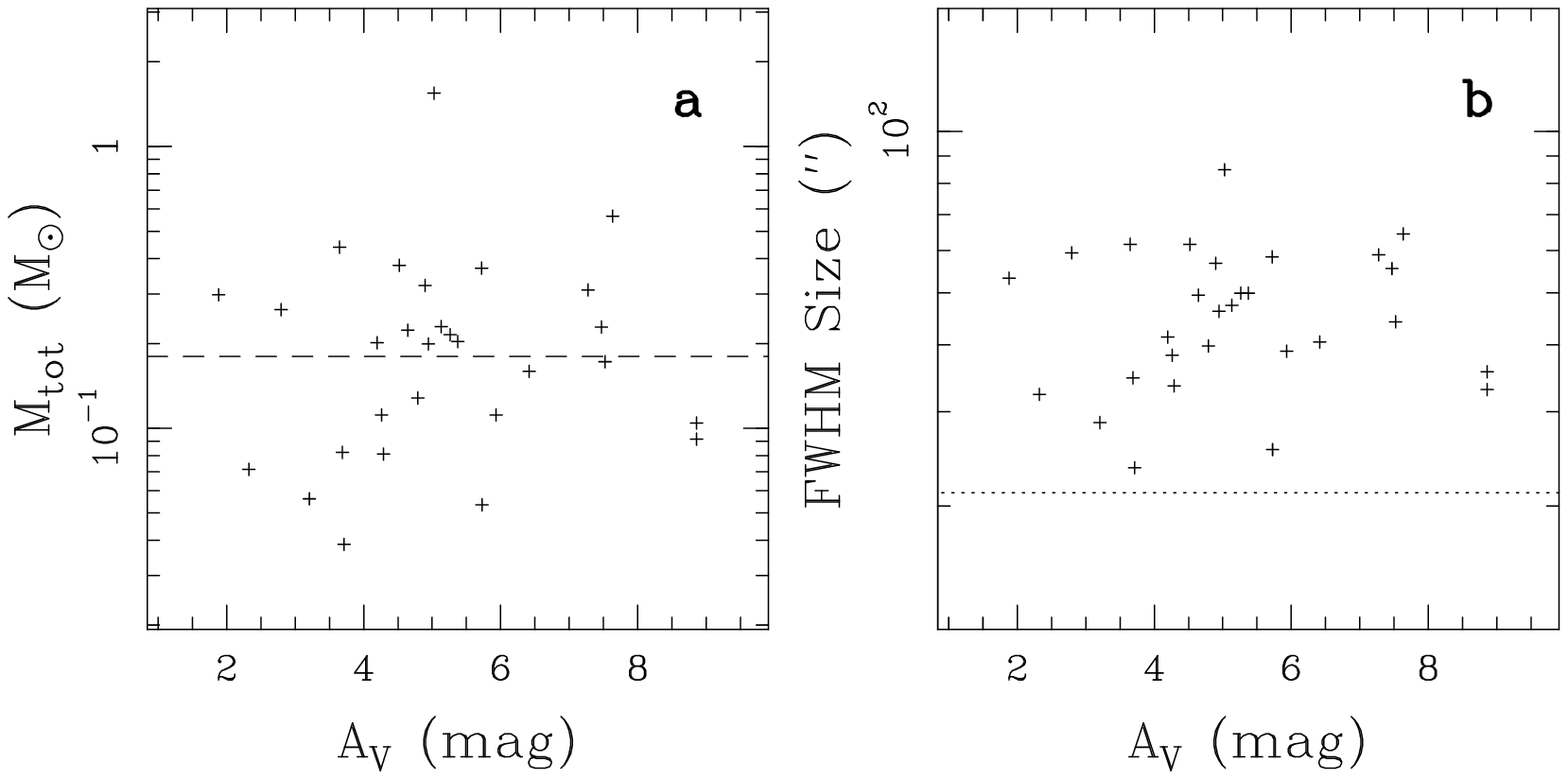}}}
\caption{\textbf{a} Total mass versus visual extinction A$_V$ for the 29 
starless sources found with \textit{Gaussclumps} in the 
\textit{sum} map of Cha~III at scale 5. The dashed line shows the estimated 
90$\%$ completeness limit (0.18 M$_\odot$).
\textbf{b} $FWHM$ size versus visual extinction. The angular resolution 
(21.2$\arcsec$) is marked by the dotted line.}
\label{f:massvsav}
\end{figure}

There is no obvious correlation between the total mass or $FWHM$ size of the 
sources and the visual extinction of the environment in which they are 
embedded (see Fig.~\ref{f:massvsav}). A similar conclusion was drawn for
Cha~I (Paper~I) and for the five nearby molecular clouds 
Ophiuchus, Taurus, Perseus, Serpens, and Orion based on SCUBA data 
\citep[][]{Sadavoy10}.

\subsubsection{Core mass distribution (CMD)}
\label{sss:cmd}

\begin{figure}
%\centerline{\resizebox{1.00\hsize}{!}{\includegraphics[angle=270]{/homes/belloche/Chamaeleon/Continuum/Cha3/Analysis/cmf_artcha3cont.eps}}}
\centerline{\resizebox{1.00\hsize}{!}{\includegraphics[angle=270]{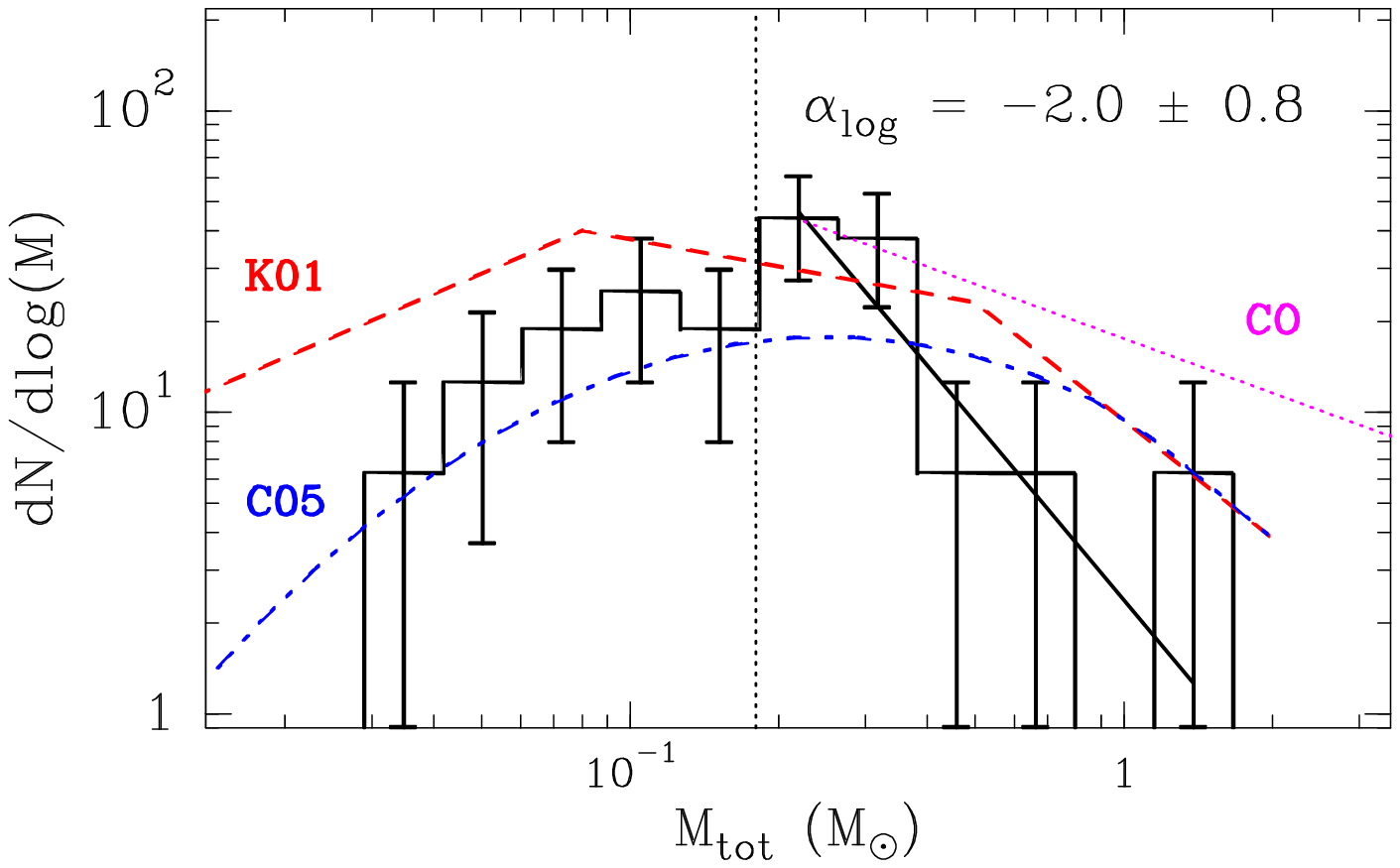}}}
\caption{Mass distribution $\mathrm{d}N/\mathrm{d}\log(M)$ of the 29 starless 
sources. The error bars represent the Poisson noise (in $\sqrt{N}$). The 
vertical dotted line is the estimated 90$\%$ completeness limit. The thick 
solid line is the best power-law fit performed on the mass bins above the 
completeness limit. The best fit exponent $\alpha_{\mathrm{log}}$
is given in the upper right corner. The IMF of 
single stars corrected for binaries \citep[][, K01]{Kroupa01} and the IMF of 
multiple systems \citep[][, C05]{Chabrier05} are shown in dashed (red) and 
dot-dashed (blue) lines, respectively. They are both vertically shifted to 
the same number at 2~M$_\odot$. The dotted (purple) curve is the 
typical mass spectrum of CO clumps \citep[][]{Blitz93,Kramer98}.}
\label{f:cmd}
\end{figure}

The mass distribution of the 29 starless sources is shown in
Fig.~\ref{f:cmd}. Its shape looks very
similar to the shape of the mass distribution found in other star forming 
regions with a power-law-like behavior at the high-mass end and a flattening 
toward the low-mass end. In our case, the flattening occurs below the 
estimated $90\%$ completeness limit (0.18~M$_\odot$) and may not be
significant. Above this limit, the distribution is consistent with a power-law
but is very noisy. The exponent of the best power-law fit 
($\alpha = -3.0 \pm 0.8$ for $\mathrm{d}N/\mathrm{d}M$, 
$\alpha_{\mathrm{log}} = -2.0 \pm 0.8$ for 
$\mathrm{d}N/\mathrm{d}\log(M)$) is consistent, within the uncertainty, with 
the value of \citet{Salpeter55} that characterizes the high-mass end of the 
stellar initial mass function ($\alpha = -2.35$). However, it is also 
consistent within 2$\sigma$ with the exponent of the typical 
mass spectrum of CO clumps \citep[$\alpha = -1.6$, see][]{Blitz93,Kramer98}.
The sample is too small to distinguish statistically between these 
two types of mass distribution.

\subsection{Spatial distribution}
\label{ss:distribution}

\begin{figure}
%\centerline{\resizebox{1.00\hsize}{!}{\includegraphics[angle=270]{/homes/belloche/Chamaeleon/Continuum/Cha3/Analysis/nearest_neighb_cha3_artcha3cont.eps}}}
\centerline{\resizebox{1.00\hsize}{!}{\includegraphics[angle=270]{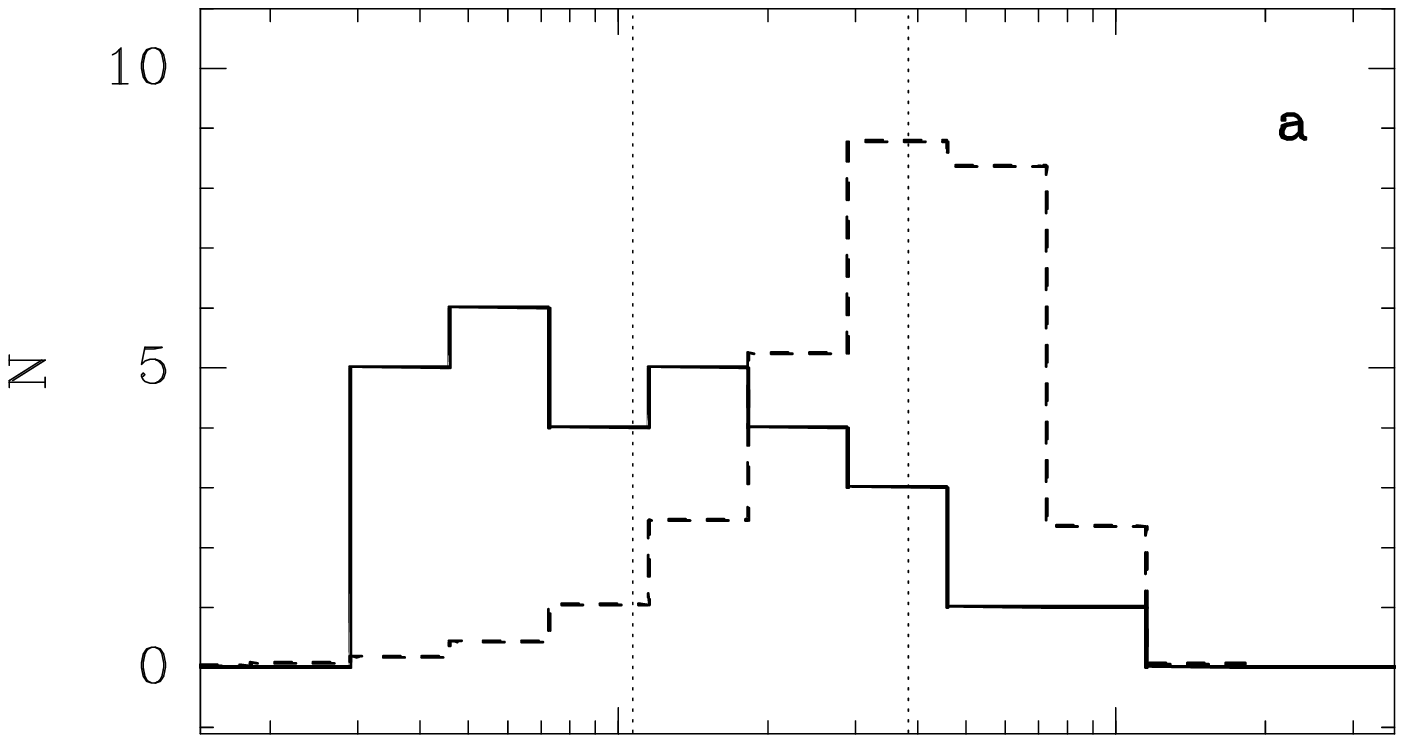}}}
\vspace*{2ex}
%\centerline{\resizebox{1.00\hsize}{!}{\includegraphics[angle=270]{/homes/belloche/Chamaeleon/Continuum/Cha3/Analysis/nearest_neighb_cha1_artcha3cont.eps}}}
\centerline{\resizebox{1.00\hsize}{!}{\includegraphics[angle=270]{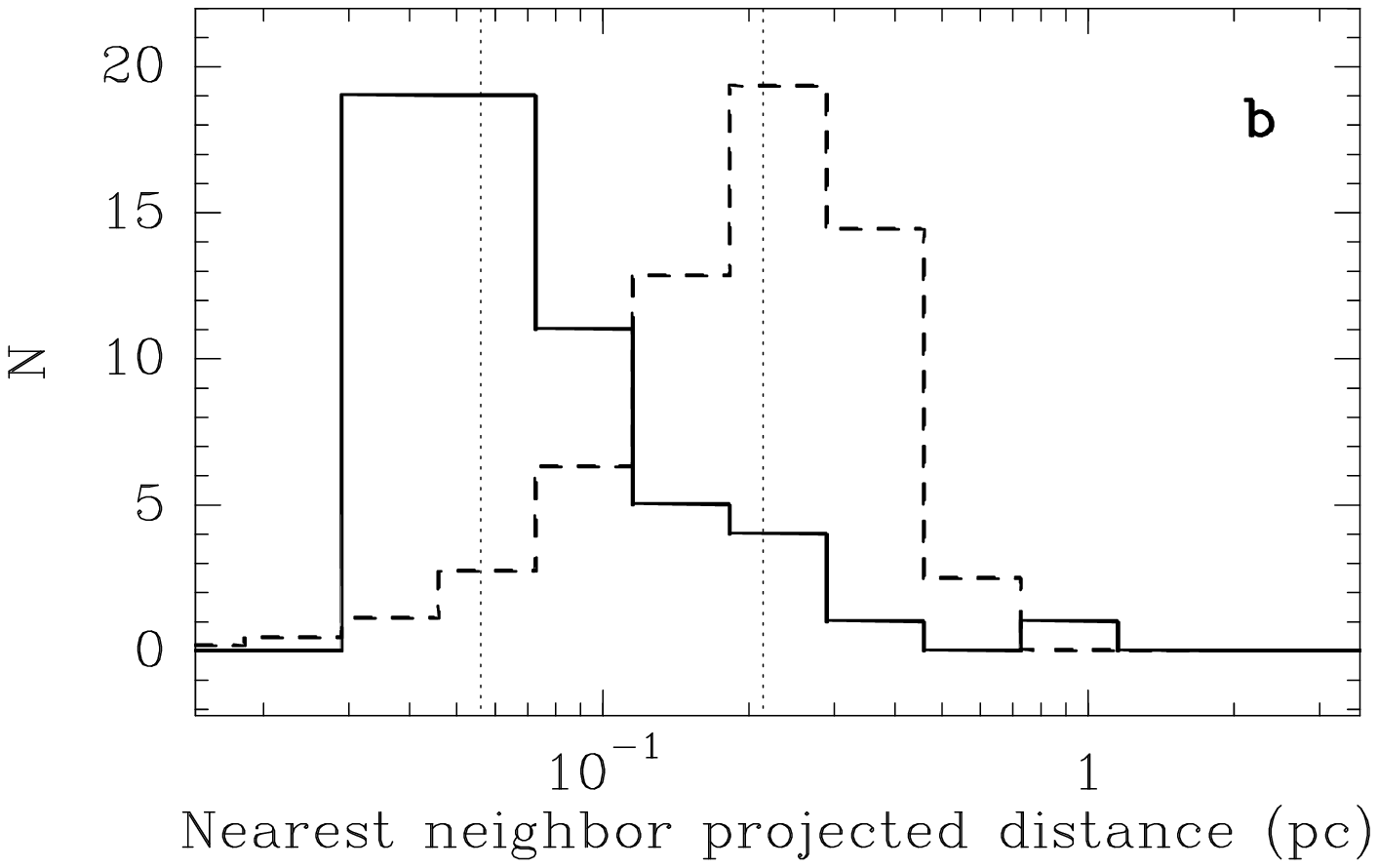}}}
\caption{Distribution of nearest-neighbor projected distance for the starless 
sources of Cha~III (\textbf{a}) and Cha~I (\textbf{b}). In each panel, the
dashed histogram shows the distribution expected for the same number of objects
randomly distributed in the same area. The median value of each distribution is
marked with a dotted line.}
\label{f:neighbor}
\end{figure}

The distribution of nearest-neighbor projected distance is presented in 
Fig.~\ref{f:neighbor} for both samples of starless sources in Cha~III and I.
The median distance $d_{\mathrm{m}}$ is 0.11~pc in Cha~III, a factor of 2 
larger than in Cha~I (0.056~pc). We follow \citet{Gomez93} to estimate the 
corresponding distribution for a sample of sources that would be randomly 
distributed in the plane of the sky over the same area, assumed to be the 
surface of a disk of diameter equal to the largest projected distance 
between two sources (4.9~pc for Cha~III and 4.0~pc for Cha~I). The median 
distances of these random distributions are 0.38~pc for Cha~III and 0.21~pc 
for Cha~I, i.e. nearly a factor of 4 larger than observed in both cases. The 
starless sources in both clouds are thus significantly clustered. 
Assuming that the nearest-neighbor pairs are randomly oriented in the 
three-dimensional space, the true separation in three dimensions is 
$\frac{4}{\pi} d_{\mathrm{m}}$ \citep[][]{Gomez93}, i.e. 0.14~pc for Cha~III 
and 0.07~pc for Cha~I.

Given that the median extinction of the local ambient medium in which the 
starless sources in Cha~III are embedded is a factor of 2 lower than in Cha~I 
(see Sect.~\ref{sss:starless_extinction}), the density of this local ambient 
medium is most likely also lower, maybe up to a factor of 2. The 
characteristic length of thermal fragmentation is inversely proportional to 
the square root of the density. The difference in median nearest-neighbor 
separation seen between Cha~III and I is therefore somewhat more pronounced 
than what could naively be expected from thermal fragmentation.

The remarkable alignment of six 870~$\mu$m sources in field Cha3-Center 
(see Fig.~\ref{f:labocamapdet}c) 
deserves some further analysis. The sources are nearly uniformly distributed
along a straight line, with a mean projected separation of $0.16 \pm 0.03$~pc. 
For a density of the intercore medium of $4 \times 10^3$~cm$^{-3}$ in this
region \citep[][]{Gahm02}, the Jeans length 
$2\pi a_{\mathrm{s}}/\sqrt{4\pi G \rho}$ is 
about 0.36~pc at 12~K. The intercore separation would match this Jeans length
if the inclination angle of the putative filament was 26$^\circ$, 
which is
statistically unlikely. However, the effective scale of fragmentation in a 
magnetized and/or rotating filament is expected to decrease with increasing 
magnetic field strength and/or rotation level 
\citep[e.g.][]{Nakamura93,Matsumoto94}. The measured core separation could 
be used to estimate the magnetic field strength and/or rotation level as was 
done for a filament in Orion~A \citep[][]{Hanawa93}. With $d_{\mathrm{FWHM}}$ 
the full width at half maximum of the filament, $\lambda$ the projected core 
separation, $\alpha$ and $\beta$ the ratios of the magnetic pressure
and centrifugal force to the thermal pressure, respectively, and $i$ the 
inclination of the filament along the line of sight, Equation 13 of 
\citet{Hanawa93} yields 
$\alpha + \beta \geq (1.75 \times \frac{5.0}{4.3} \frac{d_{\mathrm{FWHM}}}{\lambda/\sin i}+0.6)^3-1.0$. The equality holds for a pitch angle $\theta = 0^\circ$,
with $\theta$ characterizing the relative strength of the poloidal to axial
magnetic fields.
The width of the filament cannot be estimated from the LABOCA map. With the
C$^{18}$O 1--0 map shown in Fig.~9 of \citet{Gahm02}, we roughly estimate
$d_{\mathrm{FWHM}} \sim$~0.25--0.40~pc, which is larger than the typical width
of the filaments recently detected with \textit{Herschel} in three other 
nearby clouds \citep[median width 0.1~pc, see][]{Arzoumanian11}. Since our 
estimate from C$^{18}$O is rather uncertain, we also consider below (in 
parentheses) the case where the width of the filament is 0.1~pc. If the 
filament is in the plane of the sky ($i = 90^\circ$) and $\theta = 0^\circ$, we 
obtain $\alpha + \beta \sim$~53--180 (6 with 0.1~pc). Since there is 
apparently no significant level of rotation in this filament 
\citep[][]{Gahm02}, we derive $\alpha \sim$~53--180 (6). If the filament is 
not in the plane of the sky, $\alpha$ would be less than 1 for 
$i \la$~7.5--12$^\circ$ (31$^\circ$). An inclination to the line of sight of 
7.5--12$^\circ$ is statistically unlikely, so we are left with a very high 
level of magnetic pressure or an overestimated filament width. If this 
alignment of 6 regularly spaced cores is really the result of fragmentation in 
a filament, then we conclude that either the filament is strongly magnetized, 
or it is much thinner than it appears in C$^{18}$O 1--0, or the model of 
\citet{Nakamura93} of a magnetized, self-gravitating, isothermal filament in 
equilibrium does not apply to that filament. Alternatively, the observed 
regular structure may have nothing to do with fragmentation and simply 
represent transient periodic overdensities produced by 
gravitational-magnetoacoustic waves that will be damped away 
\citep[][]{Langer78}.

%
%________________________________________________________________

\section{Discussion}
\label{s:discussion}

\subsection{A puzzling population of starless cores in Cha~I and III}
\label{ss:nature}

\begin{figure}
%\centerline{\resizebox{1.00\hsize}{!}{\includegraphics[angle=270]{/homes/belloche/Chamaeleon/Continuum/Cha3/Analysis/alphabevsmass_artcha3cont.eps}}}
\centerline{\resizebox{1.00\hsize}{!}{\includegraphics[angle=270]{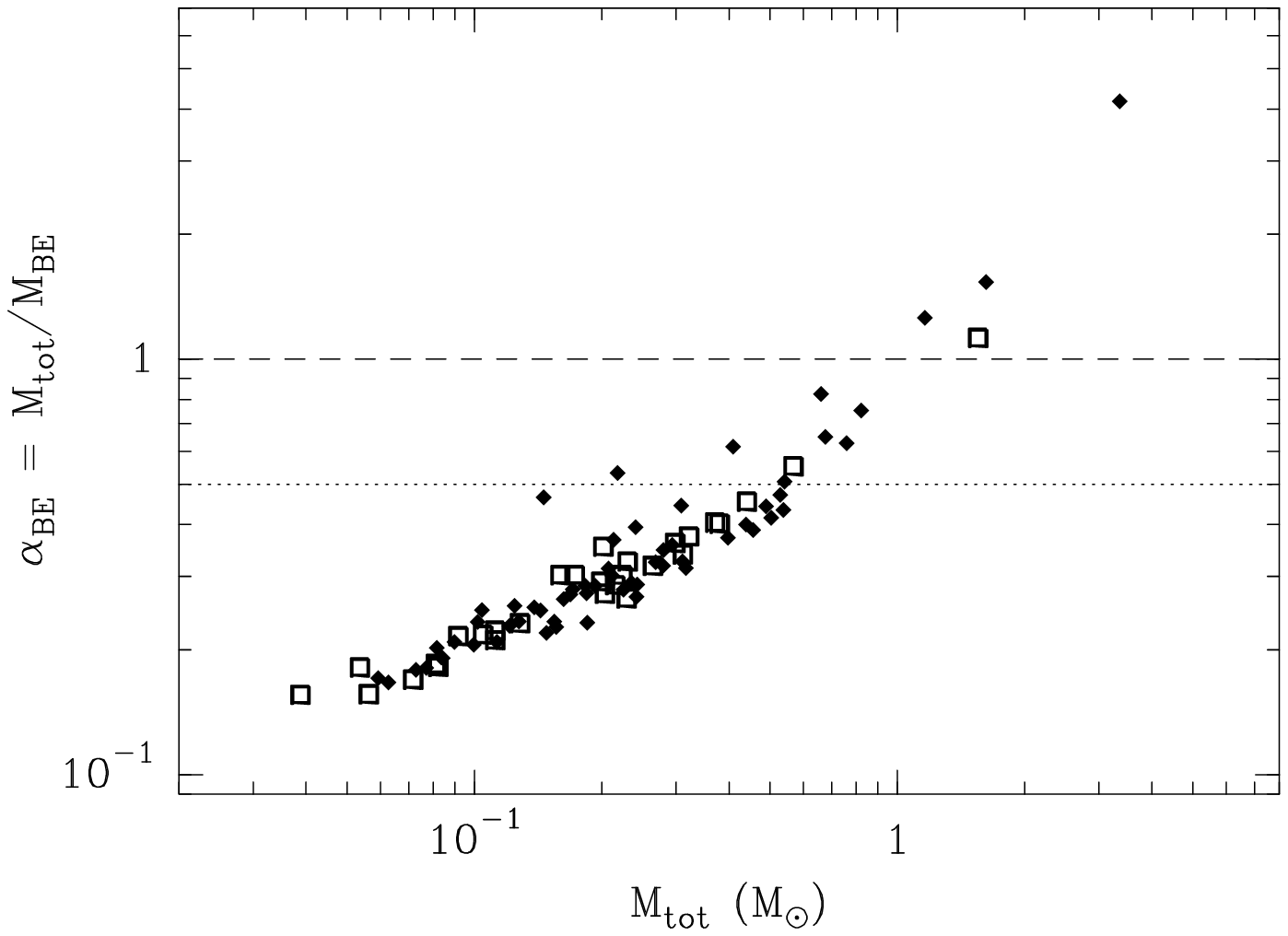}}}
\caption{Ratio of total mass to Bonnor-Ebert critical mass as a function of 
total mass. The filled diamonds and open squares show the Cha~I and III 
sources, respectively. The dashed line shows the limit above which a source
is gravitationally unstable. The gravitationally bound sources are located 
above the dotted line, provided they are supported by thermal pressure only and
the external pressure is negligible.}
\label{f:alphabe}
\end{figure}

Based on the comparison to the Bonnor-Ebert mass limit, we estimate that only
one (or at most two) source(s) out of 29 is a candidate prestellar core in 
Cha~III (see Sect.~\ref{sss:starless_massvssize}). This yields a fraction of 
candidate prestellar cores of 3--7$\%$, a factor of 2 lower than in Cha~I 
(5--17$\%$, see Paper~I). Apart from the few 
candidate prestellar cores, the population of starless cores in Cha~III is 
very similar to the one in Cha~I since they have nearly the same median peak, 
aperture, and total masses as well as the same median size and aspect ratio 
(compare Figs.~\ref{f:histo} and \ref{f:histom} to Figs.~7 and 8 
of Paper~I). They also follow the same correlation in an 
$\alpha_{\mathrm{BE}}$ versus $M_{\mathrm{tot}}$ diagram (see 
Fig.~\ref{f:alphabe}). The main striking difference is that the visual 
extinction of the medium in which the Cha~III sources are embedded is on 
average a factor of 2 lower than in Cha~I. Although we a priori cannot exclude 
that the extinction laws of both clouds may differ or that there may be more 
contamination by foreground stars toward Cha~III leading to an underestimate 
of the extinction, we rather consider that this difference may come from the 
density structure or the physical processes at work in the clouds. As 
mentioned in Sect.~\ref{s:intro}, Cha~III looks much more filamentary in cold 
dust emission at 100~$\mu$m than Cha~I \citep[see Fig.~7 of][]{Boulanger98}. 
If this is also the case on scales smaller than the resolution of our 
extinction maps (3$\arcmin$), one could expect lower extinctions for the 
``ambient'' medium in Cha~III compared to Cha~I. Alternatively, this difference
in ambient extinction (and thus ambient density) for two otherwise similar 
populations of starless cores may indicate that the structuring of the 
interstellar medium into seeds of cores does not depend much on the local 
gravity and may be dominated by other processes such as turbulence and 
magnetic fields \citep[see also][]{Andre10}. 

\begin{figure}
%\centerline{\resizebox{1.00\hsize}{!}{\includegraphics[angle=270]{/homes/belloche/Chamaeleon/Continuum/Cha3/Analysis/cmf_cha1-3_artcha3cont_10p0.eps}}}
\centerline{\resizebox{1.00\hsize}{!}{\includegraphics[angle=270]{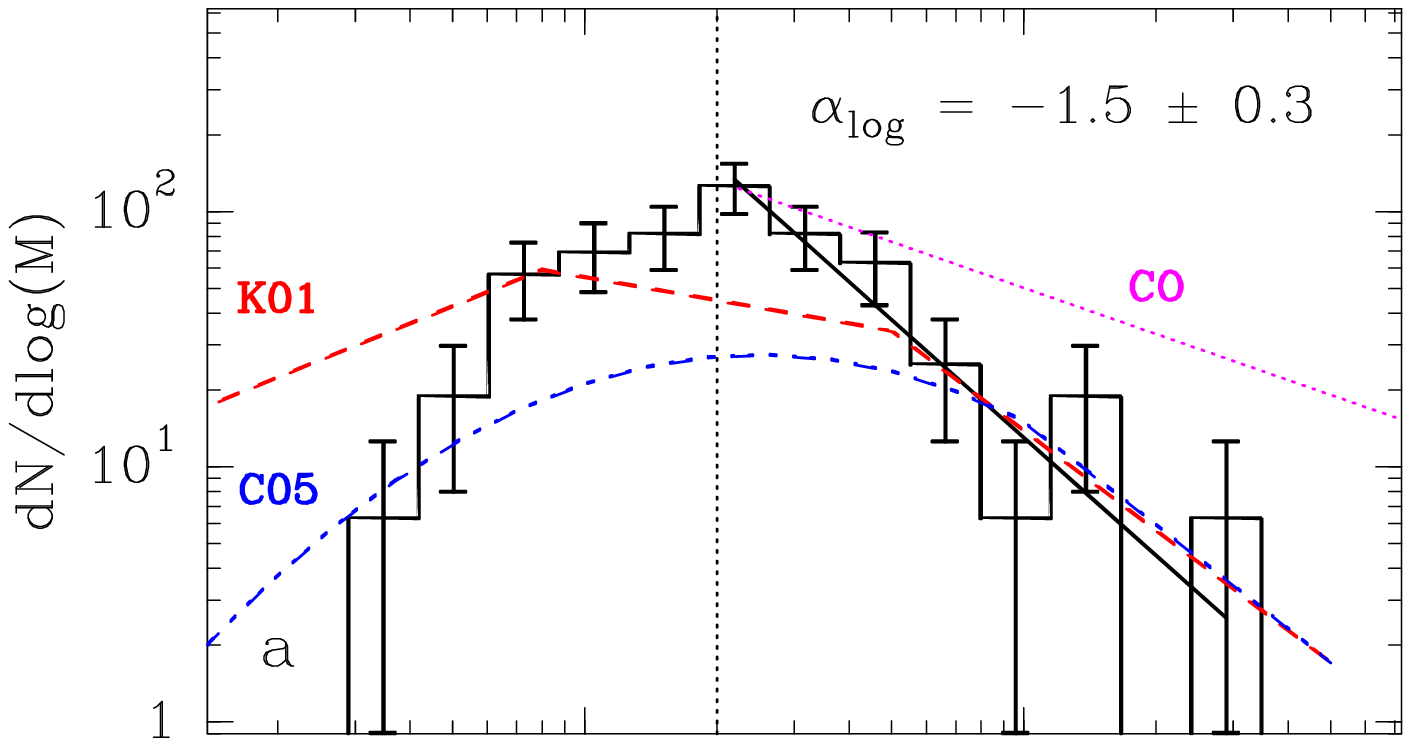}}}
\vspace*{3ex}
%\centerline{\resizebox{1.00\hsize}{!}{\includegraphics[angle=270]{/homes/belloche/Chamaeleon/Continuum/Cha3/Analysis/cmf_cha1-3_artcha3cont_1p0.eps}}}
\centerline{\resizebox{1.00\hsize}{!}{\includegraphics[angle=270]{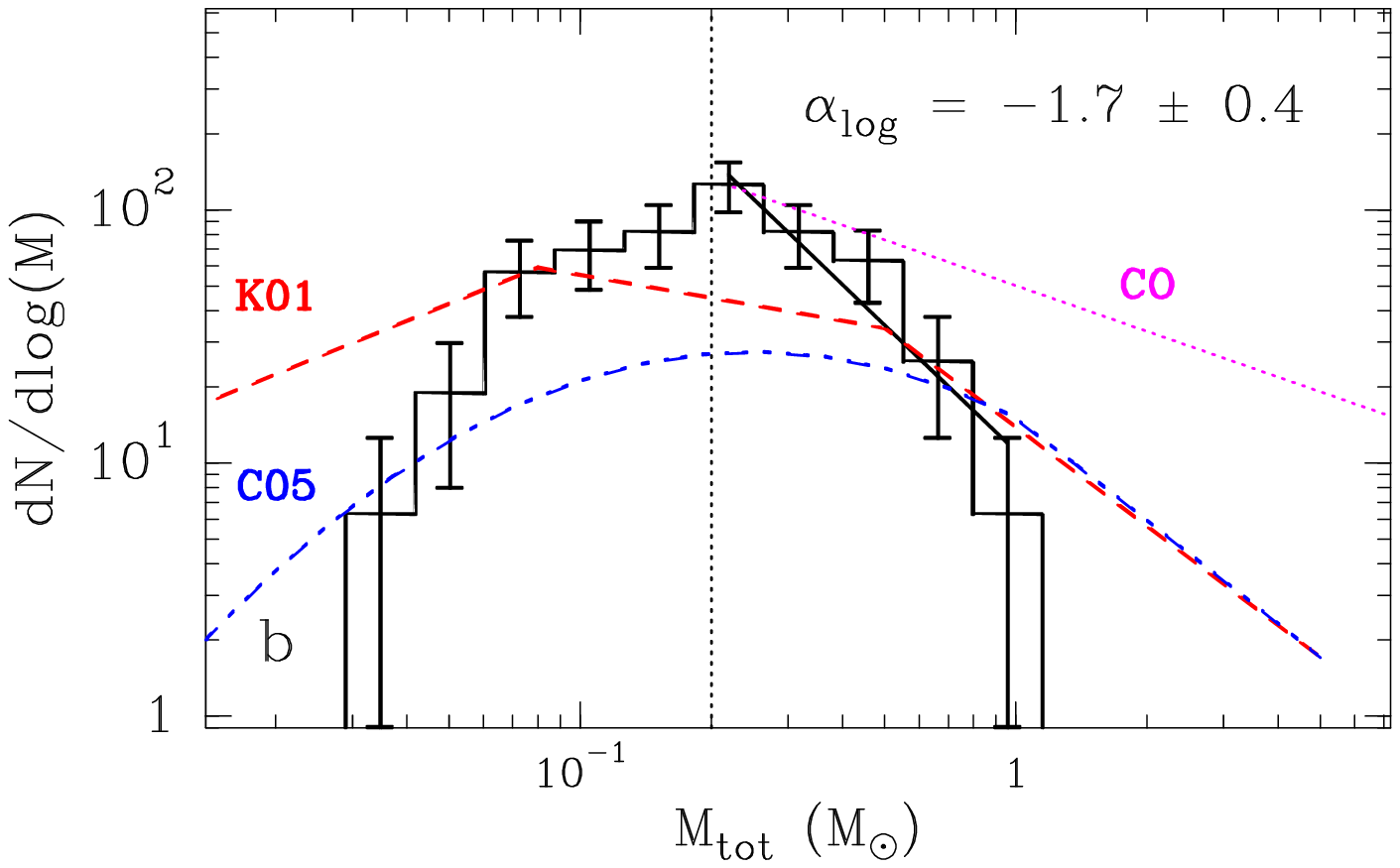}}}
\caption{Mass distribution $\mathrm{d}N/\mathrm{d}\log(M)$ of the starless 
sources of Cha~I and III. Panel \textbf{a} shows all sources while panel 
\textbf{b} displays only those with $M/M_{\mathrm{BE}} < 1.0$. The error 
bars represent the Poisson noise (in $\sqrt{N}$). The vertical dotted line is 
the average (0.20~M$_\odot$) of the estimated 90$\%$ completeness limits for 
Cha~I and III. The thick 
solid line is the best power-law fit performed on the mass bins above the 
completeness limit. See caption of Fig.~\ref{f:cmd} for other details.
The K01 and C05 IMFs are both vertically shifted to the same number at 
5~M$_\odot$.} 
\label{f:cmd_cha1-3}
\end{figure}

The sample of starless sources in Cha~III being small, the accuracy of the 
CMD is not sufficient to compare it with the IMF and the CO 
clump mass spectrum (see Sect.~\ref{sss:cmd}). To enlarge the statistics, the 
source samples of Cha~I and III can be merged, both surveys having nearly the 
same mass completeness limit and their populations of starless cores having
very similar properties. The full sample contains 89 sources (60 in Cha~I and
29 in Cha~III) and its mass distribution is shown in Fig.~\ref{f:cmd_cha1-3}a.
We can also consider only the starless sources of this enlarged sample that 
are not candidate prestellar cores, i.e. those with $\alpha_{\mathrm{BE}} < 1$. 
This yields a sample of 85 sources (57 in Cha~I and 28 in Cha~III), the 
excluded sources being Cha1-C1-3 and Cha3-C1. The mass distribution of this 
slightly smaller sample is presented in Fig.~\ref{f:cmd_cha1-3}b. Above the 
90\% completeness limit, both distributions are well fitted by a single 
power-law with an exponent $\alpha_{\mathrm{log}} = -1.5 \pm 0.3$ and 
$-1.7 \pm 0.4$, respectively. This is steeper than, but still consistent 
within $1\sigma$ with, the Salpeter exponent of the high-mass end of the 
stellar IMF, and it is definitely much steeper than the CO clump mass 
distribution that is characterized by $\alpha_{\mathrm{log}} = -0.6$ 
\citep[][]{Blitz93,Kramer98}\footnote{We note that \citet{Ikeda11} find 
$\alpha = -2.1 \pm 0.2$ with C$^{18}$O 1--0 in S140, the same cloud for which
\citet{Kramer98} derived $\alpha = -1.65 \pm 0.18$ with C$^{18}$O 2--1. They 
argue that the study of \citet{Kramer98} is limited to the central part of 
the cloud and is likely biased toward high-mass cores (hence yielding a 
flatter CMD). They also show that a poor spatial resolution leads to 
underestimating $|\alpha|$ if the resolution is worse than 0.1~pc. The 
``classical'' index $\alpha = -1.6$ is therefore most likely valid for larger 
CO clumps only.}. Even if the combined sample of Cha~I and III starless sources 
detected with LABOCA is large (89 sources), there are only about 50 sources 
above the completeness limit. We thus expect that the ongoing sensitive 
survey of the Chamaeleon clouds performed with \textit{Herschel} in the frame 
of the Gould Belt Survey \citep[][]{Andre10} will provide a more complete 
sample of starless cores and thus a more robust CMD. In particular, the shape 
of the mass distribution of the starless sources that are not candidate 
prestellar cores will be of prime importance.

The population of starless cores detected with LABOCA in Cha~I and III is very 
puzzling: although most of these sources do not appear to be prestellar based 
on the Bonnor-Ebert criterion, their mass distribution seems to be consistent 
with the stellar IMF at the high-mass end. For Cha~I, we argued that a loss of 
thermal support via cooling would not be sufficient to bring these cores above 
the Bonnor-Ebert mass limit, hence that they are unlikely to form stars in the 
future (Paper~I). Along with other arguments, this suggested that 
star formation is over in Cha~I. In the same way, we could a priori conclude 
(but see the next sections) that, apart from the northern part of the cloud 
where one candidate prestellar core is found, Cha~III does not seem to be able 
to start the process of star formation. 

The level of turbulence is very similar in both clouds, as measured via the 
linewidths of the $J=$~1--0 transitions of CO and C$^{18}$O 
(see Sect.~\ref{ss:massaccum} below). Taken at face value, turbulence thus does 
not seem to be the key parameter promoting star formation since Cha~I formed 
many stars in the past while Cha~III did not. However, the present level of 
turbulence in Cha~I may not reflect the initial conditions when the first 
stars were formed. On the one hand, the level of turbulence in this cloud may 
have been lower in the past and have been raised as a result of stellar 
feedback via, e.g., molecular outflows, possibly preventing the process of star 
formation to continue at the present epoch. On the other hand, if the feedback 
of the Cha~I (low-mass) YSOs has not been sufficient, the turbulence may have 
decayed with time. In this case, the same behaviour would a priori be expected 
for both clouds, which would not explain their difference in terms of past 
star formation activity.

\subsection{Are the Chamaeleon cores similar to the Pipe cores?}
\label{ss:pipe}

At this stage, it is very instructive to compare the population of starless
cores in Cha~I and III to the one found in the Pipe nebula via extinction maps
\citep[][]{Alves07,Lada08,Rathborne09}. The Pipe nebula is the only other
nearby star forming region in the pre-\textit{Herschel} era for which a
population of dense cores that are mostly gravitationally unbound has been 
found, with a mass distribution still consistent with
the stellar IMF at the high-mass end. In addition, its CMD
departs from a single power law and has a break like the IMF but at a higher 
mass of $2.7 \pm 1.3$~M$_\odot$, suggesting a star formation efficiency of 
$\sim 20\%$ in this cloud \citep[][]{Rathborne09}. Interestingly, only the 
sources with a mass higher than this characteristic mass appear to be 
prestellar based on the Bonnor-Ebert criterion 
\citep[see Fig.~9 of][]{Lada08}. \citet{Lada08} suggest that the 
gravitationally 
unbound cores are pressure confined, the external pressure being most likely 
provided by the weight of the cloud itself. They thus do not appear to be 
transient structures. \citet{Lada08} consider two mechanisms that may turn 
these stable but unbound cores into prestellar cores: either an increase of 
the external pressure produced by the contraction of the whole cloud, or a 
mass increase if these cores have not obtained their final mass yet. The 
former mechanism would require a decrease of a factor of 2 in cloud radius, 
which may be possible via the dissipation of its supersonic turbulence but 
casts a potential timescale issue since the more massive unstable cores would 
form stars much more rapidly than the less massive ones 
\citep[see][ for details]{Lada08}.

The starless cores in Cha~I and III that are presently unstable according to 
the Bonnor-Ebert criterion are also the most massive ones, like in the Pipe 
nebula, but the transition occurs at a lower mass ($\sim 1$ versus 
$\sim 3$~M$_\odot$, see Fig.~\ref{f:alphabe}). 
With a median mass of 0.9 M$_\odot$ and a median radius of 0.07~pc 
\citep[computed from Table~2 of][]{Rathborne09}, the Pipe cores are bigger than
the Chamaeleon ones by a factor of $\sim 4$ in mass and $\sim 2.5$ in radius
(but see the caveats about the radius in Sect.~\ref{sss:starless_sizes}).
While this could be an intrinsic property, we believe that it results from
the different tracers used to extract the cores: extinction maps are more 
sensitive to extended, low-density material than 870~$\mu$m dust emission 
maps. The factor of 3 difference between the transition masses mentioned above 
may therefore simply result from the fact that the LABOCA masses do not include 
the low-density material surrounding each core.

A major difference exists however between the CMDs of the Pipe nebula and the 
Chamaeleon~I and III clouds: there is no evidence for a break in the 
Chamaeleon CMD down to the 90\% completeness limit of 0.2~M$_\odot$, while such
a break is seen around 2.7~M$_\odot$ in the Pipe nebula. Even the possible 
mass scaling factor of 4 mentioned above would not be sufficient to explain 
this difference. However, the low-density material possibly missing in the
LABOCA masses could affect the \textit{shape} of the CMD rather than 
contributing as just a scaling factor. On the one hand, if the Chamaeleon 
cores are seeds of prestellar cores in the process of accumulating mass 
(see Sect.~\ref{ss:massaccum} below), we may conjecture that the break in the 
CMD could arise from this mass accumulation process. On the other hand, a large
fraction of the lowest mass cores may never become prestellar and they may 
currently hide the true shape of the \textit{prestellar} CMD in Chamaeleon.

\subsection{Can the starless cores in Cha~I and III turn prestellar?}
\label{ss:massaccum}

The possibility that the unbound starless cores of the Pipe and Cha~I/III 
clouds are still accumulating mass has been mentioned in Sect.~\ref{ss:pipe} 
and is very attractive in light of the results obtained by recent numerical 
simulations. In this section, we compare the properties of Cha~I 
and III to several kinds of numerical simulation.

\citet{Gomez07} study the formation and collapse of quiescent 
cloud cores induced by focused compressions (or ``convergent flows'') in a 
cloud of diameter 1~pc that initially has a constant sub-Jeans density of 
113~cm$^{-3}$ and a uniform temperature of 11.4~K. The velocity amplitude of 
the compressing impulse is 0.4 km~s$^{-1}$ (Mach number of 2). A mild shock 
propagates inwards, the material left behind it (the envelope) being set into 
infall motions. The shock bounces off the center and expands outwards, leaving 
a quiescent core in the inner part. The structure of the core+envelope system 
then resembles a truncated Bonnor-Ebert sphere with a flat inner density 
profile and a density falling as $\sim$ r$^{-2}$ in the (infalling) envelope. 
Depending on the initial conditions (position of the impulse within the 
cloud), the core can gain enough mass from the envelope to become 
gravitationally unstable and start collapsing. In this model, the inner 
quiescent core with a density of $\sim 10^5$~cm$^{-3}$ is initially unbound 
and confined by the ram pressure of the inflowing gas. It grows in size and 
mass until it gets dominated by gravity. Interestingly, there is a significant 
time delay of $\sim 5 \times 10^5$~yr between the formation of the inner core 
(when the system starts to look like a pseudo Bonnor-Ebert sphere) and the 
onset of gravitational collapse, due to the growth in mass.

The C$^{18}$O 1--0 linewidths in Cha~I and III are on the order of 
0.8--0.9~km~s$^{-1}$ at an angular resolution of 2.7$\arcmin$ 
\citep[][]{Mizuno99}, implying an rms (turbulent) velocity dispersion of 
0.3--0.4~km~s$^{-1}$. The CO 1--0 linewidths, tracing even lower-density 
material, are a factor of $\sim 2.5$ times larger 
\citep[][]{Boulanger98,Mizuno01}. The initial conditions chosen by 
\citet{Gomez07} are therefore plausible for Chamaeleon from a 
kinematic point of view. The typical densities of the Chamaeleon starless 
cores are also 
similar to the density of the inner core formed in these simulations. If this
scenario holds for Chamaeleon, a fraction of these presently unbound cores 
could turn prestellar in the future (in less than $\sim 5 \times 10^5$~yr) if 
they gain enough mass to become unstable. We note, however, that there is no 
evidence for significantly flattened inner density profiles in the Chamaeleon 
sample (see Sect.~\ref{sss:starless_masses} and Paper~I), but this
may not be a shortcoming for this scenario, the spherically symmetric model 
and initial conditions of \citet{Gomez07} being highly idealised. Finally, 
since some of the simulations of \citet{Gomez07} do not lead to the formation 
of an unstable core prone to collapse, a fraction of the Chamaeleon starless 
cores may never reach the critical mass and simply get dispersed in the future 
(see their simulation S1).

Another fruitful approach is the one followed by \citet{Clark05}. These 
authors 
examine the formation of bound coherent cores in molecular clouds supported by 
(decaying) turbulence. They model a cloud of mass 32.6~M$_\odot$ and diameter 
$\sim 0.3$~pc at 10~K with an initially uniform density of 
$5.6 \times 10^4$~cm$^{-3}$ and turbulent velocities characterized by an 
initial effective Mach number of 5.3 and a power spectrum 
$P(k) \propto k^{-4}$ (their simulation 2). A large number of 
fragments\footnote{\citet{Clark05} 
use the word ``clumps'' for these objects, but we prefer to call them 
``fragments'' since, in the context of observations, the word ``clump'' is 
usually given to larger-scale structures that may contain several cores 
\citep[see, e.g.,][]{Williams00}.} 
are formed, most of them being initially unbound. Their mass distribution is 
well fitted by a Salpeter slope at the high-mass end when star formation sets 
in (at $t \sim t_{\mathrm{ff}}$, with $t_{\mathrm{ff}}$ the initial free-fall 
time of the cloud). Interestingly, the mass distribution of these 
mostly unbound fragments is slightly steeper at the high-mass end at an 
earlier stage \citep[$t = 0.6t_{\mathrm{ff}}$, see Fig.~3 of][]{Clark05}. A 
second important point is that only the most massive fragments are 
gravitationally bound (see their Fig.~5). In this simulation, the unbound 
fragments grow in mass, as a result of coagulation, but only a small fraction 
of them become gravitationally unstable and can start collapsing. The
number of gravitationally unstable cores formed in this way (11 for their 
simulation 2) is on the same order of magnitude as the initial number of 
mean Jeans masses in the cloud (33).

The population of starless cores in Cha~I and III share similar properties with
the simulated fragments of \citet{Clark05}: their CMD resembles the IMF at the 
high-mass end, most cores are unbound, and only the most massive ones are 
gravitationally unstable. The slope of the Chamaeleon CMD at the high-mass end 
is even slightly steeper (but only at the $1\sigma$ level) than the Salpeter 
value, like for the simulated fragments before star formation sets in. However, 
most objects qualified as fragments in the simulation are much too small and 
too faint to be detected in our continuum maps. There are more than 1000
fragments formed in the simulated cloud of projected area 0.053~pc$^2$, while 
we detect only 89 cores in 17~pc$^2$. Because of the limited sensitivity and
angular resolution of our maps, the low number of detected starless cores 
does not really rule out this scenario of turbulence-generated, small, unbound 
fragments that coagulate with time. We may be seeing only those fragments that
have already grown enough by coagulation to be detected. One issue with this 
analogy may be that the initial level of turbulence assumed in the simulations 
of \citet{Clark05} is significantly higher than the one characterizing the
regions of Cha~I and III that have similar densities of 
$\sim 5 \times 10^4$~cm$^{-3}$ 
\citep[Mach number $< 2$ based on C$^{18}$O 1--0, see][]{Mizuno99,Gahm02}.

The comparison of the properties of the Chamaeleon starless cores to those 
produced in the simulations of \citet{Clark05} and \citet{Gomez07} leaves open 
the hypothesis that a fraction of these starless cores may become unstable in 
the future by accumulating mass from larger scales. Since most Chamaeleon 
starless cores are located in filamentary structures, the next question is 
whether this gain in mass could simply occur along the filaments or would 
require accumulation of material from larger scales. In Cha~I, the mass of the 
filaments traced with LABOCA is, without counting the cores embedded within 
them, roughly equal to the mass of the starless cores. If this LABOCA 
mass is representative of the real mass of the filaments, then the mass gain 
could not be larger than a factor of 2 and would not be sufficient to bring 
most starless cores above the Bonnor-Ebert limit, provided the product of 
their temperature and radius does not at the same time decrease by more than a 
factor of 2 (see the definition of $M_{\mathrm{BE}}(R)$ in 
Sect.~\ref{sss:starless_massvssize}). In their simulations of dense core 
formation in supersonic turbulent converging flows, \citet{Gong11} report that 
filaments are formed in post-shock regions at the same time as  overdense 
regions within these filaments condense into cores. It is thus tempting to 
conclude that the starless cores in Cha~I and III may grow in mass at the same 
time as the filaments do under the influence of turbulence and self-gravity. 
Interestingly, the simulations of \citet{Gong11} can produce alignments of 
nearly regularly spaced cores. For a Mach number of 5, the core separation is 
on the order of 0.2 code units, i.e. 0.15--0.47~pc for a mean density of free 
particles of 4000--400~cm$^{-3}$ (see their Fig.~7, top left and bottom right 
panels). This is very much reminiscent of the remarkable alignment of 
regularly spaced cores in Cha3-Center (see Sect.~\ref{ss:distribution} and 
Fig.~\ref{f:labocamapdet}c). The CO linewidths measured
on large scales in Cha~III are on the order of 2--2.5~km~s$^{-1}$, i.e. a Mach 
number of 4--5 \citep[][]{Boulanger98,Mizuno01}. Depending on the scale and the
assumptions made, the estimates of the mean density range between 400 and 
4000~cm$^{-3}$ for Cha3-Center 
\citep[see Sect.~\ref{ss:cha3masses} and][]{Gahm02}. Given the unknown 
inclination of the filamentary structure, the projected separation of the 
aligned cores in Cha3-Center could be in rough agreement with the one produced 
in the simulations of \citet{Gong11}. Since some of the cores become 
gravitationally unstable in these simulations, this may also be the fate of 
some of the starless cores in Cha3-Center, and by extension in the full
sample of Chamaeleon starless sources.

\subsection{A large fraction of ``failed'' cores in Cha~I and III?}
\label{ss:failed}

In the simulations of \citet{Gomez07}, the phase of mass growth, from the time 
of central core formation on, lasts about as long as the collapse phase. 
In their 1D spherically symmetric simulations of converging supersonic flows,
\citet{Gong09} find a duration of the core-building phase about 9 times longer
than the duration of the collapse phase, which is also confirmed by their 3D
simulations \citep[][]{Gong11}. However only a fraction of this core-building 
phase is observable. \citet{Gong09} estimate that 10--20$\%$ of the core 
building phase is observable if one takes as observability criterion a density 
contrast between the center and the edge of a core, 
$\rho_{\mathrm{c}}/\rho_{\mathrm{edge}}$, higher than 5. This means a duration of 
the \textit{observable} core-building phase 1--2 times longer than the 
duration of the collapse phase, which is rougly consistent with the result of 
\citet{Gomez07}. 

If we associate the four\footnote{We exclude Cha1-C1, i.e. the candidate first 
hydrostatic core Cha-MMS1, but we include Cha1-C4 that has 
$\alpha_{\mathrm{BE}} > 1$ based on the pressure.} starless cores of Cha~I and 
III that have $\alpha_{\mathrm{BE}} > 1$ with the collapse phase and the other 
starless cores (84) with the observable core-building phase, then we would 
expect only 4--8 additional cores (3--6 in Cha~I and 1--2 in Cha~III) to 
become gravitationally unstable in the future, provided the process of star 
formation occurs at a constant rate and the model prediction of the duration 
of the core-building phase mentioned in the previous paragraph is correct. If 
the LABOCA masses are underestimated by a factor of 2, then there are about 10 
and 2 candidate collapsing cores at the present time, plus 10--20 and 2--4 
additional starless cores that could become gravitationally unstable in Cha~I 
and III, respectively. Given these numbers, at least 29--39 and 23--25 
starless cores, i.e. 49--66$\%$ and 79--86$\%$ of the population of starless 
cores detected with LABOCA, will most likely \textit{never} become 
gravitationally unstable in Cha~I and III, respectively. These fractions may 
even be underestimated if the cores detected with LABOCA have a ratio 
$\rho_{\mathrm{c}}/\rho_{\mathrm{edge}}$ significantly higher than 5, which may 
well be the case since $c_n$ in Table~\ref{t:starless} (and Table~6 of 
Paper~I) is likely smaller than the true ratio 
$\rho_{\mathrm{c}}/\rho_{\mathrm{edge}}$. However, these large 
fractions of ``failed'' cores hold only if the simulations of converging flows 
mentioned above are appropriate descriptions of the physical processes at work 
in the Chamaeleon molecular clouds. If the observable core-building phase 
lasts significantly longer than the collapse phase, then the actual fraction 
of ``failed'' cores could be smaller.

In light of this new analysis, it cannot be excluded that Cha~I will go on
forming stars. However, even if mass growth occurs and the number of 
prestellar cores increases by a factor 2--3 as suggested in the previous
paragraphs, this number will still be too low by a factor 2--4 (if the
LABOCA masses are correctly estimated) to be consistent with a constant star 
formation rate since the time when the current pre-main-sequence stars were 
formed (2~Myr ago, see Paper~I and references therein). In addition, the lack 
of Class 0 protostars by a factor 
maybe as large as 10 in Cha~I (see Paper~I) remains a strong indication that 
the star formation rate has decreased with time in this cloud. Cha~I may not 
be at the end of the process of star formation as was suggested in Paper~I, 
but it is at least secure to conclude that its star formation rate has 
decreased with time by a factor $> 2$--4 over the last 2 Myr. On top of 
this overall decrease, it may have significantly fluctuated, by a factor 2--5 
to account for the lack of Class~0 protostars.

\subsection{A prime target to study the formation of prestellar cores}
\label{ss:cha3}

With one candidate prestellar core and one or two additional starless cores
that could turn prestellar (see Sect.~\ref{ss:failed}), Cha~III seems to be 
able to form stars, even if presently at a very low rate. This is somewhat in 
contrast with the Polaris flare region, another presently non-star-forming 
molecular cloud. A large population of starless cores was recently uncovered 
with \textit{Herschel} in that cloud but the cores typically lie 
one to two orders of magnitude below the Bonnor-Ebert mass limit 
\citep[see Fig.~4 of][]{Andre10}. Since the bulk of the population of starless 
cores detected with LABOCA in Cha III lies a factor 3--4 below the 
Bonnor-Ebert mass limit, the tip of the Polaris population has 
$\alpha_{\mathrm{BE}}$ lower by a factor $\sim 3$ compared to the Cha~III 
starless cores. This suggests that Cha~III may be closer to form prestellar 
cores than Polaris \citep[but see][ for possible evidence of gravitationally 
bound cores in the latter]{Heithausen02}.

With a large fraction of starless cores presently not prestellar but still some
evidence that star formation can occur, Cha~III becomes a prime target 
to study the \textit{formation} of prestellar cores, i.e. the core-building 
phase when mass is accumulated, and thus the onset of star formation. In 
particular, it will be essential to get observational constraints on the 
duration of the core-building phase prior to the phase of gravitational 
collapse. In the previous section, we relied on 
predictions of numerical simulations to estimate this duration and derive the
fraction of ``failed'' cores, but this needs to be tested observationally.

Finally, on a very speculative level, the factor of 2 difference between the 
median nearest-neighbor distances of the starless cores in Cha~I and III 
(0.07~pc versus 0.14~pc, respectively, see Sect.~\ref{ss:distribution} and 
Fig.~\ref{f:neighbor}) could be interpreted as fragmentation being more 
efficient or more advanced in Cha~I compared to Cha~III, which could in turn 
have played a role in the fact that Cha~I has been more efficient than Cha~III 
in forming stars. In that respect, it would be instructive to investigate 
whether significant differences in, e.g., magnetic field structure exist 
between both clouds that could explain this speculative difference in their 
ability to fragment.

%
%________________________________________________________________

\section{Conclusions}
\label{s:conclusions}

We performed a deep, unbiased, 870~$\mu$m dust continuum survey for starless 
and protostellar cores in Chamaeleon~III with the bolometer array 
LABOCA at APEX. The resulting 0.9~deg$^2$ map was compared with a map of dust 
extinction. The analysis was performed by carefully taking into account the 
spatial filtering properties of the data reduction process following the 
prescriptions of Paper~I. The extracted sources were compared to
those found in Cha~I (Paper~I) and other molecular clouds. Our main results and 
conclusions are the following:

\begin{enumerate}
  \item The mass detected with LABOCA (23~M$_\odot$) represents only
  \hbox{6$\%$} of the cloud mass traced by the dust extinction, and about 
  $54\%$ of the mass traced by the C$^{18}$O 1--0 emission.
  \item 29 sources were extracted from the 870~$\mu$m map, all of them being
  starless. No unresolved source is detected, which is consistent with the
  absence of any known young stellar object in this cloud.
  \item The starless sources are found down to a visual extinction of 1.9~mag.
  Unexpectedly, about half of the sources are found below 5~mag, which is in
  marked contrast with other molecular clouds, including Cha~I, where starless 
  cores are only found above this threshold. Since the LABOCA surveys toward
  Cha~I and III have the same sensitivity and were analysed in the same way, 
  this result may point to an intrinsic structural difference between the two 
  clouds.
  \item The 90$\%$ completeness limit of our 870~$\mu$m starless core survey is 
  0.18~M$_\odot$. Only 50$\%$ of the detected starless cores are above this 
  limit, suggesting that we may miss a significant fraction of the existing 
  starless cores.
  \item Although the distribution of starless sources suggests the 
  existence of filaments, these filaments are not detected with 
  LABOCA in 
  Cha~III, while the LABOCA map of Cha~I shows clear evidence of filamentary 
  structures with the same sensitivity.
  \item There is a remarkable alignment of 6 nearly equally spaced sources in 
  Cha~III which may have been produced by turbulent fragmentation of a 
  filament or simply represent transient periodic overdensities.
  \item Apart from their distribution of ambient extinction, the Cha~III 
  starless cores share very similar properties with those found in Cha~I. They
  are less dense than those in Perseus, Serpens, Ophiuchus, and Taurus by a 
  factor of a few on average.
  \item At most two sources ($< 7\%$) are above the critical Bonnor-Ebert mass 
  limit in Cha~III, which suggests that a large fraction of the starless cores 
  may not be prestellar. Only the most massive cores in Cha~I and III turn out 
  to be candidate prestellar cores according to the Bonnor-Ebert mass 
  criterion, in agreement with the correlation observed in the Pipe nebula.
  \item The mass distribution of the 85 starless cores of Cha~I and III that 
  are not candidate prestellar cores is consistent with a single power law 
  down to the 90$\%$ completeness limit, with an exponent close to the Salpeter
  value. There is no evidence for a break such as the one seen in the core mass
  distribution of the Pipe nebula, and in the stellar IMF.
  \item A fraction of the Cha~I and III starless cores that are presently not
  candidate prestellar cores may still be growing in mass and turn 
  prestellar in the future. Based on predictions of numerical simulations of 
  turbulent molecular clouds concerning the duration of the core-building 
  phase, we estimate that at most 50$\%$ and 20$\%$ of the starless cores 
  detected with LABOCA in Cha~I and III, respectively, may form stars.
\end{enumerate}

Given their large fraction of starless cores that do not appear to
be prestellar yet, Cha~I and III turn out to be excellent sites to study the 
\textit{formation} of prestellar cores. Since we find some evidence that star 
formation can start in Cha~III, this cloud even becomes a prime target to 
investigate the onset of star formation. The main next step to make further 
progress will be to get observational constraints on the duration of the 
core-building phase that precedes the gravitational collapse.

\begin{acknowledgements}
We thank the APEX staff for their support during the observing sessions
and Axel Wei\ss, Arturo G{\'o}mez-Ruiz, Tobias Troost, and Tomasz Kami{\'n}ski 
for carrying out part of the observations. AB also thanks Patrick Hennebelle and
Steve Longmore for enlightening discussions. This research has made use of the 
SIMBAD database, 
operated at CDS, Strasbourg, France. BP acknowledges her support by the German
\textit{Deutsche Forschungsgemeinschaft, DFG} Emmy Noether number PA~1692/1-1.
\end{acknowledgements}

\begin{appendix}
\section{LABOCA data reduction: spatial filtering and convergence}
\label{s:reduction_appendix}

\subsection{Spatial filtering due to the correlated noise removal}
\label{ss:filtering}

\begin{figure*}
%\centerline{\resizebox{1.0\hsize}{!}{\includegraphics[angle=270]{/homes/belloche/Chamaeleon/Continuum/Cha3/Reduce/EvalFiltering/Figures/fit_artifsources1-4_cha3_it20_3_artcha3cont.eps}}}
\centerline{\resizebox{1.0\hsize}{!}{\includegraphics[angle=270]{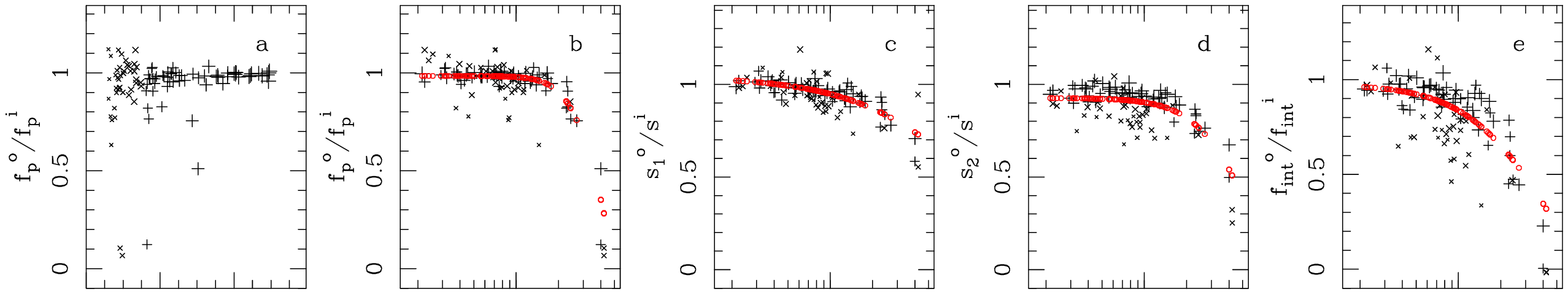}}}
\vspace*{1ex}
%\centerline{\resizebox{1.0\hsize}{!}{\includegraphics[angle=270]{/homes/belloche/Chamaeleon/Continuum/Cha3/Reduce/EvalFiltering/Figures/fit_artifsources1-4_cha3_inicheck_3_artcha3cont.eps}}}
\centerline{\resizebox{1.0\hsize}{!}{\includegraphics[angle=270]{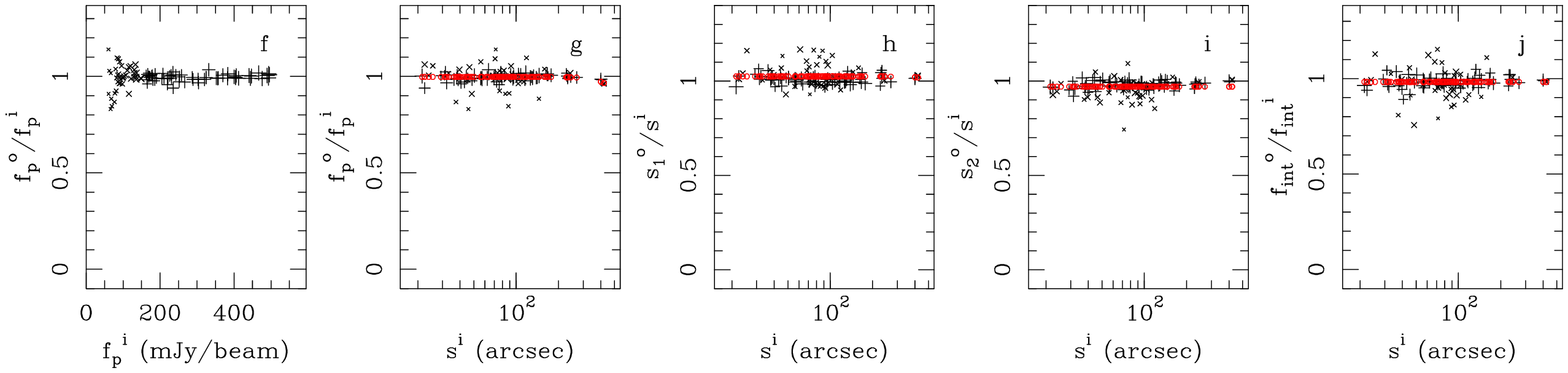}}}
\caption{Statistical properties of 100 artificial, circular, Gaussian sources 
inserted into the raw time signals before data reduction (\textbf{a} to 
\textbf{e}) or directly inserted into the final unsmoothed continuum emission 
map of Cha~I (\textbf{f} to \textbf{j}). \textbf{a} and \textbf{f}: 
ratio of output to input peak flux density as a function of input peak flux 
density. \textbf{b} and \textbf{g}: ratio of output to input peak flux density 
as a function of input size ($FWHM$). \textbf{c} and \textbf{h}: ratio of 
output major size to input size as a function of input size. \textbf{d} and 
\textbf{i}: ratio of output minor size to input size as a function of input 
size. \textbf{e} and \textbf{j}: ratio of output to input total flux as a 
function of input size. The size of the black symbols scales with the input 
peak flux density of the sources. In addition, crosses and plus symbols are 
for input peak 
flux densities below and above 150 mJy/19.2$\arcsec$-beam, respectively. The 
red circles in panels \textbf{b} to \textbf{e} and \textbf{g} to \textbf{j} 
show a fit to the data points using the arbitrary 3-parameter function 
$y = \log(\alpha/(1+(x/\beta)^\gamma))$.}
\label{f:filtering}
\end{figure*}

In a previous work, we characterized in detail the spatial filtering of LABOCA 
data due to the correlated noise removal and the high-pass filter
(Paper~I). Since these observations were performed using a 
combination of rectangular and spiral scanning patterns, the latter being more
compact (diameter 6$\arcmin$), the properties of the spatial filtering 
affecting the Cha~III dataset, that was obtained with rectangular scanning 
patterns only, could a priori differ: the correlated noise removal depends on
the size of the camera only, but the baseline removal is performed scan- 
(or subscan-)wise, i.e. on scales that can be larger than the camera, so it 
depends on the scanning pattern. We performed Monte Carlo simulations to 
evaluate the filtering properties following exactly the same procedure as 
described in Appendix~A of Paper~I, with four sets of 25 circular, 
artificial, Gaussian sources each. Figures~\ref{f:filtering}a--e and 
Table~\ref{t:filtering} present the results. As a control check, we show in 
Figs.~\ref{f:filtering}f--j the results of the fits performed on the 
artificial sources directly inserted in the final unsmoothed continuum 
emission map of Cha~III. 

To first order, the results listed in Table~\ref{t:filtering} are very similar 
to those obtained for Cha~I. The difference between strong and weak sources 
concerning the peak flux density is less pronounced for Cha~III than for Cha~I 
but this may not be statistically significant because the estimate for the weak 
artificial sources in Cha~III is based on very few extended sources: there are 
only 3 such sources between 130$\arcsec$ and 260$\arcsec$, two of them 
behaving like the sample of stronger sources and one being significantly more 
affected by the filtering (see the outlier point at (145$\arcsec$, 0.63) in 
Fig.~\ref{f:filtering}b). However, the size and integrated flux suffer more 
for Cha~III than for Cha~I. This is surprising since one would a priori expect 
the rectangular scanning pattern to be less sensitive to spatial filtering 
than the more compact spiral scanning pattern. The reasons for this difference 
are still unclear.  It may be due to a reduction of the number of well-working 
pixels in 2010 compared to 2007--2008: on average, the Cha~III scans had 
$224 \pm 2$ well-working pixels while the Cha~I scans had $232 \pm 18$. 
Alternatively, this could suggest that mixing rectangular and spiral patterns 
is more robust to spatial filtering than scanning with rectangular patterns 
only.

\begin{table}
 \caption{Filtering of artificial Gaussian sources due to the 
correlated noise removal depending on the input size ($FWHM$).}
 \label{t:filtering}
 \begin{tabular}{lccc}
  \hline\hline
  \multicolumn{1}{c}{Source sample} & \multicolumn{3}{c}{Size where} \\
  \cline{2-4}
  \noalign{\smallskip}
  \multicolumn{1}{c}{} & \multicolumn{1}{c}{$\frac{f_p^o}{f_p^i} = 85\%$} & \multicolumn{1}{c}{$\frac{\sqrt{s_1^os_2^o}}{s^i} = 85\%$} & \multicolumn{1}{c}{$\frac{f_{int}^o}{f_{int}^i} = 85\%$} \\
  \multicolumn{1}{c}{(1)} & \multicolumn{1}{c}{(2)} & \multicolumn{1}{c}{(3)} & \multicolumn{1}{c}{(4)} \\
  \hline\\[-1.5ex]
  \multicolumn{4}{l}{\textit{Circular sources:}} \\[0.3ex]
  strong sources & 230$\arcsec$ & 210$\arcsec$ & 140$\arcsec$ \\[-0.4ex]
 {\hspace*{2ex} \scriptsize ($>150$~mJy/beam)} & & & \\[0.6ex]
  weak sources   & 230$\arcsec$ & 110$\arcsec$ &  60$\arcsec$ \\[-0.4ex]
 {\hspace*{2ex} \scriptsize ($<150$~mJy/beam)} & & & \\[0.6ex]
  all sources    & 230$\arcsec$ & 180$\arcsec$ & 100$\arcsec$ \\
  \hline\\[-1.5ex]
 \end{tabular}
 \tablefoot{$f_p^o$ and $f_p^i$ are the output and input peak flux densities, 
$s^i$ the input size ($FWHM$), $s_1^o$ and $s_2^o$ the output sizes 
($FWHM$), and $f_{int}^o$ and $f_{int}^i$ the total output and input fluxes 
over an aperture of \textit{radius} equal to the input $FWHM$. 
}
\end{table}

\subsection{Convergence of the iterative data reduction}
\label{ss:convergence}

\begin{figure}
%\centerline{\resizebox{1.00\hsize}{!}{\includegraphics[angle=270]{/homes/belloche/Chamaeleon/Continuum/Cha3/Analysis/plot_totflux_artcha3cont.eps}}}
\centerline{\resizebox{1.00\hsize}{!}{\includegraphics[angle=270]{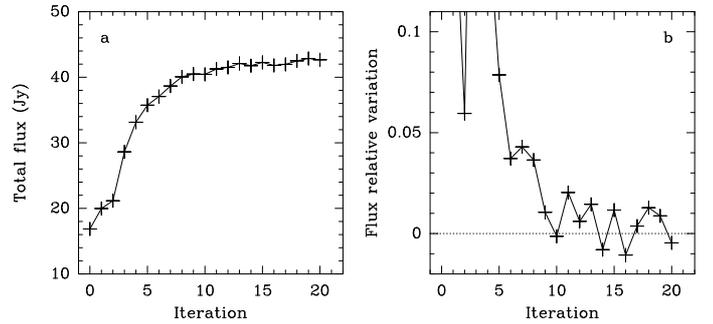}}}
\caption{\textbf{a} Convergence of the total 870~$\mu$m flux recovered in 
Cha~III by the iterative process of data reduction. \textbf{b} Relative flux 
variation between iterations $i$ and $i-1$ as a function of iteration number 
$i$.}
\label{f:total_flux}
\end{figure}

\begin{figure*}[!ht]
%\centerline{\resizebox{0.9\hsize}{!}{\includegraphics[angle=270]{/homes/belloche/Chamaeleon/Continuum/Cha3/Reduce/EvalFiltering/Figures/converg_artif2_artcha3cont.eps}}}
\centerline{\resizebox{0.9\hsize}{!}{\includegraphics[angle=270]{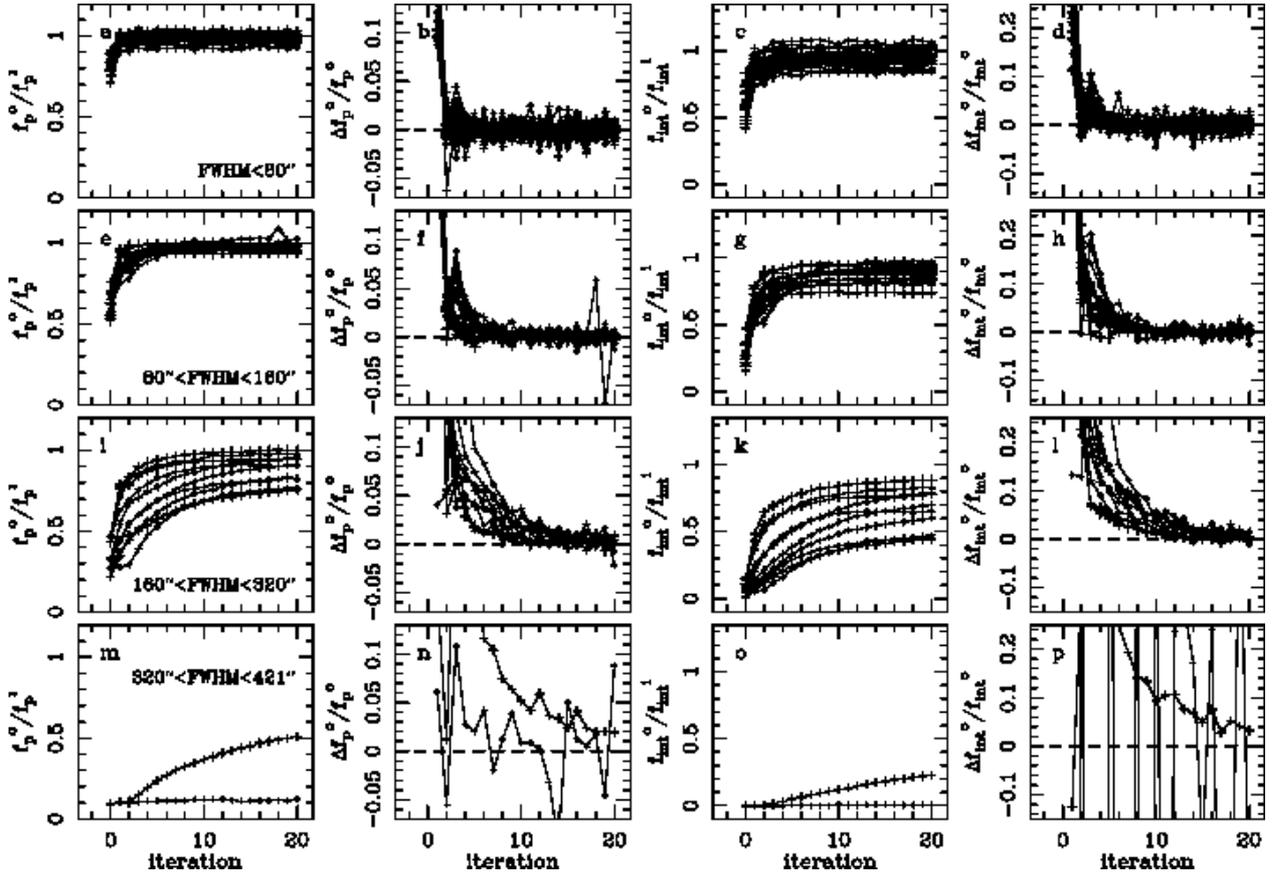}}}
\caption{Convergence of the iterative data reduction for artificial, circular, 
Gaussian sources with a peak flux density larger than 150 mJy~beam$^{-1}$. The 
first and second columns show the ratio of the output to the input peak flux 
densities, and its relative variations, respectively. The third and fourth 
columns show the same for the total flux. The different rows show sources of
different widths, as specified in each panel of the first column. The size of
the symbols increases with the input peak flux density. In panels \textbf{n} 
and \textbf{p}, the
noisier curve corresponds to a very large source with a peak flux twice as weak 
as the other one that has a similar size (see Figs.~\ref{f:filtering}a and b).}
\label{f:convergence}
\end{figure*}

Like for Cha~I, the iterative process of the data reduction improves the 
recovery of extended emission. Figure~\ref{f:total_flux} shows that the 
convergence of the total astronomical flux is reached within 1$\%$ at 
iteration 20. Figure~\ref{f:convergence} shows how the speed of convergence 
depends on the source size. The convergence in terms of peak flux density and 
integrated flux for the sources is faster for compact sources than for 
extended ones. For sources smaller than 320$\arcsec$, the convergence is 
reached within 1$\%$ in peak flux density and 2$\%$ in total flux. For the few 
larger sources, the convergence is reached within 2$\%$ and 4$\%$ at iteration 
20, respectively. These results are very similar to those obtained with the 
Cha~I dataset (see Paper~I).

\end{appendix}

\listofobjects

\end{document}